\documentclass[final,onefignum,onetabnum]{siamart220329}

\usepackage{braket,amsfonts,nccmath}
\usepackage{algorithm, algorithmic}
\usepackage{array}
\usepackage{mathtools}

\usepackage{subcaption}
\usepackage{pgfplots}

\newsiamthm{claim}{Claim}
\newsiamremark{remark}{Remark}
\newsiamremark{hypothesis}{Hypothesis}
\newsiamthm{assumption}{Assumption}
\crefname{hypothesis}{Hypothesis}{Hypotheses}

\DeclareMathOperator{\Tr}{Tr}
\DeclareMathOperator{\diag}{diag}

\usepackage{algorithmic}

\usepackage{graphicx,epstopdf}

\Crefname{ALC@unique}{Line}{Lines}

\usepackage{amsopn}

\usepackage{xspace}
\usepackage{bold-extra}
\usepackage[most]{tcolorbox}

\colorlet{texcscolor}{blue!50!black}
\colorlet{texemcolor}{red!70!black}
\colorlet{texpreamble}{red!70!black}
\colorlet{codebackground}{black!25!white!25}

\lstdefinestyle{siamlatex}{%
  style=tcblatex,
  texcsstyle=*\color{texcscolor},
  texcsstyle=[2]\color{texemcolor},
  keywordstyle=[2]\color{texemcolor},
  moretexcs={cref,Cref,maketitle,mathcal,text,headers,email,url},
}

\tcbset{%
  colframe=black!75!white!75,
  coltitle=white,
  colback=codebackground, 
  colbacklower=white, 
  fonttitle=\bfseries,
  arc=0pt,outer arc=0pt,
  top=1pt,bottom=1pt,left=1mm,right=1mm,middle=1mm,boxsep=1mm,
  leftrule=0.3mm,rightrule=0.3mm,toprule=0.3mm,bottomrule=0.3mm,
  listing options={style=siamlatex}
}

\newtcblisting[use counter=example]{example}[2][]{%
  title={Example~\thetcbcounter: #2},#1}

\newtcbinputlisting[use counter=example]{\examplefile}[3][]{%
  title={Example~\thetcbcounter: #2},listing file={#3},#1}

\DeclareTotalTCBox{\code}{ v O{} }
{ 
  fontupper=\ttfamily\color{black},
  nobeforeafter,
  tcbox raise base,
  colback=codebackground,colframe=white,
  top=0pt,bottom=0pt,left=0mm,right=0mm,
  leftrule=0pt,rightrule=0pt,toprule=0mm,bottomrule=0mm,
  boxsep=0.5mm,
  #2}{#1}

\patchcmd\newpage{\vfil}{}{}{}
\begin{tcbverbatimwrite}{tmp_\jobname_header.tex}
\title{Learning Zero-Sum Linear Quadratic Games with Improved Sample Complexity and Last-Iterate Convergence\thanks{Accepted to SIAM Journal on Control and Optimization on June 12, 2025. A preliminary version of this work appeared in the Conference on Decision and Control 2023 \cite{wu-et-al23cdc}. In this version, we have included new and stronger results on last-iterate convergence (Section~\ref{subsec:last-it-cv}).
\funding{The work is supported by ETH AI Center doctoral fellowship, ETH Foundations of Data Science (ETH-FDS), ETH Research Grant funded via ETH Zurich Foundation, NCCR Automation, and Swiss National Science Foundation (SNSF) Project Funding No. 200021-207343.}}}

\author{Jiduan Wu\thanks{The authors are affiliated with the Department of Computer Science, ETH Z\"urich, Switzerland. Correspondence to: J. W. (jiduwu@student.ethz.ch).
}
\and Anas Barakat\footnotemark[2] \and Ilyas Fatkhullin\footnotemark[2] \and Niao He\footnotemark[2] 
}
\headers{Learning Zero-Sum Linear Quadratic Games}{Jiduan Wu, Anas Barakat, Ilyas Fatkhullin, and Niao He}
\end{tcbverbatimwrite}
\input{tmp_\jobname_header.tex}

\ifpdf
\hypersetup{ pdftitle={Learning in Zero-Sum Linear Quadratic Games with Last-Iterate Convergence} }
\fi

\begin{document}
\maketitle

\begin{tcbverbatimwrite}{tmp_\jobname_abstract.tex}
\begin{abstract}
Zero-sum Linear Quadratic (LQ) games are fundamental in optimal control and can be used (i)~as a dynamic game formulation for risk-sensitive or robust control and (ii)~as a benchmark setting for multi-agent reinforcement learning with two competing agents in continuous state-control spaces. In contrast to the well-studied single-agent linear quadratic regulator problem, zero-sum LQ games entail solving a challenging nonconvex-nonconcave min-max problem with an objective function that lacks coercivity. Recently, Zhang et al.~\cite{kaiqing_finite_horizon} showed that an~$\epsilon$-Nash equilibrium (NE) of finite horizon zero-sum LQ games can be learned via nested model-free Natural Policy Gradient (NPG) algorithms with poly$(1/\epsilon)$ sample complexity. In this work, we propose a simpler nested Zeroth-Order (ZO) algorithm improving sample complexity by several orders of magnitude and guaranteeing convergence of the last iterate. Our main results are two-fold: (i) in the deterministic setting, we establish the first global last-iterate linear convergence result for the nested algorithm that seeks NE of zero-sum LQ games; (ii) in the model-free setting, we establish a~$\widetilde{\mathcal{O}}(\epsilon^{-2})$ sample complexity using a single-point ZO estimator. For our last-iterate convergence results, our analysis leverages the Implicit Regularization (IR) property and a new gradient domination condition for the primal function. Our key improvements in the sample complexity rely on a more sample-efficient nested algorithm design and a finer control of the ZO natural gradient estimation error utilizing the structure endowed by the finite-horizon setting.

\end{abstract}



\end{tcbverbatimwrite}
\input{tmp_\jobname_abstract.tex}

\section{Introduction}
\label{sec:introduction}

While policy optimization has a long history in control for unknown and parameterized system models (see for e.g., \cite{long_history}), recent successes in reinforcement learning and continuous control tasks have renewed the interest in direct policy search thanks to its flexibility and scalability to high-dimensional problems. Despite these desirable features, theoretical guarantees for policy gradient methods have remained elusive until very recently because of the nonconvexity of the induced optimization landscape. In particular, in contrast to control-theoretic approaches which are often model-based and estimate the system dynamics first before designing optimal controllers, the computational and sample complexities of model-free policy gradient methods were only recently analyzed. We refer the interested reader to a nice recent survey on policy optimization methods for learning control policies~\cite{review_hu}. For instance, while the classic Linear Quadratic Regulator~(LQR) problem induces a nonconvex optimization problem over the set of stable control gain matrices, the gradient domination property~\cite{Polyak_1963} and the coercivity of the cost function respectively allow to derive global convergence to optimal policies for policy gradient methods and ensure stable feedback policies at each iteration~\cite{global_con_pg_lq}. 
As exact gradients are often unavailable when system dynamics are unknown, derivative-free optimization techniques using cost values have been employed to design model-free policy gradient methods to solve LQR problems~\cite{global_con_pg_lq}.
Alternative approaches to solve LQR include system identification \cite{Fiechter_1997}, iterative solution of Algebraic Riccati Equation~\cite{hewer_algorithm, Lancaster_ARE_algebraic_solution} and convex semi-definite program formulations~\cite{Balakrishnan_SDP_LTI_2003}. However, such methods are not easily adaptable to the simulation-based model-free setting. 

Besides the desired stability constraint, other requirements such as robustness and risk sensitivity constraints also play an important role in the design of controllers for safety-critical control systems. Indeed, system perturbations, modeling imprecision, and adversarial uncertainty are ubiquitous in control systems and may lead to severe degradation in performance \cite{bhattacharyya_robust_control_intro, Campi_nonlinear_risk_sensitive_control}.
 Robustness constraints can be incorporated into control design via different approaches including using statistical models for disturbances such as for linear quadratic Gaussian design, adopting a game theory perspective via designing  `minimax' controllers and incorporating an~$\mathcal{H}_{\infty}$ norm bound of input-output operators as in~$\mathcal{H}_{\infty}$ control~\cite{linear_sol}. Classical linear models for robust control include the LQ disturbances attenuation problem and the linear exponential quadratic Gaussian problem which are well-known to be equivalent to zero-sum LQ games \cite{linear_sol, infinite_horizon, Mageirou_Decentralized_1977_first}. Besides its relevance for robust control problem formulation, zero-sum LQ games also constitute a benchmark problem for multi-agent continuous control problems involving two competing agents.  
 However, solving this problem faces (at least) two distinct challenges 
 requiring to deal with
 (a)~a constrained nonconvex-nonconcave problem
 and~(b)~lack of coercivity, unlike for the classic LQR problem for which descent over the objective ensures feasibility and stability of the iterates during learning.

 While the formulation of zero-sum LQ games dates back at least to the seventies~\cite{Mageirou_Decentralized_1977_first}\footnote{This formulation is under the continuous-time setting.}, the sample complexity analysis of model-free policy gradient algorithms solving this problem was only recently explored in the literature \cite{kaiqing_finite_horizon}. 
 More precisely, Zhang et al.~\cite{kaiqing_finite_horizon} 
 showed that an~$\epsilon$-NE of finite horizon zero-sum LQ games can be learned via  
 nested model-free Natural Policy Gradient (NPG) algorithms with polynomial sample complexity in~$1/\epsilon$. 
 Interestingly, 
 the aforementioned 
 algorithms enjoy an Implicit Regularization (IR) property which maintains the robustness of the controllers during learning~\cite{kaiqing_finite_horizon,infinite_horizon}.
 In particular, the iterates of the algorithms are guaranteed to stay in some feasible set where the worst-case cost is finite without using any explicit regularization or projection operation.  
In the present work, we show that we can actually achieve last-iterate linear convergence in terms of the objective function value gap. We further show that  significantly fewer samples are required to guarantee the IR property while only having access to zeroth-order information.  
Our contributions can be summarized as follows:

\paragraph{Contributions} We establish a last-iterate linear convergence result for our algorithm in the deterministic setting for finite horizon zero-sum LQ games\footnote{We do not address the infinite horizon setting which requires to investigate the additional question of stability.}. To achieve this result, we prove that the primal function satisfies a gradient domination inequality. To the best of our knowledge, this is the first global last-iterate linear convergence result using the objective function-value gap for zero-sum LQ games. In the stochastic setting, our main result states that our derivative-free nested policy gradient algorithm requires~$\widetilde{\mathcal{O}}(\epsilon^{-2})$ samples to reach an~$\epsilon$-NE\footnote{The $\epsilon$-NE in this work is defined using the last-iterate cost difference while \cite{kaiqing_finite_horizon} used average gradient norms, see Remark \ref{remark:total_sample_complexity} for a more detailed comparison.} compared to $\widetilde{\mathcal{O}}(\epsilon^{-9})$ in \cite{kaiqing_finite_horizon}.
We also show that our algorithm enjoys the IR property upon choosing adequate values for zeroth-order estimation parameters. In particular, our requirements in terms of batch sizes and perturbation radius are less restrictive compared to prior work~\cite{kaiqing_finite_horizon}. 
 Our sample complexity improvement follows from~(a) a simpler algorithm design reducing the number of calls to the inner-loop maximizing procedure, (b) a better sample complexity to solve the inner maximization problem and (c)~an improved sample complexity for solving the resulting minimization problem in our outer-loop procedure using a careful decomposition of the estimation error caused by policy gradient estimation.

\paragraph{Paper organization} 
The rest of this paper is structured as follows. In Section~\ref{sec:related_work}, we discuss related work. In Section~\ref{sec:preliminary}, we introduce the stochastic zero-sum LQ games problem together with useful background. We present our model-free nested natural policy gradient algorithm to solve the problem in Section~\ref{sec:algos} and Section~\ref{sec:results} presents our main results along with a proof sketch to highlight the key steps leading to our last-iterate convergence results in both deterministic and stochastic settings.
We conclude this paper with possible future directions. 

\section{Related work}
\label{sec:related_work}

\paragraph{Policy optimization for LQ problems} Compared to zero-sum LQ games, policy optimization for single-agent LQ problems is a well-understood topic. Theoretical guarantees for model-based and model-free algorithms searching for the optimal policy were established in~\cite{global_con_pg_lq} for the discrete-time infinite-horizon setting. 
Several subsequent works improved over the polynomial sample complexity in \cite{global_con_pg_lq} 
using single and two-point ZO estimation~\cite{local_lipschitz,Mohammadi_linear_2p_discrete}. Additionally, the LQ model has been studied under different settings including finite-horizon~\cite{Hambly_PG_LQR_Finite_Horizon_2021} and continuous-time~\cite{Ilyas_lqr,sample_complexity_improvement_continuous_single, Giegrich_2022_LQ_finite_horizon_continuous_time}. First-order methods have also been recently investigated for solving LQR \cite{Ju_LQR_FO_AC_2023,Yang_Actor_Critic_2019}. Bu et al.~\cite{Bu_indefinite_2020} provided convergence analysis for 
possibly indefinite infinite-horizon LQR problems. Guo et al.~\cite{Guo_2022_global_Goldstein} designed Goldstein subdifferential algorithms to solve the nonsmooth $\mathcal{H}_{\infty}$ control problem and 
left sample complexity analysis in the model-free setting as an important future direction. Other related problems include Markovian jump systems~\cite{Sun_Fazel_Global_Convex_Parametrization_2021}, output control design~\cite{Furieri_Distributed_Output_Local_PL_2020,Ilyas_lqr,Zhao_Global_Output_2022}, decentralized control~\cite{Feng_Decentralized_2019,Li_Distributed_RL_2020}, receding-horizon policy gradient methods~\cite{xiangyuan_2023_LQR_RHPG}, and nonlinear dynamics~\cite{Han_Nonlinear_LQR}.

\paragraph{Zero-sum LQ games} 
Recent research efforts have
been devoted to studying the more challenging zero-sum LQ games problem~\cite{infinite_horizon,LQ_PG_NE,improved_kaiqing,kaiqing_finite_horizon}. Zhang et al.~\cite{LQ_PG_NE} proposed projected\footnote{The projection step was not implemented in their simulations, yet essential for theoretical analysis.} nested gradient-based algorithms with global sublinear and local linear convergence rates for infinite-horizon zero-sum LQ games. Later, Bu et al.~\cite{improved_kaiqing} removed the projection step, but their analysis requires access to the exact solution of the inner maximization problem and cannot be easily extended to the model-free case. Zhang et al.~\cite{infinite_horizon, kaiqing_finite_horizon} then introduced a nested natural gradient-based algorithm that ensures the IR property in the model-based case, where they utilized the equivalent zero-sum game formulation and designed model-free algorithms without sample complexity analysis. In the model-free setting, Al-Tamimi et al.~\cite{Al_Tamimi_ZsLQR_Qlearning_2007} proposed a Q-learning-based method to solve zero-sum LQ games without providing a sample complexity analysis.

\paragraph{Other LQ formulations} It is well-known that mixed $\mathcal{H}_2/\mathcal{H}_{\infty}$ problems can be formulated as risk-sensitive control problems or zero-sum dynamical games~\cite{glover1988state_mix_design_risk_sensitive_connection}, and the solutions of these two classes of problems oftentimes inspire each other~\cite{infinite_horizon}. A comprehensive discussion of the connection among them can be found in~\cite{infinite_horizon, linear_sol}. Borrowing ideas from robust control theory, Zhang et al.~\cite{zhang2020stability_LQ_case_study} identified the stability issue of the robust adversarial reinforcement learning problem on LQ systems and proposed a double-loop algorithm using proper initialization as a solution. For LQ Gaussian control, 
Cui et al.~\cite{cui2023reinforcement} designed a different dual-loop algorithm based on approximately solving a Generalized Algebraic Riccati Equation (GARE). Their algorithm enjoys a last-iterate linear convergence in the deterministic case and the continuous-time counterpart was studied in~\cite{cui2022mixed, molumixed_improved}. In the present paper, we focus on the nested natural gradient algorithm and reveal more insights into its convergence and sample complexity properties. Recently, an $N$-player general-sum game formulation of LQR was studied in \cite{mazumdar_LQG_no_guarantee,Hambly_PG_N_Player_LQR_2022, Yang_Phd_LQG_N_Player}. However, such a problem in the 2-player case is different from our zero-sum formulation. In the context of mean-field games, counterparts of LQR and zero-sum LQ games were developed in~\cite{lq_zo_mfg_pg, Carmona_LQR_MFRL_2019}.

\section{Preliminaries}
\label{sec:preliminary}

\paragraph{Notations} 
For any matrix~$M\in\mathbb R^{n\times n}$ where~$n$ is a nonzero integer, we denote by~$M^{\top}$ its transpose and~$\Tr(M)$ its trace. We use the notations~$\|M\|$ and $\|M\|_F$ for its  operator and Frobenius norms respectively. The spectral radius of a matrix~$M$ is denoted by~$\rho(M)$ and a matrix is said to be (Schur) stable if $\rho(M)<1$, i.e., all the absolute values of the eigenvalues of the matrix~$M$ are (strictly) smaller than~$1$. The smallest eigenvalue of a symmetric matrix~$M$ is denoted by~$\lambda_{\min}(M)$. For~$N$ diagonal matrices~$X_i$ for~$i \in \{0, \cdots N-1\}$ for some integer~$N \geq 1$, the block-diagonal matrix with diagonal entries $X_0,\cdots,X_{N-1}$ is denoted by~$\diag(X_{0-(N-1)})$. The uniform distribution over a measurable compact subset~$S$ of~$\mathbb{R}^n$ is denoted as~$\text{Unif}(S)$.

\paragraph{Stochastic zero-sum linear quadratic dynamic games}
We consider a zero-sum LQ game~(following the exposition in~\cite{kaiqing_finite_horizon}) where the system state 
evolves as follows: 
\begin{align}
\label{eq:system-dynamics}
x_{h+1}=A_hx_h+B_hu_h+D_hw_h+\xi_h, \, h \in \{0, \cdots, N-1\},
\end{align}
where~$N$ is a finite nonzero horizon, $x_0 \in \mathbb{R}^m$ is an initial random state and where for any stage~$h \in \{0, \cdots, N-1\},$ $x_h\in\mathbb{R}^m$ is the system state, $u_h \in\mathbb{R}^d$ and~$w_h \in \mathbb{R}^n$ are the control inputs of the min and max players respectively\footnote{These controls depend on the history of state-control pairs at each time step~$h$ for now, stationary control policies will be sufficient as will be mentioned later on.} and~$\xi_h$ is a random variable describing noisy perturbations to the system while~$A_h, B_h, D_h$ are (possibly) time-dependent system matrices with appropriate dimensions.

\begin{assumption}
\label{assumption:noises}
    The initial state~$x_0$ and the noise~$\xi_h$ for $h\in\{0,\cdots,N-1\}$ are independent random variables following a distribution with zero-mean and  positive-definite covariance. Moreover, there exists a positive scalar $\vartheta$ such that for all $h \in\{0,\cdots,N-1\},$ $\|x_0\| \leq\vartheta$ and~$\|\xi_h\|\leq\vartheta$ almost surely.\footnote{The almost sure boundedness can be relaxed when considering sub-Gaussian distributions as noticed in prior work \cite{local_lipschitz, Furieri_Distributed_2020}.}
\end{assumption}
Our goal is to solve the following zero-sum game:
\begin{equation}
    \label{eq:pb-zslqr}
    \underset{(u_h)}{\inf}\, \underset{(w_h)}{\sup} \mathbb{E}_{\pmb{\xi}}\Biggl[\sum_{h=0}^{N-1}(x_h^{\top} Q_h x_h+u_h^{\top}R_h^u u_h
    -w_h^{\top}R_h^w w_h) + c_N\Biggr]
\end{equation}
where~$c_N\coloneqq x_N^{\top}Q_N x_N$ and $\pmb{\xi}\coloneqq\begin{bmatrix*}
        x_0^{\top},\xi_0^{\top},\cdots,\xi_{N-1}^{\top}
        \end{bmatrix*}^{\top}$ 
and the system states follow the linear time-varying system dynamics described in~\eqref{eq:system-dynamics} and for every~$h \in \{0, \cdots, N-1\}, Q_h \succeq 0, R_h^{u}, R_h^{w} \succ 0$ are symmetric matrices defining the quadratic objective. 
In view of our robust control motivation, the two players can be seen as a min controller and a max disturbance. Under standard assumptions \cite{kaiqing_finite_horizon,linear_sol} reported in Appendix~\ref{appendix:basic_lemmas_LQ} in details (see also Remark~\ref{rem:existence-condition}), the saddle-point control policies solving~\eqref{eq:pb-zslqr} are unique and have the linear state-feedback form. Thus, we can restrict our search to gain matrices~$K_h \in \mathbb{R}^{d\times m}$ and~$L_h \in \mathbb{R}^{n\times m}$ such that the controls are given by~$u_h = - K_h x_h, w_h = - L_h x_h$ for~$h \in \{0, \cdots, N-1\}$. Therefore, we will mainly focus on solving the following min-max policy optimization problem resulting from~\eqref{eq:pb-zslqr}: 
\begin{align}
\label{eq:pb-minmax}
\min_{(K_h)} \max_{(L_h)} \mathbb{E}_{\pmb{\xi}}\Biggl[\sum_{h=0}^{N-1} &x_h^{\top} M_h x_h +c_N\Biggr]\,,
\end{align}
where $M_h\coloneqq Q_h+K_h^{\top}R_h^u K_h - L_h^{\top}R_h^w L_h$ and the system state follows the dynamics~$x_{h+1} = (A_h - B_h K_h - D_h L_h) x_h + \xi_h$ for~$h \in \{0, \cdots, N-1\}\,.$

\paragraph{Compact reformulation} To simplify the exposition and our analysis, we rewrite problem~\eqref{eq:pb-minmax} under a more compact form following the reformulation proposed in~\cite{kaiqing_finite_horizon}. Consider the following notations: 
\begin{align}
    &\pmb{x} := [x_0^{\top}, \cdots, x_N^{\top}]^{\top}, 
    \pmb{u} := [u_0^{\top}, \cdots, u_{N-1}^{\top}]^{\top},\pmb{w} := [w_0^{\top}, \cdots, w_{N-1}^{\top}]^{\top},\notag\\
    &\pmb{\xi} := [x_0^{\top}, \xi_0^{\top}, \cdots, \xi_{N-1}^{\top}]^{\top},\,\pmb{A}:=\begin{bmatrix*}
        \pmb{0}_{m\times mN}&\pmb{0}_{m\times m}\\
        \diag(A_{0-(N-1)})&\pmb{0}_{mN\times m}
    \end{bmatrix*},\,\notag\\
    &\pmb{B}:=\begin{bmatrix*}
        \pmb{0}_{m\times dN}\\
        \diag(B_{0-(N-1)})
    \end{bmatrix*},\,\pmb{D}:=\begin{bmatrix*}
        \pmb{0}_{m\times nN}\\
        \diag(D_{0-(N-1)})
    \end{bmatrix*},\notag\\
     &\pmb{Q}:=\diag(Q_{0-N}),\,\pmb{R}^u:=\diag(R_{0-(N-1)}^u),\,\pmb{R}^w:=\diag(R_{0-(N-1)}^w),  \notag\\
    &\pmb{K}:=\begin{bmatrix*}
        \diag(K_{0-(N-1)})&\pmb{0}_{dN\times m}
    \end{bmatrix*},
    \pmb{L}:=\begin{bmatrix*}
        \diag(L_{0-(N-1)})&\pmb{0}_{nN\times m}
    \end{bmatrix*}.\label{def:k_l}
\end{align}
We denote by~$\mathcal{S}_1 \subset \mathbb{R}^{dN\times m(N+1)}$ and~$\mathcal{S}_2 \subset \mathbb{R}^{nN\times m(N+1)}$ the matrix subspaces induced by the sparsity patterns described in~\eqref{def:k_l} for the gain matrices~$\pmb{K}$ and~$\pmb{L}$ respectively. The subspaces $\mathcal{S}_1$, $\mathcal{S}_2$ where we search for a NE solution~$(\pmb{K}^*,\pmb{L}^*)$, are of dimensions $d_{\pmb{K}}\coloneqq dmN$ and $d_{\pmb{L}}\coloneqq nmN$ respectively. Then, problem~\eqref{eq:pb-minmax} can be rewritten as:
\begin{equation}
\label{eq:pb-minmax_compact}
\min_{\pmb{K} \in \mathcal{S}_1} \max_{\pmb{L} \in \mathcal{S}_2}  \mathcal{G}(\pmb{K},\pmb{L}):=   \mathbb{E}_{\pmb{\xi}} [ \pmb{x}^{\top} (\pmb{Q} + \pmb{K}^{\top} \pmb{R}^{u} \pmb{K} - \pmb{L}^{\top} \pmb{R}^{w} \pmb{L} )\, \pmb{x} ]\,,
\end{equation}
where the transition dynamics are described by~$\pmb{x} = \pmb{A}\pmb{x} + \pmb{B}\pmb{u} + \pmb{D}\pmb{w} + \pmb{\xi} = \pmb{A}_{\pmb{K},\pmb{L}}\pmb{x} + \pmb{\xi}$ where $\pmb{A}_{\pmb{K},\pmb{L}}\coloneqq\pmb{A}-\pmb{BK}-\pmb{DL}$. Notice that our search for gain matrices~$\pmb{K}, \pmb{L}$ is restricted to the matrices of the form described in~\eqref{def:k_l} as this set of sparse matrices is sufficient to find the NE we are looking for. For any gain matrices~$\pmb{K}$ and~$\pmb{L}$, we can rewrite the objective function value~$\mathcal{G}(\pmb{K}, \pmb{L})$ as follows:
\begin{align*}
 \mathcal{G}(\pmb{K},\pmb{L}) &=  \mathbb{E}_{\pmb{\xi}}[\mathcal{G}_{\pmb{\xi}}(\pmb{K},\pmb{L})]=\Tr(\pmb{P}_{\pmb{K},\pmb{L}}\pmb{\Sigma}_0)=\Tr\bigl((\pmb{Q}+\pmb{K}^{\top}\pmb{R}^u\pmb{K}-\pmb{L}^{\top}\pmb{R}^w\pmb{L})\pmb{\Sigma}_{\pmb{K},\pmb{L}}\bigr),
\end{align*}
where~$\mathcal{G}_{\pmb{\xi}}(\pmb{K},\pmb{L}) := \pmb{\xi}^{\top}\pmb{P}_{\pmb{K},\pmb{L}}\pmb{\xi}$, $\pmb{\Sigma}_0 := \mathbb{E}_{\pmb{\xi}}[ \pmb{\xi} \pmb{\xi}^{\top}] \succ 0$ (see Assumption~\ref{assumption:noises}) and let $\phi\coloneqq \lambda_{\min}(\pmb{\Sigma}_0)$. Matrices~$\pmb{P}_{\pmb{K},\pmb{L}}$, $\pmb{\Sigma}_{\pmb{K},\pmb{L}} := \mathbb{E}_{\pmb{\xi}}[\diag(x_0 x_0^{\top}, \cdots, x_N x_N^{\top})]$ are the unique solutions to the recursive Lyapunov equations 
\begin{eqnarray}
    \label{eq:lyapunov1}
    \pmb{P}_{\pmb{K},\pmb{L}}&=&\pmb{A}_{\pmb{K},\pmb{L}}^{\top}\pmb{P}_{\pmb{K},\pmb{L}}\pmb{A}_{\pmb{K},\pmb{L}}  +\pmb{Q}+\pmb{K}^{\top}\pmb{R}^u\pmb{K}-\pmb{L}^{\top}\pmb{R}^w\pmb{L}, \\
\pmb{\Sigma}_{\pmb{K},\pmb{L}}&=&\pmb{A}_{\pmb{K},\pmb{L}}\pmb{\Sigma}_{\pmb{K},\pmb{L}}\pmb{A}_{\pmb{K},\pmb{L}}^{\top} +\pmb{\Sigma}_0\,. \label{eq:lyapunov2}
\end{eqnarray}
The objective~$\mathcal{G}(\pmb{K},\pmb{L})$ is nonconvex-nonconcave in general (see Lemma~3.1 in \cite{kaiqing_finite_horizon}). 
We define some quantities that will be frequently used: $\pmb{A}_{\pmb{K}}\coloneqq \pmb{A}-\pmb{BK}$, $\pmb{A}_{\pmb{L}}\coloneqq \pmb{A}-\pmb{DL}$.
\paragraph{Policy gradients} The gradients of~$\mathcal{G}$ w.r.t. $\pmb{K},\pmb{L}$ are given in the following:
\begin{align}
    \label{pg:K-L}
    &\nabla_{\pmb{K}}\mathcal{G}(\pmb{K},\pmb{L}) =2\pmb{F}_{\pmb{K},\pmb{L}}\pmb{\Sigma}_{\pmb{K},\pmb{L}}\,,\, \pmb{F}_{\pmb{K},\pmb{L}} \coloneqq (\pmb{R}^u+\pmb{B}^{\top}\pmb{P}_{\pmb{K},\pmb{L}}\pmb{B})\pmb{K}-\pmb{B}^{\top}\pmb{P}_{\pmb{K},\pmb{L}}\pmb{A}_{\pmb{L}}\,,\\ &\nabla_{\pmb{L}}\mathcal{G}(\pmb{K},\pmb{L})=2\pmb{E}_{\pmb{K},\pmb{L}}\pmb{\Sigma}_{\pmb{K},\pmb{L}}\,,\,\pmb{E}_{\pmb{K},\pmb{L}} \coloneqq (-\pmb{R}^w+\pmb{D}^{\top}\pmb{P}_{\pmb{K},\pmb{L}}\pmb{D})\pmb{L}-\pmb{D}^{\top}\pmb{P}_{\pmb{K},\pmb{L}}\pmb{A}_{\pmb{K}}\,.
\end{align}

\begin{remark} 
\label{rem:existence-condition}
If $\pmb{P}_{\pmb{K},\pmb{L}}\succeq 0$ and $\pmb{R}^w-\pmb{D}^{\top}\pmb{P}_{\pmb{K},\pmb{L}}\pmb{D} \succ 0$ for a stationary point~$(\pmb{K},\pmb{L})$ of~$\mathcal{G}$, then this stationary point is the unique NE of the game (see Lemma 3.2 in \cite{kaiqing_finite_horizon}).
\end{remark}

\section{Nested Derivative-Free Natural Policy Gradient (NPG) Algorithm}
\label{sec:algos}

In this section, we present our model-free and derivative-free nested NPG algorithm inspired by the recent work~\cite{kaiqing_finite_horizon}.

\subsection{Exact nested NPG algorithm}
\label{subsec:exact-nested-npg}

To prepare the stage for the model-free setting, we briefly introduce the nested NPG algorithm in the deterministic setting, i.e., when we have access to the policy gradients w.r.t. both control variables~$\pmb{K}, \pmb{L}$ as reported in~\eqref{pg:K-L}. This algorithm was considered for example in~\cite{kaiqing_finite_horizon} and we follow a similar exposition in this subsection. We first solve the inner maximization problem in~\eqref{eq:pb-minmax_compact} for any fixed control gain matrix~$\pmb{K}\in\mathcal{S}_1$ to obtain an approximate $\pmb{L}\in\mathcal{S}_2$ of the exact solution $\pmb{L}(\pmb{K})$ before solving the outer-loop minimization problem with the resulting objective~$\mathcal{G}(\pmb{K},\pmb{L})$. The following proposition that we report here from Lemma 3.3 in \cite{kaiqing_finite_horizon} guarantees that there exists a unique solution~$\pmb{L}(\pmb{K})$ to the inner maximization problem whenever the control gain matrix~$\pmb{K}$ lies in a set which is known to contain the optimal control gain matrix solving the min-max problem. 
\begin{lemma}
\label{lemma:optimality}
    (Inner-loop well-definedness 
    \cite{kaiqing_finite_horizon}) 
    Consider the Riccati equation 
    \begin{align}
    \label{eq:riccati}
    \pmb{P}_{\pmb{K},\pmb{L}(\pmb{K})}&=\pmb{Q}+\pmb{K}^{\top}\pmb{R}^u\pmb{K}+\pmb{A}_{\pmb{K}}^{\top}\widetilde{\pmb{P}}_{\pmb{K},\pmb{L}(\pmb{K})}\pmb{A}_{\pmb{K}},
\end{align}
where~$\widetilde{\pmb{P}}_{\pmb{K},\pmb{L}(\pmb{K})}\coloneqq \pmb{P}_{\pmb{K},\pmb{L}(\pmb{K})}
    +\pmb{P}_{\pmb{K},\pmb{L}(\pmb{K})}\pmb{D}(\pmb{R}^w-\pmb{D}^{\top}\pmb{P}_{\pmb{K},\pmb{L}(\pmb{K})}\pmb{D})^{-1}\pmb{D}^{\top}\pmb{P}_{\pmb{K},\pmb{L}(\pmb{K})}$,
and define the set 
\begin{align}
    \label{eq:def-K}
    \mathcal{K}\coloneqq \bigl\{\pmb{K}\in\mathcal{S}_1\mid 
    \pmb{P}_{\pmb{K},\pmb{L}(\pmb{K})}\succeq 0, \pmb{R}^w-\pmb{D}^{\top}\pmb{P}_{\pmb{K},\pmb{L}(\pmb{K})}\pmb{D}\succ 0\bigr\}\,.
\end{align} where~$\pmb{P}_{\pmb{K},\pmb{L}(\pmb{K})}$ is a solution to~\eqref{eq:riccati} when it exists. 
Then, for any~$\pmb{K} \in \mathcal{K}$
, there exists a unique solution~$\pmb{L}(\pmb{K})\in\mathcal{S}_2$\footnote{It can be shown that $\pmb{L}(\pmb{K})$ always lies in $\mathcal{S}_2$ via simple calculations using the sparsity pattern.} to the inner maximization problem in~(\ref{eq:pb-minmax_compact}) given by $
    \pmb{L}(\pmb{K})=(-\pmb{R}^w+\pmb{D}^{\top}\pmb{P}_{\pmb{K},\pmb{L}(\pmb{K})}\pmb{D})^{-1}\pmb{D}^{\top}\pmb{P}_{\pmb{K},\pmb{L}(\pmb{K})}\pmb{A}_{\pmb{K}}.$
Moreover, for any~$\pmb{K}\in\mathcal{K}$ and any~$\pmb{L} \in \mathcal{S}_2$, $\pmb{P}_{\pmb{K},\pmb{L}}\preceq \pmb{P}_{\pmb{K},\pmb{L}(\pmb{K})}.$ 
\end{lemma}

We are now ready to introduce the nested NPG algorithm which can be written as follows using positive step-sizes~$\tau_1, \tau_2$ for the inner and outer loops respectively: 
\noindent\begin{minipage}{0.5\textwidth}
\begin{equation}
\pmb{L}_{k+1}=\pmb{L}_k+\tau_1\pmb{E}_{\pmb{K}_t,\pmb{L}_k},\label{eq:det-inner-loop-npg} \text{(inner)}
\end{equation} 
    \end{minipage}
    and 
    \begin{minipage}{0.5\textwidth}
\begin{equation}
\pmb{K}_{t+1}=\pmb{K}_t-\tau_2 \pmb{F}_{\pmb{K}_t,\pmb{L}_{T_{in}}},\label{eq:det-outer-loop-npg} \text{(outer)} 
\end{equation}
    \end{minipage}
where $k= 0,\ldots,T_{in}-1$, $t= 0,\ldots,T-1$ and 
$T_{in},T$ are the inner and outer loop iteration counts respectively.  The output~$\pmb{L}_{T_{in}}$ of the inner loop satisfies $\mathcal{G}(\pmb{K}_t,\pmb{L}(\pmb{K}_t))-\mathcal{G}(\pmb{K}_t,\pmb{L}_{T_{in}})\leq\epsilon_1$, $\|\pmb{L}(\pmb{K}_t)-\pmb{L}_{T_{in}}\|_F \leq \sqrt{\lambda_{\min}^{-1}(\pmb{H}_{\pmb{K}_t,\pmb{L}(\pmb{K}_t)})\cdot \epsilon_1}$~\cite{kaiqing_finite_horizon} where~$\pmb{H}_{\pmb{K},\pmb{L}} \coloneqq \pmb{R}^w-\pmb{D}^{\top}\pmb{P}_{\pmb{K},\pmb{L}}\pmb{D}$.

\subsection{Derivative-free nested NPG algorithm} 

In this subsection, we describe our algorithm to solve problem~\eqref{eq:pb-minmax_compact} in the model-free setting where we do not have access to exact gradients. In this setting for which system parameters are unknown, namely $\pmb{A},\pmb{B},\pmb{D},\pmb{Q}, \pmb{R}^u, \pmb{R}^w$, we can simulate system trajectories, $(x_h)_{h=0,\cdots,N}$, using a pair of control gain matrices~$(\pmb{K}, \pmb{L})$ and we have access to zeroth-order (ZO) information consisting of the (stochastic) cost $\mathcal{G}_{\pmb{\xi}}(\pmb{K},\pmb{L})$ incurred by this pair of controllers. In Algorithms~\ref{alg:inner_nested_npg} and~\ref{alg:outer_nested_npg}, we apply the single-point ZO estimation procedures.

\noindent\textbf{Inner loop ZO-NPG algorithm (see Algorithm~\ref{alg:inner_nested_npg}).}
In the light of the update rule~\eqref{eq:det-inner-loop-npg} in the deterministic exact setting, for any fixed matrix~$\pmb{K}$ and any time index~$k$, we replace the gradient~$\nabla_{\pmb{L}}\mathcal{G}(\pmb{K},\pmb{L}_k)$ and the covariance matrix~$\pmb{\Sigma}_{\pmb{K},\pmb{L}_k}$ by ZO estimates denoted as~$\widetilde{\nabla}_{\pmb{L}}\mathcal{G}(\pmb{K},\pmb{L}_k)$ and~$\widetilde{\pmb{\Sigma}}_{\pmb{K},\pmb{L}_k}$ respectively. By sampling two independent trajectories at each sample step, we firstly obtain an unbiased estimate of the gradient w.r.t. $\pmb{L}$ of the smoothed objective $\mathcal{G}_{r_1}(\pmb{K},\pmb{L})$ 
in the sense that: $\mathbb{E}[\widetilde{\nabla}_{\pmb{L}}\mathcal{G}(\pmb{K},\pmb{L}_k)]=\nabla_{\pmb{L}}\mathcal{G}_{r_1}(\pmb{K},\pmb{L}_k)$, $\mathcal{G}_{r_1}(\pmb{K},\pmb{L})\coloneqq\mathbb{E}[\mathcal{G}(\pmb{K},\pmb{L}+r_1\pmb{U})]$, where $\pmb{U}$ is uniformly sampled on a unit ball in $\mathcal{S}_1$. Secondly, we obtain an unbiased estimate of the covariance matrix, i.e., $\mathbb{E}[\widetilde{\pmb{\Sigma}}_{\pmb{K},\pmb{L}_k}]=\pmb{\Sigma}_{\pmb{K},\pmb{L}_k}$. For any given $\pmb{K}\in\mathcal{K}$, Algorithm \ref{alg:inner_nested_npg} outputs $\pmb{L}_{T_{in}}$ that satisfies the accuracy requirement described after~(\ref{eq:det-outer-loop-npg})  with proper choices of parameters. The detailed sampling and computation procedures can be found in Algorithm 1 of~\cite{kaiqing_finite_horizon}, we report it in Algorithm~\ref{alg:inner_nested_npg} for completeness. 
\begin{algorithm}[!htbp]
\label{detailed_algo}
    \begin{algorithmic}[1]
    \caption{(Algorithm 1 in \cite{kaiqing_finite_horizon}) Inner-loop Zeroth-Order Maximization Oracle}\label{alg:inner_nested_npg}
    \REQUIRE $\pmb{K}\in\mathcal{K}$, $\pmb{L}_0$, number of iterations $T_{in}$, sample size $M_1$, perturbation radius $r_1$, stepsize $\tau_1$, horizon $N$, dimension $d_{\pmb{L}}=nmN$.
    \ENSURE $\pmb{L}_{\text{out}} = \pmb{L}_{T_{in}}$.
    \FOR{$k=0,1,\ldots,T_{in}-1$}
    \FOR{$i=0,1,\ldots,M_1-1$}
        \STATE Sample $\pmb{L}_k^i=\pmb{L}_k+r_1\pmb{U}_i$ where $\pmb{U}_i \sim \text{Unif}(\mathcal{S}_2)$  
        with $\|\pmb{U}_i\|_F=1$.
        \STATE Simulate a first trajectory using control  $(\pmb{K},\pmb{L}_k^i)$ for horizon $N$ starting from~$x_{0,i}$ under one realization of noises~$\pmb{\xi}_i$ and collect the cost $\mathcal{G}_{\pmb{\xi}_i}(\pmb{K},\pmb{L}_k^i)$.
        \STATE Simulate another independent trajectory using control $(\pmb{K},\pmb{L}_k)$ for horizon $N$ starting from $x_{0,i}'$ and compute
           $ \widetilde{\pmb{\Sigma}}_{\pmb{K},\pmb{L}_k}^i= \text{diag}\bigl(x_{0,i}' {x_{0,i}'}^{\top},\cdots,x_{N,i}' {x_{N,i}'}^{\top}\bigr).$
    \ENDFOR
    \STATE Update $\pmb{L}_{k+1}=\pmb{L}_k+\tau_1\widetilde{\nabla}_{\pmb{L}}\mathcal{G}(\pmb{K},\pmb{L}_k)\widetilde{\pmb{\Sigma}}_{\pmb{K},\pmb{L}_k}^{-1}$ where

    $\widetilde{\nabla}_{\pmb{L}}\mathcal{G}(\pmb{K},\pmb{L}_k)=\frac{1}{M_1}\sum_{i=0}^{M_1-1}\frac{d_{\pmb{L}}}{r_1}\mathcal{G}_{\pmb{\xi}_i}(\pmb{K},\pmb{L}_k^i)\pmb{U}_i,\, \widetilde{\pmb{\Sigma}}_{\pmb{K},\pmb{L}_k}=\frac{1}{M_1}\sum_{i=0}^{M_1-1}\widetilde{\pmb{\Sigma}}_{\pmb{K},\pmb{L}_k}^i.$
    \ENDFOR
    \end{algorithmic}
\end{algorithm}
\noindent\textbf{Outer loop ZO-NPG (see Algorithm~\ref{alg:outer_nested_npg}).}
Similarly to the inner loop procedure, we now replace the unknown quantities~$\nabla_{\pmb{K}}\mathcal{G}(\pmb{K}_t,\pmb{L}(\pmb{K}_t))$ and $\pmb{\Sigma}_{\pmb{K}_t,\pmb{L}(\pmb{K}_t)}$ in~$\eqref{eq:det-outer-loop-npg}$ by ZO estimates. As for the exact solution~$\pmb{L}(\pmb{K}_t)$ to the inner maximization problem, we use the output of the inner loop ZO-NPG algorithm instead. Notice that the ZO single-point estimate~$\widetilde{\nabla}_{\pmb{K}}\mathcal{G}(\pmb{K}_t,\pmb{L}_t)$ as defined in Algorithm~\ref{alg:outer_nested_npg} is an unbiased estimate of the gradient w.r.t. $\pmb{K}$ of the smoothed objective~$\mathcal{G}_{r_2}(\pmb{K},\pmb{L})$ (evaluated at the pair~$(\pmb{K}_t,\pmb{L}_t)$) in the sense that:
$
    \mathbb{E}[\widetilde{\nabla}_{\pmb{K}}\mathcal{G}(\pmb{K}_t,\pmb{L}_t)]=\nabla_{\pmb{K}}\mathcal{G}_{r_2}(\pmb{K}_t,\pmb{L}_t)$, $\mathcal{G}_{r_2}(\pmb{K},\pmb{L})\coloneqq \mathbb{E}[\mathcal{G}(\pmb{K}+r_2\pmb{V},\pmb{L})],
$
where $\pmb{V}$ is uniformly sampled on a unit ball in $\mathcal{S}_2$.

\noindent\textbf{Comparison to the derivative free NPG algorithm in~\cite{kaiqing_finite_horizon}.} We point out here an important difference between our proposed algorithm and the zeroth-order NPG algorithm in~\cite{kaiqing_finite_horizon} which inspired this work. This difference lies in the outer loops of the algorithms: namely comparing Algorithm~\ref{alg:outer_nested_npg} and Algorithm~2 in~\cite{kaiqing_finite_horizon}. In their work, at each time step~$t$ of the outer loop, Algorithm~1 (which provides an approximate solution of the maximization problem) is called for each perturbation~$\pmb{K}_t^{m}$ (for~$m = 0, \cdots, M_2-1$) of the control gain matrix~$\pmb{K}_t$ (see step 6: in their Algorithm~2) in order to control the gradient estimation error. In contrast to their work, observe that we only call Algorithm~\ref{alg:inner_nested_npg} once at each outer loop iteration~$t$ in Algorithm~\ref{alg:outer_nested_npg} and use the approximate maximizer~$\pmb{L}_t$ to compute our ZO estimates for updating the control gain matrix sequence~$(\pmb{K}_t)_{t\geq 0}$. This observation is crucial for our sample complexity improvement discussed in the next section.

\begin{algorithm}[!htbp]
    \begin{algorithmic}[1]
    \caption{Outer-loop Nested Natural Policy Gradient}\label{alg:outer_nested_npg}
    \REQUIRE $\pmb{K}_0\in\mathcal{K}$, number of iterations $T$, sample size $M_2$, perturbation radius $r_2$, stepsize $\tau_2$, horizon $N$, dimension $d_{\pmb{K}}=dmN$.
    \ENSURE $\pmb{K}_{\text{out}} = \pmb{K}_T$.
    \FOR{$t=0,1,\ldots,T-1$}
    \STATE Call Algorithm~\ref{alg:inner_nested_npg} to obtain~$\pmb{L}_t$.
    \FOR{$m=0,1,\ldots,M_2-1$}
            \STATE Sample $\pmb{K}_t^m=\pmb{K}_t+r_2\pmb{V}_m$ where $\pmb{V}_m \sim \text{Unif}(\mathcal{S}_2)$ 
            with $\|\pmb{V}_m\|_F=1$. 
            \STATE Simulate a first trajectory using control $(\pmb{K}_t^m,\pmb{L}_t)$ for horizon $N$ under $\pmb{\xi}_m=[x_{0,m}^{\top},\xi_{0,m}^{\top},\ldots,\xi_{N-1,m}^{T-1}]^{\top}$ and collect the cost $\mathcal{G}_{\pmb{\xi}_m}(\pmb{K}_t^m,\pmb{L}_t)$.
        \STATE Simulate another independent trajectory using control $(\pmb{K}_t,\pmb{L}_t)$ for horizon $N$ starting from $x_{0,m}'$ and compute $\widetilde{\pmb{\Sigma}}_{\pmb{K}_t,\pmb{L}_t}^m= \text{diag}\bigl(x_{0,m}' {x_{0,m}'}^{\top},\cdots,x_{N,m}' {x_{N,m}'}^{\top}\bigr)$.
    \ENDFOR
    \STATE Update $\pmb{K}_{t+1}=\pmb{K}_t-\tau_2\widetilde{\nabla}_{\pmb{K}}\mathcal{G}(\pmb{K}_t,\pmb{L}_t)\widetilde{\pmb{\Sigma}}_{\pmb{K}_t,\pmb{L}_t}^{-1}$ where
    
    $\widetilde{\nabla}_{\pmb{K}}\mathcal{G}(\pmb{K}_t,\pmb{L}_t)=\frac{1}{M_2}\sum_{m=0}^{M_2-1}\frac{d_{\pmb{K}}}{r_2}\mathcal{G}_{\pmb{\xi}_m}(\pmb{K}_t^m,\pmb{L}_t)\pmb{V}_m,\,\widetilde{\pmb{\Sigma}}_{\pmb{K}_t,\pmb{L}_t}=\frac{1}{M_2}\sum_{m=0}^{M_2-1}\widetilde{\pmb{\Sigma}}_{\pmb{K}_t,\pmb{L}_t}^m.$
    \ENDFOR
    \end{algorithmic}
\end{algorithm}

\begin{remark}
In Algorithms~\ref{alg:inner_nested_npg} (steps 4-5) and~\ref{alg:outer_nested_npg} (steps 5-6), we can also use a single trajectory sampled using the perturbed controls sampled in steps~3 and~4 respectively. This would induce an additional bias in the covariance matrix estimates due to the control perturbations. Our analysis can be slightly adapted to cover this case without changing our overall sample complexity using a similar analysis as in \cite[Lemma 32]{global_con_pg_lq}. We use two trajectories for simplicity of the analysis.
\end{remark}

\section{Convergence Analysis and Sample Complexity}
\label{sec:results}

In this section, we analyze the iteration complexity of the exact nested NPG algorithm in \eqref{eq:det-outer-loop-npg} and show a global last-iterate linear convergence rate. Then we derive the sample complexity of Algorithm \ref{alg:outer_nested_npg} to reach an approximate NE. In preparation for the main results, we establish a critical foundation by answering how is implicit regularization guaranteed.

\subsection{Implicit regularization} In this subsection, we first state one of our key technical results, which ensures that the iterates $(\pmb{K}_t,\pmb{L}_t)$ generated by both deterministic and stochastic nested algorithms will remain in some compact set. In the stochastic setting, our key technical improvement over the analogous result in Theorem 4.2 of \cite{kaiqing_finite_horizon} is that we require a much smaller number of samples for achieving this. This improvement is crucial for achieving our better total sample complexity stated in Theorem~\ref{theorem:last_iterate_stochastic}.

When assuming exact inner-loop solutions, i.e., $\epsilon_1=0$, and noiseless gradients, it was shown in Theorem~3.7 in~\cite{kaiqing_finite_horizon} that~(a) the sequence~$(\pmb{P}_{\pmb{K}_t,\pmb{L}(\pmb{K}_t)})_{t\geq 0}$ is well-defined, satisfies the conditions in~\eqref{eq:def-K} for every~$t \geq 0$ and is (most importantly) non-increasing and bounded below in the sense of positive definiteness; and as a consequence~(b) for every~$t \geq 0, \pmb{K}_t \in \mathcal{K}$ when~$\pmb{K}_0 \in \mathcal{K}$. They refer to this property as \textit{implicit regularization}. See also our similar result below in Lemma~\ref{lemma:ori_descent_inequality} and Remark~\ref{rem:special-cases-descent-ineq} for the special deterministic case with exact inner maximization solution. 
When using approximate inner-loop solutions and estimated outer-loop natural gradients, the descent property of the sequence $(\pmb{P}_{\pmb{K}_t,\pmb{L}(\pmb{K}_t)})_{t \geq 0}$ is violated 
and we need to show that the iterates~$(\pmb{K}_t)$ remain confined in a suitable compact set by other means. In the following, we consider a subset~$\hat{\mathcal{K}}$ of $\mathcal{K}$ for which we prove that IR holds. Consider an initial point $\pmb{K}_0\in\mathcal{K}$ and define 
\begin{equation}
    \label{eq:def-hat-K}
    \hat{\mathcal{K}}\coloneqq \biggl\{ \pmb{K} \in \mathcal{S}_1  \mid 
    0\preceq\pmb{P}_{\pmb{K},\pmb{L}(\pmb{K})}\preceq \pmb{P}_{\pmb{K}_0,\pmb{L}(\pmb{K}_0)}+\frac{\lambda_{\min}(\pmb{H}_{\pmb{K}_0,\pmb{L}(\pmb{K}_0)})}{2\|\pmb{D}\|^2}\pmb{I}\biggr\}\,, 
\end{equation}
where $\pmb{P}_{\pmb{K},\pmb{L}(\pmb{K})}$ is a solution to~\eqref{eq:riccati} when it exists.

As can be observed from~\eqref{eq:def-hat-K}, we need to control the error induced by the inner-loop solver which provides an approximation of~$\pmb{L}(\pmb{K})$ (as well as the estimation of outer-loop natural gradients using zeroth-order information) in order to show the recurrence of the iterates~$\pmb{K}_t$ in the set~$\hat{\mathcal{K}}$ (with high probability).

This inner maximization problem which takes the form of an LQR problem has been previously addressed~\cite{global_con_pg_lq,kaiqing_finite_horizon, local_lipschitz}.
It follows from Theorem 4.1 in \cite{kaiqing_finite_horizon} that any control gain matrix~$\pmb{L}$ produced by the inner-loop solver lies in the following bounded set (with high probability): 
\begin{align}
    \label{eq:def-hat-L}
    \hat{\mathcal{L}} &\coloneqq \Bigl\{\pmb{L} \in \mathcal{S}_2 \mid \|\pmb{L}(\pmb{K})-\pmb{L}\|_F\leq H,\, \forall\pmb{K}\in\hat{\mathcal{K}}\Bigr\}, \\ H &\label{ineq:h_bound}\coloneqq \sup_{\pmb{K}\in\hat{\mathcal{K}}}\lambda_{\min}^{-1}(\pmb{H}_{\pmb{K},\pmb{L}(\pmb{K})})\leq 2\lambda_{\min}^{-1}(\pmb{H}_{\pmb{K}_0,\pmb{L}(\pmb{K}_0)}) . 
\end{align}
Using the sets~$\hat{\mathcal{K}}$ and~$\hat{\mathcal{L}}$ respectively defined in~\eqref{eq:def-hat-K} and~\eqref{eq:def-hat-L}, we are now ready to state the IR of our deterministic and stochastic nested NPG algorithm w.r.t. both control gain matrices~$\pmb{K}$ and~$\pmb{L}$. More specifically, we  prove that the pair of iterates~$(\pmb{K}_t,\pmb{L}_t)$ generated by \eqref{eq:det-outer-loop-npg} (Algorithm~\ref{alg:outer_nested_npg}) are maintained in the bounded set~$\hat{\mathcal{K}}\times\hat{\mathcal{L}}$ (with high probability) for every~$t$ if we properly choose the inner-loop accuracy~$\epsilon_1$ (as well as the batch sample size~$M_2$ and the smoothing radius~$r$). Before stating the IR result, we state some nice Lipschitzness properties over the set $\hat{\mathcal{K}}\times\hat{\mathcal{L}}$ that will contribute to our analysis. First of all, we present the following result of sensitivity analysis in preparation of the Lipschitzness properties: For any stable matrix~$F$ and discrete-time Lyapunov equation: $M=X-F^{\top}XF$,
we have: 
\begin{lemma}
\label{lemma:sensitivity_analysis} (Theorem 2.6 in \cite{lyapunov_sensitivity})
Consider two stable Lyapunov equations that admit unique solutions: $M=X-F^{\top} X F$, $M+\Delta M=X+\Delta X-(F+\Delta F)^{\top}(X+\Delta X)(F+\Delta F)$. If $\|U\|\|\Delta F\|(2\|F\|+\|\Delta F\|)<1$, then we have $\|\Delta X\|\leq \bigl(1-\|U\|\|\Delta F\|(2\|F\|+\|\Delta F\|)\bigr)^{-1}\cdot\bigl(\|U\|\|\Delta M\|+\|U\|\|\Delta F\|(2\|F\|+\|\Delta F\|)\|X\|\bigr)$ where $U$ is the unique solution of $X-F^{\top}UF=I$.
\end{lemma}

With the above tool for sensitivity analysis, we obtain the following result.
\begin{proposition}
\label{propostion:curvature_properties} (Proposition 1 in \cite{cdc_version}) Let $\pmb{K}_0\in\mathcal{K}$ and consider the corresponding set $\hat{\mathcal{K}}$. For any $(\pmb{K},\pmb{L})\in \hat{\mathcal{K}}\times\hat{\mathcal{L}}$, $\pmb{K}'\in\mathcal{K}$, there exist $D_1,D_2,l_1,l_2>0$ such that if we let $
    \|\pmb{K}-\pmb{K}'\|\leq D_1$, $\|\pmb{L}-\pmb{L}'\|\leq D_2$ then 
$
    \|\pmb{F}_{\pmb{K}',\pmb{L}}-\pmb{F}_{\pmb{K},\pmb{L}}\|\leq l_1\|\pmb{K}-\pmb{K}'\|$, and $\|\pmb{F}_{\pmb{K},\pmb{L}'}-\pmb{F}_{\pmb{K},\pmb{L}}\|\leq l_2\|\pmb{L}-\pmb{L}'\|$.
Similar results also hold when replacing~$\pmb{F}_{\pmb{K},\pmb{L}}$ by~$\pmb{\Sigma}_{\pmb{K},\pmb{L}}$, and $\pmb{P}_{\pmb{K},\pmb{L}}$, see Lemmas~\ref{lemma:ef_continuity}, \ref{lemma:sigma_continuity} and~\ref{lemma:p_continuity} in Appendix \ref{appendix:preparation} for the proofs and explicit expressions of the corresponding constants.

\end{proposition} 
The smoothness and continuity over the set $\hat{\mathcal{K}}\times\hat{\mathcal{L}}$ naturally motivate us to borrow ideas from stochastic optimization. In particular, it is tempting to follow the analysis of stochastic nested algorithms for global Lipschitz smooth functions, see for instance~\cite{Lin_gdmax_inner_loop_length, zeng2022regularized}.
Unfortunately, such analysis is not directly applicable since the properties stated in Proposition~\ref{propostion:curvature_properties}, only hold locally within the set $\hat{\mathcal{K}}\times\hat{\mathcal{L}}$, therefore one needs to ensure that the iterates of \eqref{eq:det-outer-loop-npg} (Algorithm~\ref{alg:outer_nested_npg}) remain in this set. This can be achieved by controlling the value matrix~$\pmb{P}_{\pmb{K},\pmb{L}(\pmb{K})}$ along the iterations. However, in the case when the estimated gradients from approximate inner-loop solutions (and ZO estimation) are used, the situation is more challenging. Such sequence is no longer non-increasing and the deviation must be controlled.

Our next technical lemma upperbounds the value matrix difference $\pmb{P}_{\pmb{K}',\pmb{L}(\pmb{K}')}-\pmb{P}_{\pmb{K},\pmb{L}(\pmb{K})}$ for any two controllers~$\pmb{K}, \pmb{K}'$. 
The upper bound consists of a negative drift term driven by the natural gradient~$\pmb{F}_{\pmb{K},\pmb{L}(\pmb{K})}$ and an error term originating from three different source errors. The first one~($\pmb{e}_{1,\pmb{K},\pmb{K}'}$, below) is a bias due to zeroth-order estimation using the smoothing technique, the second one~$(\pmb{e}_{2,\pmb{K},\pmb{K}'})$ stems from using an approximate solution computed in the inner-loop instead of an exact solution~$\pmb{L}(\pmb{K})$ and the third one~$(\pmb{e}_{3,\pmb{K},\pmb{K}'})$ results from the fluctuations of the natural gradient estimators around their means. 
Our error decomposition is more refined compared to prior work~\cite{kaiqing_finite_horizon} and contributes to our sample complexity improvements. This result is also crucial to show that $\pmb{K}'$ remains in $\hat{\mathcal{K}}$.
For convenience, we define some quantities that will be frequently used: $\pmb{G}_{\pmb{K},\pmb{L}(\pmb{K})}\coloneqq\pmb{R}^u+\pmb{B}^{\top}\widetilde{\pmb{P}}_{\pmb{K},\pmb{L}(\pmb{K})}\pmb{B}$ and $G\coloneqq \sup_{\pmb{K}\in\hat{\mathcal{K}}}\|\pmb{G}_{\pmb{K},\pmb{L}(\pmb{K})}\|$.
\begin{lemma}
\label{lemma:ori_descent_inequality}
    Let $\pmb{K}_0\in\mathcal{K}$, $\pmb{K}\in\hat{\mathcal{K}}$, and consider $\pmb{K}'=\pmb{K}-\tau_2\widetilde{\pmb{F}}_{\pmb{K},\pmb{L}}$ where $\widetilde{\pmb{F}}_{\pmb{K},\pmb{L}}\coloneqq \frac{1}{2}\widetilde{\nabla}_{\pmb{K}}\mathcal{G}(\pmb{K},\pmb{L})\widetilde{\pmb{\Sigma}}_{\pmb{K},\pmb{L}}^{-1}$ and $\pmb{L}$ is the output of Algorithm \ref{alg:inner_nested_npg} given $\pmb{K}$. If we choose $\tau_2\leq \min\{1/(4G), 1\}$, then we have the following inequality 
    \begin{equation*}\pmb{P}_{\pmb{K}',\pmb{L}(\pmb{K}')}-\pmb{P}_{\pmb{K},\pmb{L}(\pmb{K})}\preceq {\sum}_{i=0}^{N}(\pmb{A}_{\pmb{K}',\pmb{L}(\pmb{K}')}^{\top})^i\pmb{e}_{\pmb{K},\pmb{K}'}(\pmb{A}_{\pmb{K}',\pmb{L}(\pmb{K}')})^i -\frac{\tau_2}{4}\pmb{F}_{\pmb{K},\pmb{L}(\pmb{K})}^{\top}\pmb{F}_{\pmb{K},\pmb{L}(\pmb{K})}.\end{equation*}
Here~$\pmb{e}_{\pmb{K},\pmb{K}'}\coloneqq \pmb{e}_{1,\pmb{K},\pmb{K}'}+\pmb{e}_{2,\pmb{K},\pmb{K}'}+\pmb{e}_{3,\pmb{K},\pmb{K}'}$, $\pmb{e}_{1,\pmb{K},\pmb{K}'}\coloneqq (4\tau_{2}+4\tau_{2}^2\|\pmb{G}_{\pmb{K},\pmb{L}(\pmb{K})}\|)(\pmb{F}_{\pmb{K},\pmb{L}}^r-\pmb{F}_{\pmb{K},\pmb{L}})^{\top}(\pmb{F}_{\pmb{K},\pmb{L}}^r-\pmb{F}_{\pmb{K},\pmb{L}})$, $\pmb{e}_{2,\pmb{K},\pmb{K}'}\coloneqq \tau_{2}(\pmb{F}_{\pmb{K},\pmb{L}(\pmb{K})}-\pmb{F}_{\pmb{K},\pmb{L}})^{\top}(\pmb{F}_{\pmb{K},\pmb{L}(\pmb{K})}-\pmb{F}_{\pmb{K},\pmb{L}})$, $\pmb{e}_{3,\pmb{K},\pmb{K}'}\coloneqq (2\tau_{2}+2\tau_{2}^2\|\pmb{G}_{\pmb{K},\pmb{L}(\pmb{K})}\|)V(\widetilde{\pmb{F}}_{\pmb{K},\pmb{L}})$, 
    where $\pmb{F}_{\pmb{K},\pmb{L}}^r\coloneqq\mathbb{E}[\widetilde{\pmb{F}}_{\pmb{K},\pmb{L}}]$ and $V(\widetilde{\pmb{F}}_{\pmb{K},\pmb{L}})\coloneqq(\widetilde{\pmb{F}}_{\pmb{K},\pmb{L}}-\pmb{F}_{\pmb{K},\pmb{L}}^r)^{\top}(\widetilde{\pmb{F}}_{\pmb{K},\pmb{L}}-\pmb{F}_{\pmb{K},\pmb{L}}^r)$. 
\end{lemma}
\begin{proof}
    Computing the difference between the two Lyapunov equations satisfied by~$\pmb{P}_{\pmb{K}',\pmb{L}(\pmb{K}')}$ and~$\pmb{P}_{\pmb{K},\pmb{L}(\pmb{K})}$ respectively and reorganizing the terms, we obtain 
    \begin{align}
&\pmb{P}_{\pmb{K}',\pmb{L}(\pmb{K}')}-\pmb{P}_{\pmb{K},\pmb{L}(\pmb{K})}=\pmb{A}_{\pmb{K}',\pmb{L}(\pmb{K}')}^{\top}(\pmb{P}_{\pmb{K}',\pmb{L}(\pmb{K}')}-\pmb{P}_{\pmb{K},\pmb{L}(\pmb{K})})\pmb{A}_{\pmb{K}',\pmb{L}(\pmb{K}')}\label{eq:above_eq}+X\\
        &X\coloneqq \pmb{\mathcal{R}}_{\pmb{K},\pmb{K}'}-\pmb{\Xi}_{\pmb{K},\pmb{K}'}^{\top}(\pmb{R}^w-\pmb{D}^{\top}\pmb{P}_{\pmb{K},\pmb{L}(\pmb{K})}\pmb{D})^{-1}\pmb{\Xi}_{\pmb{K},\pmb{K}'}\notag\nonumber
    \end{align}
    \begin{align}
    \pmb{\mathcal{R}}_{\pmb{K},\pmb{K}'}\coloneqq \diag(\mathcal{R}_{K_0,K_0'}, \cdots ,\mathcal{R}_{K_{N-1},K_{N-1}'},\pmb{0}_{m\times m}),\label{def:r_def}
    \end{align}
    \begin{align}
        &\pmb{\Xi}_{\pmb{K},\pmb{K}'}\coloneqq \begin{bmatrix}
            \pmb{0}_{m\times nN}\notag\\
            \diag(\Xi_{K_0,K_0'},\cdots,\Xi_{K_{N-1},K_{N-1}'})
        \end{bmatrix}.\notag\\
        &\mathcal{R}_{K_h,K_h'}\coloneqq (K_h'-K_h)^TF_{K_h,L(K_h)}+F_{K_h,L(K_h)}^T(K _h'-K_h)\notag\\
        & +(K_h'-K_h)^T(R_h^u+B_h^T\widetilde{P}_{K_{h+1},L(K_{h+1})}B_h)(K_h'-K_h).\notag\\
        &\widetilde{P}_{K_{h+1},L(K_{h+1})}\coloneqq P_{K_{h+1},L(K_{h+1})}\notag\\
        &+P_{K_{h+1},L(K_{h+1})}D_h(R_h^w-D_h^TP_{K_{h+1},L(K_{h+1})}D_h)^{-1}D_h^TP_{K_{h+1},L(K_{h+1})},\notag\\
        &\Xi_{K_h,K_h'}\coloneqq -(R_h^w-D_h^{\top}P_{K_{h+1},L(K_{h+1})}D_h)L(K_h')\notag\\& -D_h^{\top}P_{K_{h+1},L(K_{h+1})}(A_h-B_hK_h'),\, h=0,\cdots, N-1.\notag
    \end{align}
    To bound the solution of \eqref{eq:above_eq}, we first upperbound $X$. Recall the definition of $\mathcal{R}_{\pmb{K},\pmb{K}'}$ in \eqref{def:r_def} and plug in $\pmb{K}'=\pmb{K}-\tau_2\widetilde{\pmb{F}}_{\pmb{K},\pmb{L}}$ to obtain 
    \begin{align*}
        X&\preceq\pmb{\mathcal{R}}_{\pmb{K},\pmb{K}'}=-\tau_{2}\widetilde{\pmb{F}}_{\pmb{K},\pmb{L}}^{\top}\pmb{F}_{\pmb{K},\pmb{L}(\pmb{K})} -\tau_{2}\pmb{F}_{\pmb{K},\pmb{L}(\pmb{K})}^{\top}\widetilde{\pmb{F}}_{\pmb{K},\pmb{L}}+\tau_{2}^2\widetilde{\pmb{F}}_{\pmb{K},\pmb{L}}^{\top}\pmb{G}_{\pmb{K},\pmb{L}(\pmb{K})}\widetilde{\pmb{F}}_{\pmb{K},\pmb{L}}\\
        &= -\tau_2(\widetilde{\pmb{F}}_{\pmb{K},\pmb{L}}-\pmb{F}_{\pmb{K},\pmb{L}}^r+\pmb{F}_{\pmb{K},\pmb{L}}^r)^{\top}\pmb{F}_{\pmb{K},\pmb{L}(\pmb{K})}-\tau_2\pmb{F}_{\pmb{K},\pmb{L}(\pmb{K})}^{\top}(\widetilde{\pmb{F}}_{\pmb{K},\pmb{L}}-\pmb{F}_{\pmb{K},\pmb{L}}^r+\pmb{F}_{\pmb{K},\pmb{L}}^r)\\
        &\quad +\tau_2^2(\widetilde{\pmb{F}}_{\pmb{K},\pmb{L}}-\pmb{F}_{\pmb{K},\pmb{L}}^r+\pmb{F}_{\pmb{K},\pmb{L}}^r)^{\top}\pmb{G}_{\pmb{K},\pmb{L}(\pmb{K})}(\widetilde{\pmb{F}}_{\pmb{K},\pmb{L}}-\pmb{F}_{\pmb{K},\pmb{L}}^r+\pmb{F}_{\pmb{K},\pmb{L}}^r)\\
        &\overset{(a)}{\preceq} (2\tau_2+2\tau_2^2\|\pmb{G}_{\pmb{K},\pmb{L}(\pmb{K})}\|)(\widetilde{\pmb{F}}_{\pmb{K},\pmb{L}}-\pmb{F}_{\pmb{K},\pmb{L}}^r)^{\top}(\widetilde{\pmb{F}}_{\pmb{K},\pmb{L}}-\pmb{F}_{\pmb{K},\pmb{L}}^r)+\frac{\tau_2}{2}\pmb{F}_{\pmb{K},\pmb{L}(\pmb{K})}^{\top}\pmb{F}_{\pmb{K},\pmb{L}(\pmb{K})}\\
        &\quad -\tau_{2}(\pmb{F}_{\pmb{K},\pmb{L}}^r-\pmb{F}_{\pmb{K},\pmb{L}}+\pmb{F}_{\pmb{K},\pmb{L}})^{\top}\pmb{F}_{\pmb{K},\pmb{L}(\pmb{K})}-\tau_{2}\pmb{F}_{\pmb{K},\pmb{L}(\pmb{K})}^{\top}(\pmb{F}_{\pmb{K},\pmb{L}}^r-\pmb{F}_{\pmb{K},\pmb{L}}+\pmb{F}_{\pmb{K},\pmb{L}})\\
        &\quad +2\tau_2^2\|\pmb{G}_{\pmb{K},\pmb{L}(\pmb{K})}\|(\pmb{F}_{\pmb{K},\pmb{L}}^r)^{\top}\pmb{F}_{\pmb{K},\pmb{L}}^r\\
        &\overset{(b)}{\preceq} (2\tau_2+2\tau_2^2\|\pmb{G}_{\pmb{K},\pmb{L}(\pmb{K})}\|)(\widetilde{\pmb{F}}_{\pmb{K},\pmb{L}}-\pmb{F}_{\pmb{K},\pmb{L}}^r)^{\top}(\widetilde{\pmb{F}}_{\pmb{K},\pmb{L}}-\pmb{F}_{\pmb{K},\pmb{L}}^r)+\frac{\tau_2}{2}\pmb{F}_{\pmb{K},\pmb{L}(\pmb{K})}^{\top}\pmb{F}_{\pmb{K},\pmb{L}(\pmb{K})}  \\
        &\quad +4\tau_2(\pmb{F}_{\pmb{K},\pmb{L}}^r-\pmb{F}_{\pmb{K},\pmb{L}})^{\top}(\pmb{F}_{\pmb{K},\pmb{L}}^r-\pmb{F}_{\pmb{K},\pmb{L}})+\frac{\tau_2}{4}\pmb{F}_{\pmb{K},\pmb{L}(\pmb{K})}^{\top}\pmb{F}_{\pmb{K},\pmb{L}(\pmb{K})}\\
        &\quad -\tau_2(\pmb{F}_{\pmb{K},\pmb{L}(\pmb{K})}^{\top}\pmb{F}_{\pmb{K},\pmb{L}(\pmb{K})}+\pmb{F}_{\pmb{K},\pmb{L}}^{\top}\pmb{F}_{\pmb{K},\pmb{L}}-(\pmb{F}_{\pmb{K},\pmb{L}}-\pmb{F}_{\pmb{K},\pmb{L}(\pmb{K})})^{\top}(\pmb{F}_{\pmb{K},\pmb{L}}-\pmb{F}_{\pmb{K},\pmb{L}(\pmb{K})}))\\
        &\quad +4\tau_2^2\|\pmb{G}_{\pmb{K},\pmb{L}(\pmb{K})}\|(\pmb{F}_{\pmb{K},\pmb{L}}^r-\pmb{F}_{\pmb{K},\pmb{L}})^{\top}(\pmb{F}_{\pmb{K},\pmb{L}}^r-\pmb{F}_{\pmb{K},\pmb{L}})+4\tau_2^2\|\pmb{G}_{\pmb{K},\pmb{L}(\pmb{K})}\|\pmb{F}_{\pmb{K},\pmb{L}}^{\top}\pmb{F}_{\pmb{K},\pmb{L}}   \\
        &= \pmb{e}_{\pmb{K},\pmb{K}'} -\frac{\tau_{2}}{4}\pmb{F}_{\pmb{K},\pmb{L}(\pmb{K})}^{\top}\pmb{F}_{\pmb{K},\pmb{L}(\pmb{K})}+(4\tau_{2}^2\|\pmb{G}_{\pmb{K},\pmb{L}(\pmb{K})}\|-\tau_{2})\pmb{F}_{\pmb{K},\pmb{L}}^{\top}\pmb{F}_{\pmb{K},\pmb{L}}\\
        &\overset{(d)}{\preceq} \pmb{e}_{\pmb{K},\pmb{K}'}-\frac{\tau_{2}}{4}\pmb{F}_{\pmb{K},\pmb{L}(\pmb{K})}^{\top}\pmb{F}_{\pmb{K},\pmb{L}(\pmb{K})}
    \end{align*}
    where $(a)$, $(b)$ use $ M^{\top}N+N^{\top}M\preceq\gamma M^{\top}M+\gamma^{-1}N^{\top}N$, $A^{\top}MA\preceq A^{\top}NA$ if $M\preceq N$, $(M+N)^{\top}(M+N)\preceq 2M^{\top}M+2N^{\top}N$, and basic matrix computations whereas~$(d)$ holds since we choose $\tau_2\leq 1/(4G)$. Combining the solution of stable Lyapunov equations and the fact that $\pmb{A}_{\pmb{K}',\pmb{L}(\pmb{K}')}$ is nilpotent, we conclude
    \begin{align*}
        \pmb{P}_{\pmb{K}',\pmb{L}(\pmb{K}')}-\pmb{P}_{\pmb{K},\pmb{L}(\pmb{K})}&\preceq \sum_{i=0}^{\infty}(\pmb{A}_{\pmb{K}',\pmb{L}(\pmb{K}')}^{\top})^i(\pmb{e}_{\pmb{K},\pmb{K}'}-\frac{\tau_2}{4}\pmb{F}_{\pmb{K},\pmb{L}(\pmb{K})}^{\top}\pmb{F}_{\pmb{K},\pmb{L}(\pmb{K})})(\pmb{A}_{\pmb{K}',\pmb{L}(\pmb{K}')})^i\\
        &\preceq\sum_{i=0}^{N}(\pmb{A}_{\pmb{K}',\pmb{L}(\pmb{K}')}^{\top})^i\pmb{e}_{\pmb{K},\pmb{K}'}(\pmb{A}_{\pmb{K}',\pmb{L}(\pmb{K}')})^i-\frac{\tau_2}{4}\pmb{F}_{\pmb{K},\pmb{L}(\pmb{K})}^{\top}\pmb{F}_{\pmb{K},\pmb{L}(\pmb{K})}.
    \end{align*}
\end{proof}

\begin{remark}
\label{rem:special-cases-descent-ineq}
In the deterministic case when we have access to $\pmb{F}_{\pmb{K},\pmb{L}}$, we have $\pmb{e}_{1,\pmb{K},\pmb{K}'} = \pmb{e}_{3,\pmb{K},\pmb{K}'} = 0\,.$ If we further have access to $\pmb{F}_{\pmb{K},\pmb{L}(\pmb{K})}$, then $\pmb{e}_{\pmb{K},\pmb{K}'} = 0$ and it easily follows that $\pmb{K}'\in\hat{\mathcal{K}}\,.$ 
\end{remark}

We are now in position to establish our `descent' lemma using Lemma~\ref{lemma:ori_descent_inequality}. 
The following result indicates that the value matrices are `monotonically decreasing' up to the three errors previously described. 
We control the first error by tuning the radius~$r_2$ used in our zeroth order estimator, the second one by choosing a suitable accuracy with which the inner loop solution~$\pmb{L}(\pmb{K})$ is approximated given the controller~$\pmb{K}$ and the last one will be made small later on using a large enough sample size. In particular, using Proposition~\ref{propostion:curvature_properties} together with concentration results for estimators around their mean, we obtain the following descent lemma.

\begin{lemma}
\label{lemma:descent_inequality}
    (Descent lemma) Let $\pmb{K}_0\in\mathcal{K}$, $\pmb{K}\in\hat{\mathcal{K}}$ and $\pmb{K}'=\pmb{K}-\tau_2\widetilde{\pmb{F}}_{\pmb{K},\pmb{L}}$ where~$\pmb{L}$ is the output of Algorithm \ref{alg:inner_nested_npg} given $\pmb{K}$. Let $\tau_2$, $r_2$, $\epsilon_1$ be small enough, and $M_2$ large enough. Then with high probability, we have $\pmb{K}'\in\mathcal{K}$ and there exist positive constants $c_1, c_2, c_3$ such that
\begin{align}
\pmb{P}_{\pmb{K}',\pmb{L}(\pmb{K}')}-\pmb{P}_{\pmb{K},\pmb{L}(\pmb{K})}&\preceq \tau_2 (c_1 r_2^2+c_2 \epsilon_1+c_3 \|V(\widetilde{\pmb{F}}_{\pmb{K},\pmb{L}})\|) I -\frac{\tau_2}{4}\pmb{F}_{\pmb{K},\pmb{L}(\pmb{K})}^{\top}{\pmb{F}_{\pmb{K},\pmb{L}(\pmb{K})}}\,.
\label{ineq:descent_ineq_sto}
    \end{align}
    In particular, in the deterministic case, the descent inequality can be simplified as
    \begin{align}
        \pmb{P}_{\pmb{K}',\pmb{L}(\pmb{K}')}-\pmb{P}_{\pmb{K},\pmb{L}(\pmb{K})}&\preceq \tau_2c_2\cdot \epsilon_1\cdot I-\frac{\tau_2}{4}\pmb{F}_{\pmb{K},\pmb{L}(\pmb{K})}^{\top}{\pmb{F}_{\pmb{K},\pmb{L}(\pmb{K})}}.\label{ineq:descent_ineq_det}
    \end{align}
\end{lemma}

See Lemma \ref{lemma:detailed_descent_inequality} in the appendix for a detailed version of this lemma and its proof.
Inequality \eqref{ineq:descent_ineq_sto} follows from the Lipschitzness properties established in Proposition~\ref{propostion:curvature_properties}. The proof is inspired by the analysis of stochastic double-loop algorithms for minmax optimization problems with globally smooth functions (see e.g. \cite[appendix E]{Lin_gdmax_inner_loop_length}). 
More specifically, the proof of Lemma~\ref{lemma:descent_inequality}  shares some similarity with the proof of Theorem E.2 in \cite{Lin_gdmax_inner_loop_length} from the conceptual viewpoint. For instance, we both derive the descent-like inequality using some smoothness properties of $\mathcal{G}(\pmb{K},\pmb{L}(\pmb{K}))$ ($\Phi(x):= f(x,y^*(x))$ in~\cite{Lin_gdmax_inner_loop_length} where $f(x,y)$ is the minmax objective and~$y^*(x)$ is the unique solution to the maximization of~$f(x, \cdot)$). Moreover, the max oracle for solving the inner maximization problem can be implemented with a last-iterate convergence and the same order of sample complexity in both our works. However, \cite{Lin_gdmax_inner_loop_length} leverages strong concavity and global smoothness of $f(x, \cdot)$. Neither strong concavity nor global smoothness hold in our setting. For the inner maximization problem which is an LQR problem, we rather use the PL property and \textit{local} smoothness of $\mathcal{G}(\pmb{K},\cdot)\,.$ Note in addition that \cite{Lin_gdmax_inner_loop_length} focuses on first-order stationary guarantees for their nonconvex-strongly concave problems. We will further establish stronger last-iterate convergence guarantees in our main theorem in Section~5.2 as our nonconvex-nonconcave zero-sum LQ game enjoys a nice gradient domination property.

We are ready to present our implicit regularization result.
\begin{proposition}
\label{proposition:implicit_regularization} (Implicit regularization, Proposition 2 in \cite{cdc_version}) Let Assumption~\ref{assumption:noises} hold.  
Let~$\pmb{K}_0\in\mathcal{K}$ and consider the corresponding $\hat{\mathcal{K}}$ set. For any~$\delta_1\in(0,1), \epsilon_1>0$ and for any~$\pmb{K}\in\mathcal{K}$, the inner-loop ZO-NPG algorithm with single-point estimation outputs~$\pmb{L}$ such that $\mathcal{G}(\pmb{K},\pmb{L}(\pmb{K}))-\mathcal{G}(\pmb{K},\pmb{L})\leq \epsilon_1$ with probability at least $1-\delta_1$ using~$T_{in}M_1=\widetilde{O}(\epsilon_1^{-2})$ samples\footnote{This inner-loop sample complexity was established in \cite{kaiqing_finite_horizon}.}. 
Moreover for any~$\delta_2\in(0,1)$ and any integer~$T \geq 1$, if the estimation parameters in Algorithm~\ref{alg:outer_nested_npg} satisfy
$M_2=\tilde{\mathcal{O}}\bigl(T^2\bigr)$, $\tau_2=\mathcal{O}(1)$, $r_2=\mathcal{O}(T^{-1/2})$, $\epsilon_1=\mathcal{O}(T^{-1})$, $\delta_1=\mathcal{O}(\delta_2/T)
$,
then, it holds with probability at least $1-\delta_2$ that $\pmb{K}_t\in\hat{\mathcal{K}}$ for all $t=1,\cdots,T$. In particular, in the deterministic setting, if we choose small enough $\tau_2=\mathcal{O}(1)$, $\epsilon_1=\mathcal{O}(T^{-1})$ then $\pmb{K}_t\in\hat{\mathcal{K}}$ for all $t=1,\cdots,T$.

\end{proposition}

Note that by choosing $\epsilon_1=\mathcal{O}(T^{-1})$, $r_2=\mathcal{O}(T^{-1/2})$ and a large enough~$M_2$ such that~$V(\widetilde{\pmb{F}}_{\pmb{K}_t,\pmb{L}_t})= \mathcal{O}(1/T)$, we obtain from the descent inequality \eqref{ineq:descent_ineq_sto} that with high probability, $\pmb{P}_{\pmb{K}_{t+1},\pmb{L}(\pmb{K}_{t+1})}-\pmb{P}_{\pmb{K}_{t},\pmb{L}(\pmb{K}_{t})} \preceq \tau_2 (c_1 r_2^2+c_2 \epsilon_1+c_3 \|V(\widetilde{\pmb{F}}_{\pmb{K}_t,\pmb{L}_t})\|)\cdot I=\mathcal{O}\left(\frac{1}{T}\right)\cdot I.$ This bound then allows to show that~$\pmb{K}_{t+1}$ can be kept in~$\hat{\mathcal{K}}$ for~$t=0,\cdots,T-1$ with proper choices of constants.  See Appendix \ref{appendix:main_results} for a detailed version of Proposition~\ref{proposition:implicit_regularization} and its formal proof. 

    The above proposition implies that with the choice of natural policy gradients, a careful choice of inner-loop problem accuracy $\epsilon_1$, a good approximation of outer-loop natural gradients, and the deployment of the nested structure,  Algorithm~\ref{alg:outer_nested_npg} achieves the important IR effect: the iterates are guaranteed to remain in the feasible set defining admissible stable controls without any explicit regularization of the problem. Maintaining the feasibility of the iterates during learning is important since it translates to preserving the robustness of the controllers in the face of adversarial perturbations.

    Proposition \ref{proposition:implicit_regularization} also shows that Algorithm \ref{alg:outer_nested_npg} achieves better sample complexity via a more sample-efficient algorithm compared to \cite{kaiqing_finite_horizon} that maintains the IR property: (a) we have a looser requirement for the inner-loop problem accuracy $\epsilon_1=\mathcal{O}(T^{-1})$ while in \cite{kaiqing_finite_horizon} $\epsilon_1=\mathcal{O}(T^{-2})$; (b) we achieve a better sample complexity for the outer-loop problem using a more careful decomposition of the estimation error caused by the estimated natural gradients: we only require $r_2=\mathcal{O}(T^{-1/2})$ while \cite{kaiqing_finite_horizon} chose $r_2=\mathcal{O}(T^{-1})$; and~(c) we reduce the number of inner-loop algorithm calls with a more natural version of the model-free nested algorithm~(see the comparison at the end of Section~\ref{sec:algos}).   

    Let us now briefly compare our analysis to the proof of the corresponding result in \cite{kaiqing_finite_horizon} from the technical viewpoint.  
    In \cite{kaiqing_finite_horizon}, the value matrix difference in Lemma~\ref{lemma:descent_inequality} is upperbounded using a deterministic descent lemma combined with a norm upper bound of the deviation of the value matrix at the stochastic next iterate from the deterministic next iterate (updated with the true gradient) using a sensitivity analysis. In contrast, we deal with the upper bound directly from the descent inequality using an error decomposition akin to the analysis in the optimization literature. 
    Our approach leads to a tighter analysis as discussed above.

\subsection{Last-iterate convergence}
\label{subsec:last-it-cv}
In this subsection, we establish our last-iterate convergence result in terms of cost function values in both deterministic and stochastic settings. 
Our general proof strategy is inspired from the optimization literature in which a combination of smoothness of the objective function and a so-called Polyak-\L{}ojasiewicz inequality satisfied by the same objective are known to lead to a linear last-iterate convergence rate in the deterministic setting (see e.g.~\cite{karimi-nutini-schmidt16} for an expository paper).  
Implementing this strategy for our special zero-sum LQ game setting, one of our key contributions is to establish a gradient domination inequality for the cost objective at the optimal solution of the inner maximization problem, also known as the primal function~$\Phi(\pmb{K}) := \mathcal{G}(\pmb{K},\pmb{L}(\pmb{K}))\,.$
We first present a well-known result for Lyapunov equations that will become useful in the proof of our gradient domination inequality.  

\begin{lemma}
\label{lemma:dual_lyapunov}
   (Dual Lyapunov equations) 
   Let $A$ be a Schur stable matrix. Let $X$ be the solution to the Lyapunov equation $A^{\top} X A + W = X$ and $Y$ be the solution to the dual Lyapunov equation $A Y A^{\top} + V = Y $. Then $\Tr(XV)=\Tr(YW)$.
\end{lemma}

\begin{proposition}
\label{proposition:gradient_domination} (Gradient domination) For any~$\pmb{K},\pmb{K}'\in\hat{\mathcal{K}}$, we have 
\begin{align}
\Tr(\pmb{F}_{\pmb{K},\pmb{L}(\pmb{K})}^{\top}\pmb{F}_{\pmb{K},\pmb{L}(\pmb{K})})&\geq \mu^{-1}(\mathcal{G}(\pmb{K},\pmb{L}(\pmb{K}))-\mathcal{G}(\pmb{K}^*,\pmb{L}^*)),\label{ineq:gradient_domination}\\ 
\Tr(\nabla_{\pmb{K}} G(\pmb{K}, \pmb{L}(\pmb{K}))^{\top} \nabla_{\pmb{K}} G(\pmb{K}, \pmb{L}(\pmb{K})) &\geq \frac{\phi^2}{\mu} (\mathcal{G}(\pmb{K},\pmb{L}(\pmb{K}))-\mathcal{G}(\pmb{K}^*,\pmb{L}^*))\,,
\end{align}
where $(\pmb{K}^*,\pmb{L}^*)$ is the NE solution, $\mu\coloneqq \sigma_{\min}^{-1}(\pmb{R}^u)s_4$, $s_4\coloneqq \sup_{\pmb{K}\in\hat{\mathcal{K}}}\|\pmb{\Sigma}_{\pmb{K}^*,\widetilde{\pmb{L}}_{\pmb{K},\pmb{K}^*}}\|$, and 
\begin{align}
    \label{def:l_kk}\widetilde{\pmb{L}}_{\pmb{K},\pmb{K}'}\coloneqq\pmb{L}(\pmb{K})-(-\pmb{R}^w+\pmb{D}^{\top}\pmb{P}_{\pmb{K},\pmb{L}(\pmb{K})}\pmb{D})^{-1}\pmb{D}^{\top}\pmb{P}_{\pmb{K},\pmb{L}(\pmb{K})}\pmb{B}(\pmb{K}'-\pmb{K}).
\end{align}

\end{proposition}

\begin{proof}
    We first invoke a value matrix difference lemma from \cite{improved_kaiqing} reported in Lemma~\ref{lemma:true_matrix_difference}.
    For any~$\pmb{K}$ and $\pmb{K}'$ in $\mathcal{K}$, we have
    \begin{align}
    \label{eq:decomp-diff-PKL}
    \pmb{P}_{\pmb{K}',\pmb{L}'}-\pmb{P}_{\pmb{K},\pmb{L}(\pmb{K})}&=
        \pmb{A}_{\pmb{K}',\pmb{L}'}^{\top}(\pmb{P}_{\pmb{K}',\pmb{L}'}-\pmb{P}_{\pmb{K},\pmb{L}(\pmb{K})})\pmb{A}_{\pmb{K}',\pmb{L}'}+N\,,\nonumber\\
    N&\coloneqq\Delta_{\pmb{K}}^{\top}\pmb{F}_{\pmb{K},\pmb{L}(\pmb{K})}+\pmb{F}_{\pmb{K},\pmb{L}(\pmb{K})}^{\top}\Delta_{\pmb{K}} + \Delta_{\pmb{K}}^{\top}(\pmb{R}^u+\pmb{B}^{\top}\pmb{P}_{\pmb{K},\pmb{L}(\pmb{K})}\pmb{B})\Delta_{\pmb{K}}\\
        &\quad +(\pmb{L}'-\pmb{L}(\pmb{K}))^{\top}\pmb{D}^{\top}\pmb{P}_{\pmb{K},\pmb{L}(\pmb{K})}\pmb{B}\Delta_{\pmb{K}}+\Delta_{\pmb{K}}^{\top}\pmb{B}^{\top}\pmb{P}_{\pmb{K},\pmb{L}(\pmb{K})}\pmb{D}(\pmb{L}'-\pmb{L}(\pmb{K}))\nonumber\\
        &\quad +(\pmb{L}'-\pmb{L}(\pmb{K}))^{\top}(-\pmb{R}^w+\pmb{D}^{\top}\pmb{P}_{\pmb{K},\pmb{L}(\pmb{K})}\pmb{D})(\pmb{L}'-\pmb{L}(\pmb{K}))\nonumber.
    \end{align}
    Multiplying both sides by $\pmb{\Sigma}_0$ and taking the trace, we obtain $\mathcal{G}(\pmb{K}',\pmb{L}')-\mathcal{G}(\pmb{K},\pmb{L}(\pmb{K}))=\Tr((\pmb{P}_{\pmb{K}',\pmb{L}'}-\pmb{P}_{\pmb{K},\pmb{L}(\pmb{K})})\pmb{\Sigma}_0)=\Tr(N\pmb{\Sigma}_{\pmb{K}',\pmb{L}'})$
    where the second equality stems from \eqref{eq:lyapunov2} and Lemma \ref{lemma:dual_lyapunov} with $A = \pmb{A}_{\pmb{K}',\pmb{L}'}$, $X = \pmb{P}_{\pmb{K}',\pmb{L}'}-\pmb{P}_{\pmb{K},\pmb{L}(\pmb{K})}$, $V=\pmb{\Sigma}_0$, $Y=\pmb{\Sigma}_{\pmb{K}',\pmb{L}'}$, and $W=N$. As $-\pmb{R}^w+\pmb{D}^{\top}\pmb{P}_{\pmb{K},\pmb{L}(\pmb{K})}\pmb{D}\prec~0$, the maximum of the RHS is achieved for  $\pmb{L}'=\widetilde{\pmb{L}}_{\pmb{K},\pmb{K}'}\in\mathcal{S}_2$ and we have
    \begin{align*}
        &\mathcal{G}(\pmb{K}',\widetilde{\pmb{L}}_{\pmb{K},\pmb{K}'})-\mathcal{G}(\pmb{K},\pmb{L}(\pmb{K}))=\Tr((2\Delta_{\pmb{K}}^{\top}\pmb{F}_{\pmb{K},\pmb{L}(\pmb{K})}+\Delta_{\pmb{K}}^{\top}M\Delta_{\pmb{K}})\pmb{\Sigma}_{\pmb{K}',\pmb{L}'})\,,
    \end{align*}
    where $M\coloneqq \pmb{R}^u+\pmb{B}^{\top}\pmb{P}_{\pmb{K},\pmb{L}(\pmb{K})}\pmb{B}+\pmb{B}^{\top}\pmb{P}_{\pmb{K},\pmb{L}(\pmb{K})}\pmb{D}(\pmb{R}^w-\pmb{D}^{\top}\pmb{P}_{\pmb{K},\pmb{L}(\pmb{K})}\pmb{D})^{-1}\pmb{D}^{\top}\pmb{P}_{\pmb{K},\pmb{L}(\pmb{K})}\pmb{B}$. 
  Setting  $\pmb{K}'=\pmb{K}^*\in\hat{\mathcal{K}}$ and using the definition of a Nash equilibrium, we obtain
    \begin{align}
        &\mathcal{G}(\pmb{K}^*,\pmb{L}^*)-\mathcal{G}(\pmb{K},\pmb{L}(\pmb{K}))\geq \mathcal{G}(\pmb{K}^*,\widetilde{\pmb{L}}_{\pmb{K},\pmb{K}^*})-\mathcal{G}(\pmb{K},\pmb{L}(\pmb{K}))\\
        &=\Tr((2(\pmb{K}^*-\pmb{K})^{\top}\pmb{F}_{\pmb{K},\pmb{L}(\pmb{K})} +(\pmb{K}^*-\pmb{K})^{\top}M(\pmb{K}^*-\pmb{K}))\pmb{\Sigma}_{\pmb{K}^*,\widetilde{\pmb{L}}_{\pmb{K},\pmb{K}^*}})\notag\\
        &\overset{(a)}{\geq} -\Tr(\pmb{F}_{\pmb{K},\pmb{L}(\pmb{K})}^{\top}M^{-1}\pmb{F}_{\pmb{K},\pmb{L}(\pmb{K})}\pmb{\Sigma}_{\pmb{K}^*,\widetilde{\pmb{L}}_{\pmb{K},\pmb{K}^*}})\notag\\
        &\overset{(b)}{\geq} -\Tr(\pmb{F}_{\pmb{K},\pmb{L}(\pmb{K})}^{\top}(\pmb{R}^u)^{-1}\pmb{F}_{\pmb{K},\pmb{L}(\pmb{K})}\pmb{\Sigma}_{\pmb{K}^*,\widetilde{\pmb{L}}_{\pmb{K},\pmb{K}^*}})\notag\\
        &\geq -\lambda_{\min}^{-1}(\pmb{R}^u)\|\pmb{\Sigma}_{\pmb{K}^*,\widetilde{\pmb{L}}_{\pmb{K},\pmb{K}^*}}\|\Tr(\pmb{F}_{\pmb{K},\pmb{L}(\pmb{K})}^{\top}\pmb{F}_{\pmb{K},\pmb{L}(\pmb{K})}),\notag
     \end{align}
     where $(a)$ holds since $
         (\pmb{K}^*-\pmb{K}+M^{-1}\pmb{F}_{\pmb{K},\pmb{L}(\pmb{K})})^{\top}M(\pmb{K}^*-\pmb{K}+M^{-1}\pmb{F}_{\pmb{K},\pmb{L}(\pmb{K})})\succeq 0
     $. As for $(b)$, we use $\pmb{R}^w-\pmb{D}^{\top}\pmb{P}_{\pmb{K},\pmb{L}(\pmb{K})}\pmb{D}\succ 0$ since $\pmb{K}\in\mathcal{K}$. 
     Finally, since $\pmb{\Sigma}_{\pmb{K}^*,\tilde{\pmb{L}}_{\pmb{K},\pmb{K}^*}}$ is continuous w.r.t. $\pmb{K},\pmb{L}(\pmb{K})$ and $(\pmb{K},\pmb{L}(\pmb{K}))\in\hat{\mathcal{K}}\times\hat{\mathcal{L}}$, which is a compact set, $s_4\coloneqq \sup_{\pmb{K}\in\hat{\mathcal{K}}}\|\pmb{\Sigma}_{\pmb{K}^*,\widetilde{\pmb{L}}_{\pmb{K},\pmb{K}^*}}\|$ is finite. This concludes the proof. 
\end{proof}

\begin{remark}
When $\pmb{D}=0$ and $\pmb{R}^w=0$, our zero-sum LQ game boils down to a (single agent) LQR problem. Proposition 5.8 generalizes the gradient domination previously established in \cite[Lemma 3 and 11]{global_con_pg_lq} to our two-player game setting. 
\end{remark}

\begin{remark}
\label{remark:sparsity}
    The key insight to obtain a \textit{uniform} gradient domination inequality is to observe that $\mathcal{S}_2$ is ``closed'' in the sense that for any $\pmb{K},\pmb{K}'\in\mathcal{K}$, $\pmb{L}(\pmb{K})$ stays within~$\mathcal{S}_2$, which implies that  $\widetilde{\pmb{L}}_{\pmb{K},\pmb{K}'}$, defined in \eqref{def:l_kk}, stays in $\mathcal{S}_2$ as well. This nice sparsity structure is a consequence of the finite-horizon setting. 
    This is not necessarily the case under the infinite-horizon setting 
    in which a general control pair~$\pmb{K}$, $\pmb{L}$ does not necessarily enjoy a sparsity pattern like in the compact reformulation in \eqref{eq:pb-minmax_compact}. In particular, our analysis here to conclude a uniform bound for the gradient domination coefficient does not apply directly to the infinite horizon setting such as \cite{LQ_PG_NE}, which proves gradient domination only by directly assuming the iterates remain in the neighborhood of NE. 
    Obtaining such a uniform bound in the infinite horizon setting remains an open question.
\end{remark}
Now we present our main results on the last-iterate convergence guarantee. 
\begin{theorem}
\label{theorem:last_iterate_deterministic_gdmax}
    (Last-iterate linear convergence in the deterministic setting) Suppose $\pmb{K}_0\in\mathcal{K}$ and let $\epsilon >0$. Consider the nested NPG algorithm in the deterministic case: $\pmb{K}_{t+1}=\pmb{K}_t-\tau_2\pmb{F}_{\pmb{K}_t,\pmb{L}_t}$ where $\pmb{L}_t$ is the output of the inner-loop NPG algorithm such that $\mathcal{G}(\pmb{K}_t,\pmb{L}(\pmb{K}_t))-\mathcal{G}(\pmb{K}_t,\pmb{L}_t)\leq\epsilon_1$. Let the stepsize $\tau_2$ and accuracy requirement $\epsilon_1=\mathcal{O}(\epsilon)$\footnote{With this notation, we denote the dependence on $\epsilon$.} be small enough. Then the iterates converge linearly in the sense that it takes $T=\mathcal{O}(\log(1/\epsilon))$ iterations to achieve $\mathcal{G}(\pmb{K}_{T},\pmb{L}(\pmb{K}_{T}))-\mathcal{G}(\pmb{K}^*,\pmb{L}^*)\leq \epsilon$. 
\end{theorem}

The proof for Theorem \ref{theorem:last_iterate_deterministic_gdmax} is a special case of the proof for Theorem \ref{theorem:last_iterate_stochastic} and hence omitted here.

\begin{theorem}
\label{theorem:last_iterate_stochastic}
    (Last-iterate convergence in the stochastic setting) Let Assumption~\ref{assumption:noises} hold. Let $\pmb{K}_0\in\mathcal{K}$ and consider the corresponding set $\hat{\mathcal{K}}$ and the same inner-loop setting as Proposition \ref{proposition:implicit_regularization}. 
    Moreover, for any $\delta_2\in(0,1)$ and any accuracy requirement $\epsilon\geq 0$, if the estimation parameters in Algorithm \ref{alg:outer_nested_npg} satisfy $T=\mathcal{O}(\log(\epsilon^{-1}))$, $M_2=\widetilde{\mathcal{O}}(\epsilon^{-2})$, $\tau_2=\mathcal{O}(1)$, $r_2=\mathcal{O}(\epsilon^{-1/2})$, $\epsilon_1=\mathcal{O}(\epsilon)$, $\delta_1=\mathcal{O}(\delta_2/T)$, then it holds with probability at least $1-\delta_2$ that $\pmb{K}_t\in\hat{\mathcal{K}}$ for all $t=1,\cdots,T$ and $\mathcal{G}(\pmb{K}_T,\pmb{L}(\pmb{K}_T))-\mathcal{G}(\pmb{K}^*,\pmb{L}^*)\leq\epsilon$. The total sample complexity is given by $\mathcal{O}(T(T_{in}M_1+M_2))=\widetilde{\mathcal{O}}(\epsilon^{-2})$. See Appendix \ref{appendix:main_results} for a detailed version.
\end{theorem}
\begin{proof}
Apply Lemma~\ref{lemma:descent_inequality} and Proposition~\ref{proposition:gradient_domination} with proper choices of parameters:
\begin{align}
    &\mathcal{G}(\pmb{K},\pmb{L}(\pmb{K}))-\mathcal{G}(\pmb{K}^*,\pmb{L}^*)\leq \mu \Tr(\pmb{F}_{\pmb{K},\pmb{L}(\pmb{K})}^{\top}\pmb{F}_{\pmb{K},\pmb{L}(\pmb{K})}),\label{ineq:PL_cond_primal} \\
    &\mathcal{G}(\pmb{K}',\pmb{L}(\pmb{K}'))-\mathcal{G}(\pmb{K},\pmb{L}(\pmb{K})) \notag \\
    &\leq \tau_2 (c_1\cdot r_2^2+c_2\epsilon_1+c_3\cdot \|V(\widetilde{\pmb{F}}_{\pmb{K},\pmb{L}})\|)\Tr(\pmb{\Sigma}_0)-\frac{\tau_2\phi}{4}\Tr(\pmb{F}_{\pmb{K},\pmb{L}(\pmb{K})}^{\top}\pmb{F}_{\pmb{K},\pmb{L}(\pmb{K})}).\label{ineq:one_step_descent}
\end{align}
To obtain \eqref{ineq:one_step_descent}, we multiply \eqref{ineq:descent_ineq_sto} with the positive definite matrix $\pmb{\Sigma}_0$ and take the trace on both sides. Combining the above two inequalities, we obtain
\begin{align*}
    \mathcal{G}(\pmb{K}',\pmb{L}(\pmb{K}'))-\mathcal{G}(\pmb{K}^*,\pmb{L}^*)&\leq (1-\frac{\tau_2\phi}{4\mu})(\mathcal{G}(\pmb{K},\pmb{L}(\pmb{K}))-\mathcal{G}(\pmb{K}^*,\pmb{L}^*))\\
    &\quad +\tau_2 (c_1\cdot r_2^2+c_2\epsilon_1+c_3\cdot \|V(\widetilde{F}_{\pmb{K},\pmb{L}})\|)\Tr(\pmb{\Sigma}_0).
\end{align*}
    To control the deviation term in the above descent inequality, we apply Proposition~\ref{proposition:implicit_regularization} with $M_2=\tilde{\mathcal{O}}\bigl(T^2\bigr)$, $\tau_2=\mathcal{O}(1)$, $r_2=\mathcal{O}(T^{-1/2})$, $\epsilon_1=\mathcal{O}(T^{-1})$, $\delta_1=\mathcal{O}(\delta_2/T)
$. We can then ensure that $\pmb{K}_t\in\hat{\mathcal{K}}$ for all $t=1,\cdots,T$ w.p. at least $1-\delta_2$. Moreover, we have
\begin{align*}
    \mathcal{G}(\pmb{K}_T,\pmb{L}(\pmb{K}_T))&-\mathcal{G}(\pmb{K}^*,\pmb{L}^*)\leq (1-\frac{\phi\tau_2}{4\mu})^T(\mathcal{G}(\pmb{K}_0,\pmb{L}(\pmb{K}_0))-\mathcal{G}(\pmb{K}^*,\pmb{L}^*))\\
    &\quad +\sum_{t=0}^{T-1}(1-\frac{\phi\tau_2}{4\mu})^t\tau_2(c_1\cdot r_2^2+c_2\cdot \epsilon_1+c_3 V(\widetilde{\pmb{F}}_{\pmb{K}_t,\pmb{L}_t}))\Tr(\pmb{\Sigma}_0).
\end{align*} 
Setting $T=\mathcal{O}(\log \epsilon^{-1})$ yields $(1-\frac{\phi\tau_2}{4\mu})^T(\mathcal{G}(\pmb{K}_0,\pmb{L}(\pmb{K}_0))-\mathcal{G}(\pmb{K}^*,\pmb{L}^*))\leq \epsilon/2$. Furthermore, we choose $M_2=\widetilde{\mathcal{O}}(\epsilon^{-2})$, $\tau_2=\mathcal{O}(1)$ small enough, $r_2=\mathcal{O}(\epsilon^{-1/2})$, $\epsilon_1=\mathcal{O}(\epsilon)$, $\delta_1=\mathcal{O}(\delta_2/T)$ and via standard minibatch approximation (see Lemma \ref{lemma:bounded_variance}), we further have $\sum_{t=0}^{T-1}(1-\frac{\phi\tau_2}{4\mu})^t\tau_2(c_1\cdot r_2^2+c_2\cdot \epsilon_1+c_3 V(\widetilde{\pmb{F}}_{\pmb{K}_t,\pmb{L}_t}))\Tr(\pmb{\Sigma}_0)\leq \epsilon/2$ with probability at least $1-\delta_2$. Hence the total sample complexity is $\mathcal{O}(T(T_{in}M_1+M_2))=\widetilde{\mathcal{O}}(\epsilon^{-2})$.
\end{proof}
\begin{remark}
\label{remark:total_sample_complexity}
Theorem \ref{theorem:last_iterate_stochastic} significantly improves the convergence result from~\cite{kaiqing_finite_horizon}. We now compare to their result in detail. Zhang et al. \cite{kaiqing_finite_horizon} show convergence on the average natural gradient norm squared and can find a point $\pmb{K}$ such that $\mathbb E_t [\Tr(\pmb{F}_{\pmb{K}_t,\pmb{L}(\pmb{K}_t)}^{\top}\pmb{F}_{\pmb{K}_t,\pmb{L}(\pmb{K}_t)})] \leq \epsilon$.\footnote{E.g., selecting a point $\pmb K$ uniformly at random from the iterates of the outer loop.} In the model-free setting, combining their Theorems 4.1 and 4.3, we can compute that they require $T(T_{in}M_1M_2+T_{in}M_1)=\mathcal{O}(\epsilon^{-9})$\footnote{Notice that the total sample complexity for inner and outer loops together was not explicitely stated in~\cite{kaiqing_finite_horizon}, but can be inferred from their intermediate results.} samples to achieve this. Now using our gradient domination condition \eqref{ineq:PL_cond_primal}, it implies that the point found will be $\mathcal{O}(\epsilon)$-NE (in expectation), i.e. $\mathbb E_t [\mathcal{G}(\pmb{K}_t,\pmb{L}(\pmb{K}_t))-\mathcal{G}(\pmb{K}^*,\pmb{L}^*)] \leq \mu\epsilon$ using the same $\widetilde{\mathcal{O}}(\epsilon^{-9})$ samples. Such result is (by several orders) worse than our Theorem~\ref{theorem:last_iterate_stochastic}, which guarantees $\mathcal{G}(\pmb{K}_{T},\pmb{L}(\pmb{K}_T))-\mathcal{G}(\pmb{K}^*,\pmb{L}^*) \leq \epsilon$ with high probability using only $\widetilde{\mathcal{O}}(\epsilon^{-2})$ samples. 
\end{remark}

\section{Simulations}
\label{sec:simulations}

\begin{figure}
     \centering
     \begin{subfigure}[b]{0.3\textwidth}
         \centering
         \includegraphics[width=\textwidth]{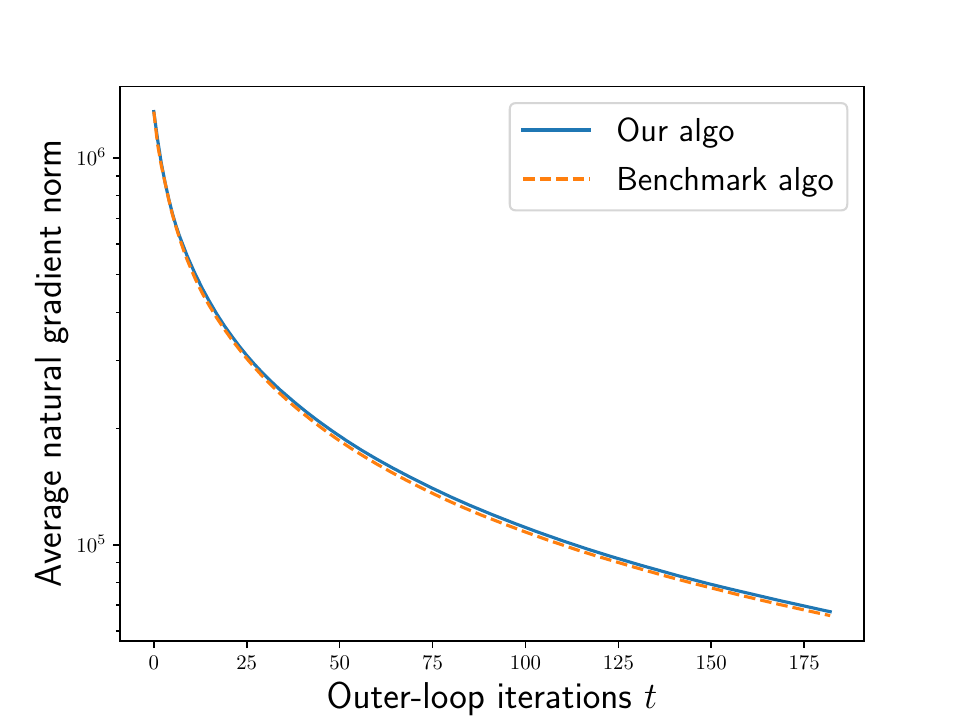}
     \end{subfigure}
     \hfill
     \begin{subfigure}[b]{0.3\textwidth}
        \centering
        \includegraphics[width=\textwidth]{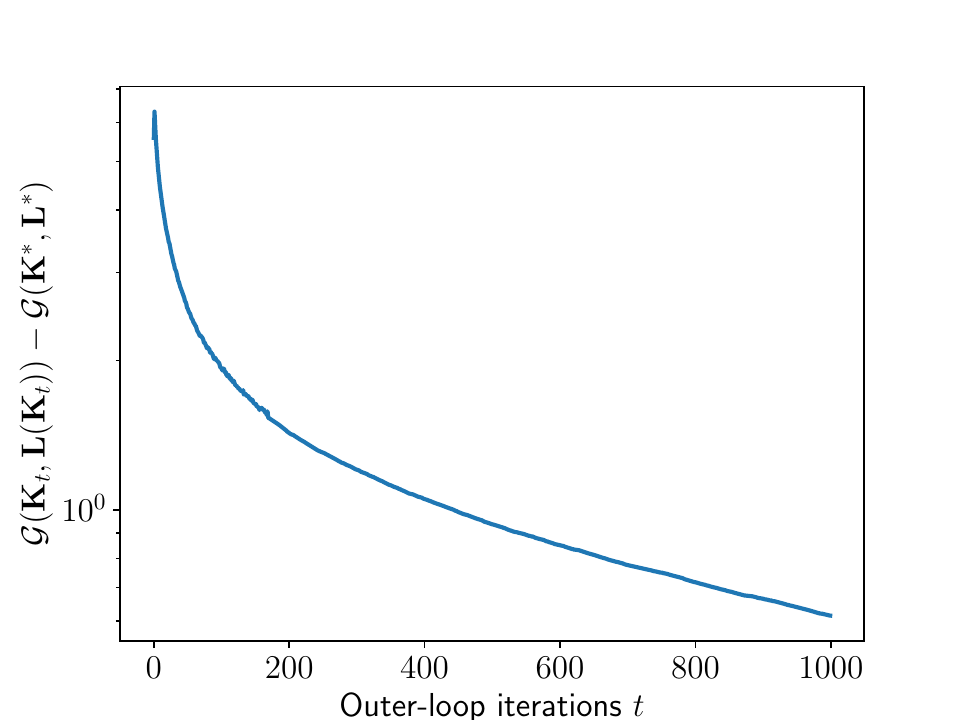}
     \end{subfigure}
     \hfill
     \begin{subfigure}[b]{0.3\textwidth}
        \centering
         \includegraphics[width=\textwidth]{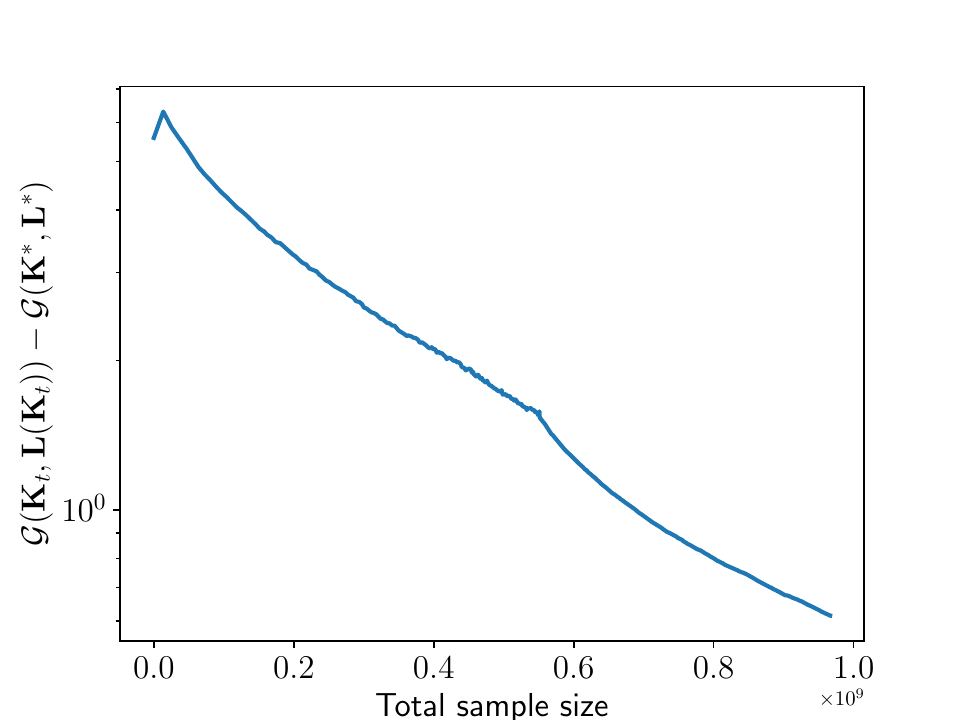}
     \end{subfigure}
     \caption{Simulation results of main results. Comparison between Algorithm \ref{alg:outer_nested_npg} and the benchmark algorithm in terms of $T^{-1}\sum_{t=0}^{T-1}\|\pmb{F}_{\pmb{K}_t,\pmb{L}(\pmb{K}_t)}\|_F^2$, exact inner-loop natural gradient, and estimated outer-loop natural gradients with $M_2=5\times 10^5$, $\tau_1=0.1$, $\tau_2=2\times 10^{-3}$, and $r_2=0.02$ (left); Last-iterate convergence results of Algorithm~\ref{alg:outer_nested_npg} when using estimated inner-loop \& outer-loop natural gradients with $r_1=0.5$, $M_1=10^6$, and $\tau_1=0.04$  (center); relationship between sample size and last-iterate convergence when using estimated inner-loop \& outer-loop natural gradients (right). 
     }
     \label{fig:simulation_res}
\end{figure}

In this section, we present simulation results\footnote{The codes can be found at \url{https://github.com/wujiduan/Zero-sum-LQ-Games.git}.} to further validate our theoretical findings. We mainly present simulation results to show 
(i) Algorithm \ref{alg:outer_nested_npg} achieves global linear last-iterate convergence.
(ii) Algorithm~\ref{alg:outer_nested_npg} is more sample-efficient compared to the benchmark algorithm.  
\paragraph{Simulation setup} All the experiments are executed with Python 3.8.5 on a high-performance computing cluster where the reserved memory for executing experiments is 2000 MB.
For the sake of comparison, we adopt the same set of model parameters as~\cite{kaiqing_finite_horizon}. We recall the setting here. The horizon length $H$ is set to 5 and $A_t=A$, $B_t=B$, $D_t=D$, $Q_t=Q$, $R_t^u=R^u$, and $R_t^w=R_w$, where $A=\begin{bmatrix}
        1 & 0 & -5\\
        -1 & 1 & 0\\
        0 & 0 & 1
    \end{bmatrix} $, $B=\begin{bmatrix}
        1 & -10 & 0\\
        0 & 3 & 1 \\
        -1 & 0 & 2
    \end{bmatrix}$, $D=\begin{bmatrix}
        0.5 & 0 & 0\\
        0 & 0.2 & 0\\
        0 & 0 & 0.2
    \end{bmatrix}$, $Q=\begin{bmatrix}
        2 & -1 & 0 \\
        -1 & 2 & -1\\
        0 & -1 & 2
    \end{bmatrix}$, $R^u=\begin{bmatrix}
        4 & -1 & 0\\
        -1 & 4 & -2\\
        0 & -2 & 3
    \end{bmatrix}$, $R^w=5\cdot I$. 
The NE solution $(\pmb{K}^*,\pmb{L}^*)$ computed by GARE yields $\mathcal{G}(\pmb{K}^*,\pmb{L}^*)\approx 3.2330$, $\lambda_{\min}(\pmb{H}_{\pmb{K}^*,\pmb{L}^*})\approx 4.2860$. For the purpose of comparison, we choose the same set of parameters for both the benchmark algorithm and Algorithm \ref{alg:outer_nested_npg} in this paper. We choose $\pmb{\Sigma}_0=0.05\cdot I$, and default values of other parameters are as follows: $r_2=0.08$, $M_2=5\times 10^5$, $\epsilon_1=10^{-4}$, $\tau_1=0.1$, $\tau_2=4.67\times 10^{-4}$, $\pmb{K}_0=\begin{bmatrix}
    \diag(K, K, K, K, K)&\pmb{0}_{15\times 3}
\end{bmatrix}$, $K\coloneqq \begin{bmatrix}
    -0.08& 0.35 & 0.62\\
    -0.21& 0.19 & 0.32\\
    -0.06 & 0.10 & 0.41
\end{bmatrix}$, $\pmb{L}_0=\pmb{0}_{15\times 18}$.
\paragraph{Last-iterate convergence} To validate the convergence results in Theorem \ref{theorem:last_iterate_deterministic_gdmax} and Theorem \ref{theorem:last_iterate_stochastic}, we conduct experiments in the stochastic setting with estimated inner-loop and estimated outer-loop natural gradients. 
In Figure \ref{fig:simulation_res} (right), the plot shows the number of trajectories sampled along the convergence of Algorithm \ref{alg:outer_nested_npg}. 
\paragraph{Sample complexity improvement} In the implementation of Algorithm \ref{alg:outer_nested_npg}, we adopt a constant number, $T_{in}$ with default value 10, of inner-loop iterations instead of assuming access to $\epsilon_1$ to determine when to terminate the inner-loop iterations\footnote{In the implementation of the benchmark algorithm, we assume exact access to the solution of the inner-loop problem given each perturbed $\pmb{K}_t^m$, i.e., $\pmb{L}(\pmb{K}_t^m)$ $m=0,\cdots,M_2-1$ for the efficiency of the simulations.}.
In Figure \ref{fig:simulation_res} (left), our algorithm shows a comparable convergence rate compared to the benchmark algorithm.
These results indicate Algorithm \ref{alg:outer_nested_npg} is more sample-efficient than the benchmark algorithm. As in the benchmark algorithm, Algorithm~\ref{alg:inner_nested_npg} is called $M_2$ times more often at each inner-loop step than Algorithm \ref{alg:outer_nested_npg}.

\section{Conclusion}
\label{sec:conclusions}

In this work, we provided a novel global last-iterate convergence result for nested algorithms for zero-sum LQ games in the deterministic setting. Moreover, we showed a~$\tilde{\mathcal{O}}(\epsilon^{-2})$ sample complexity for a derivative-free nested natural policy gradient algorithm for solving the zero-sum LQ dynamic game problem, which guarantees last-iterate convergence. Possible future research directions include~(a)~extending our analysis to continuous-time and infinite-horizon settings beyond our finite-horizon setting using techniques such as sensitivity analysis for stable continuous-time Lyapunov equations~\cite{continuous_time_sensitivity}, (b) improving the dependence on problem dimensions and considering more general noise distributions since the boundedness of noises is not required by the stability constraint under the finite-horizon setting, 
and (c) designing theoretically grounded single-loop algorithms for zero-sum LQ games.

\bibliographystyle{siamplain}
\bibliography{references}

\appendix

\vspace{5mm}
Complementary results on structural properties of Zero-sum LQ games, concentration results, and matrix inequalities are presented in Appendix \ref{appendix:preparation} in preparation for the detailed statements of the main results and their complete proofs in Appendix~\ref{appendix:main_results}. 

\section{Technical Lemma and Auxiliary Results}
\label{appendix:preparation}

The following quantities are positive and well-defined since $\hat{\mathcal{K}}$, $\hat{\mathcal{L}}$ are bounded:
    \begin{align*}
        D_1&\coloneqq \inf_{(\pmb{K},\pmb{L})\in\hat{\mathcal{K}}\times\hat{\mathcal{L}}}\min\biggl\{1,\frac{\phi}{2\Tr(\pmb{\Sigma}_{\pmb{K},\pmb{L}})\|\pmb{B}\|(2\|\pmb{A}_{\pmb{K},\pmb{L}}\|+\|\pmb{B}\|)}\biggr\},\\ D_2&\coloneqq\inf_{(\pmb{K},\pmb{L})\in\hat{\mathcal{K}}\times\hat{\mathcal{L}}}\min\biggl\{1,\frac{\phi}{2\Tr(\pmb{\Sigma}_{\pmb{K},\pmb{L}})\|\pmb{D}\|(2\|\pmb{A}_{\pmb{K},\pmb{L}}\|+\|\pmb{D}\|)}\biggr\},\\
        D_3&\coloneqq \inf_{(\pmb{K},\pmb{L})\in\hat{\mathcal{K}}\times\hat{\mathcal{L}}}\min\biggl\{1,H^{-1},H^{-1}\biggl(\frac{\phi}{2\Tr(\pmb{\Sigma}_{\pmb{K},\pmb{L}})\|\pmb{D}\|(2\|\pmb{A}_{\pmb{K},\pmb{L}}\|+\|\pmb{D}\|)}\biggr)^2\biggr\}.
    \end{align*}
\subsection{Structural Properties of Zero-sum LQ Games}\label{appendix:basic_lemmas_LQ}
This section collects some results for LQ games, some of which are similar to results in \cite{global_con_pg_lq}, \cite{LQ_PG_NE}, and~\cite{kaiqing_finite_horizon}.\\

\noindent\textbf{Existence of a unique Nash equilibrium.} 
We start by reporting from \cite{kaiqing_finite_horizon} an assumption used to guarantee the existence of a value for our zero-sum LQ game as introduced in \eqref{eq:pb-zslqr}. Based on this assumption, the existence result follows from \cite[Theorem 3.2]{linear_sol} and \cite[Theorem 6.7]{basar_noncooperative_game_1998}. 
Consider the following time-varying generalized Riccati difference equation:
	\begin{align}\label{eqn:RDE_recursive}
		P^*_h = Q_h + A^{\top}_hP^*_{h+1}\Lambda^{-1}_{h}A_h, \quad h \in \{0, \cdots, N-1\},
	\end{align}
	where $\Lambda_{h}:= I + \big(B_h(R^u_h)^{-1}B^{\top}_h - D_h(R^w_h)^{-1}D_h^{\top}\big)P^*_{h+1}$ and $P^*_N = Q_N$. 
\begin{assumption}
\label{as:existence-unique-nash}
For all $h\in\{0,\cdots,N-1\},$
$R^w_h - D^{\top}_hP^*_{h+1}D_h \succ 0$ with $P^*_{h+1} \succeq~0$.
\end{assumption}

We make this assumption throughout the paper. See also Remark~\ref{rem:existence-condition} for a more compact assumption statement. 
Under Assumption~\ref{as:existence-unique-nash}, it follows from \cite{basar_noncooperative_game_1998} that the the saddle-point control policies are linear state-feedback whenever they exist while their corresponding matrices are unique, see \cite[p. 4]{kaiqing_finite_horizon} for their expression. The value of the zero-sum LQ game exists thanks to Assumption~\ref{as:existence-unique-nash} and it is attained by the aforementioned gain matrices.  

As mentioned in \cite[remark A.5]{kaiqing_finite_horizon}, note that Assumption~\ref{as:existence-unique-nash} is also `almost' necessary besides being sufficient for the existence of a value, in the sense that if the matrices in the assumption have a negative eigenvalue, then the upper value of the game is unbounded. 
We refer the reader to \cite[remark 6.8]{basar_noncooperative_game_1998} regarding this remark.\\

The following results will be used in Appendix \ref{appendix:main_results}.

\begin{lemma}
\label{lemma:keep_in_k}
    (Lemma B.9 in \cite{kaiqing_finite_horizon}) For any $\pmb{K}\in\mathcal{K}$, there exists some $\mathcal{B}_{2,\pmb{K}}>0$ such that all $\pmb{K}'$ satisfying $\|\Delta_{\pmb{K}}\|_F\leq\mathcal{B}_{2,\pmb{K}}$ satisfy $\pmb{K}'\in\mathcal{K}$. Then for any $\pmb{K}\in\hat{\mathcal{K}}$ with $\pmb{K}_0\in\mathcal{K}$, there exists some $B_2>0$ s.t. all $\pmb{K}'$ satisfying $\|\Delta_{\pmb{K}}\|_F\leq B_2$ satisfy $\pmb{K}'\in\mathcal{K}$.
\end{lemma}

\begin{lemma}
\label{lemma:slp_continuity}
    (Lemma B.7 in \cite{kaiqing_finite_horizon}: Local Lipschitz continuity of $\pmb{\Sigma}_{\pmb{K},\pmb{L}(\pmb{K})}$
    ) For any $\pmb{K},\pmb{K}'\in\mathcal{K}$, there exist some $\mathcal{B}_{1,\pmb{K}}$,  
    $\mathcal{B}_{\pmb{\Sigma},\pmb{K}}>0$ that are continuous functions of~$\pmb{K}$ such that all $\pmb{K}'$ satisfying $\|\Delta_{\pmb{K}}\|_F\leq \mathcal{B}_{1,\pmb{K}}$ satisfy $\|\pmb{\Sigma}_{\pmb{K}',\pmb{L}(\pmb{K}')}-\pmb{\Sigma}_{\pmb{K},\pmb{L}(\pmb{K})}\|_F\leq \mathcal{B}_{\pmb{\Sigma},\pmb{K}}\cdot\|\Delta_{\pmb{K}}\|_F.$ For convenience, we define the following positive constants $B_1\coloneqq \inf_{\pmb{K}\in\hat{\mathcal{K}}}\mathcal{B}_{1,\pmb{K}}$,
        $B_{\pmb{\Sigma}}\coloneqq \sup_{\pmb{K}\in\hat{\mathcal{K}}} \mathcal{B}_{\pmb{\Sigma},\pmb{K}}.$
\end{lemma}

\begin{lemma}
\label{lemma:p_continuity}
    (Local Lipschitz continuity of $\pmb{P}_{\pmb{K},\pmb{L}}$) Let $\pmb{K}_0\in\mathcal{K}$ and consider the corresponding sets $\hat{\mathcal{K}},\hat{\mathcal{L}}$ as defined in \eqref{eq:def-hat-K},\eqref{eq:def-hat-L}. For any $(\pmb{K},\pmb{L})\in \hat{\mathcal{K}}\times\hat{\mathcal{L}}$ with structures defined in \eqref{def:k_l} and $(\pmb{K}'\in\mathcal{K},\pmb{L}')$ that satisfy $\Delta_{\pmb{K}}\leq D_1$, $\Delta_{\pmb{L}}\leq D_2$, then there exist positive constants $l_5,l_6$ such that $\|\pmb{P}_{\pmb{K}',\pmb{L}}-\pmb{P}_{\pmb{K},\pmb{L}}\|\leq l_5\|\Delta_{\pmb{K}}\|$, $\|\pmb{P}_{\pmb{K},\pmb{L}'}-\pmb{P}_{\pmb{K},\pmb{L}}\|\leq l_6\|\Delta_{\pmb{L}}\|$ where $l_5\coloneqq \sup_{(\pmb{K},\pmb{L})\in \hat{\mathcal{K}}\times\hat{\mathcal{L}}}2\Tr(\pmb{\Sigma}_{\pmb{K},\pmb{L}})/\phi \cdot((\|\pmb{K}\|+1)\|\pmb{R}^u\|+\|\pmb{R}^u\|\|\pmb{K}\|+\|\pmb{B}\|(2\|\pmb{A}_{\pmb{K},\pmb{L}}\|+\|\pmb{B}\|)\|\pmb{P}_{\pmb{K},\pmb{L}}\|)$, $l_6\coloneqq \sup_{(\pmb{K},\pmb{L})\in \hat{\mathcal{K}}\times\hat{\mathcal{L}}}2\Tr(\pmb{\Sigma}_{\pmb{K},\pmb{L}})/\phi \cdot((\|\pmb{L}\|+1)\|\pmb{R}^w\|+\|\pmb{R}^w\|\|\pmb{L}\|+\|\pmb{D}\|(2\|\pmb{A}_{\pmb{K},\pmb{L}}\|+\|\pmb{D}\|)\|\pmb{P}_{\pmb{K},\pmb{L}}\|)$
\end{lemma}
\begin{proof}
    We apply Lemma \ref{lemma:unique_sol} and the sensitivity analysis in Lemma \ref{lemma:sensitivity_analysis} with $F=\pmb{A}_{\pmb{K}',\pmb{L}}$, $M=\pmb{Q}+(\pmb{K}')^{\top}\pmb{R}^u(\pmb{K}')-\pmb{L}^{\top}\pmb{R}^v\pmb{L}$, $\pmb{H}$ being the solution of the Lyapunov equation: $I+\pmb{A}_{\pmb{K},\pmb{L}}^{\top}\pmb{H}\pmb{A}_{\pmb{K},\pmb{L}}=\pmb{H}$ and $X= \pmb{P}_{\pmb{K}',\pmb{L}}$. Then if $\|\pmb{H}\|\|\pmb{B}\Delta_{\pmb{K}}\|(2\|\pmb{A}_{\pmb{K},\pmb{L}}\|+\|\pmb{B}\Delta_{\pmb{K}}\|)<1$, we have
    \begin{align*}
        \|\pmb{P}_{\pmb{K}',\pmb{L}}&-\pmb{P}_{\pmb{K},\pmb{L}}\|\leq (1-\|\pmb{H}\|\|\pmb{B}\Delta_{\pmb{K}}\|(2\|\pmb{A}_{\pmb{K},\pmb{L}}\|+\|\pmb{B}\Delta_{\pmb{K}}\|))^{-1}\|\pmb{H}\|\\
        &\quad \cdot (\|(\pmb{K}')^{\top}\pmb{R}^u\Delta_{\pmb{K}}+\Delta_{\pmb{K}}^{\top}\pmb{R}^u\pmb{K}\|+\|\pmb{B}\Delta_{\pmb{K}}\|(2\|\pmb{A}_{\pmb{K},\pmb{L}}\|+\|\pmb{B}\Delta_{\pmb{K}}\|)\|\pmb{P}_{\pmb{K},\pmb{L}}\|)\\
        &\leq  (1-\|\pmb{H}\|\|\pmb{B}\Delta_{\pmb{K}}\|(2\|\pmb{A}_{\pmb{K},\pmb{L}}\|+\|\pmb{B}\Delta_{\pmb{K}}\|))^{-1}\|\pmb{H}\|\cdot ((\|\pmb{K}\|+\|\Delta_{\pmb{K}}\|)\|\pmb{R}^u\|\\&\quad+\|\pmb{R}^u\|\|\pmb{K}\|+\|\pmb{B}\|(2\|\pmb{A}_{\pmb{K},\pmb{L}}\|+\|\pmb{B}\|\|\Delta_{\pmb{K}}\|)\|\pmb{P}_{\pmb{K},\pmb{L}}\|)\cdot \|\Delta_{\pmb{K}}\|.
    \end{align*}
    When we apply Lemma \ref{lemma:bounded_ata} to bound $\|\pmb{H}\|$ and choose $\pmb{K}'$ such that $\|\Delta_{\pmb{K}}\|\leq D_1$, then $\|\pmb{H}\|\|\pmb{B}\Delta_{\pmb{K}}\|(2\|\pmb{A}_{\pmb{K},\pmb{L}}\|+\|\pmb{B}\Delta_{\pmb{K}}\|)\leq \|\pmb{H}\|\|\pmb{B}\|\|\Delta_{\pmb{K}}\|(2\|\pmb{A}_{\pmb{K},\pmb{L}}\|+\|\pmb{B}\|\|\Delta_{\pmb{K}}\|)<1$. We can then upperbound $\|\pmb{P}_{\pmb{K}',\pmb{L}}-\pmb{P}_{\pmb{K},\pmb{L}}\|$ with
    \begin{align*}
 2\Tr(\pmb{\Sigma}_{\pmb{K},\pmb{L}})/\phi \cdot((\|\pmb{K}\|+1)\|\pmb{R}^u\| +\|\pmb{R}^u\|\|\pmb{K}\|+\|\pmb{B}\|(2\|\pmb{A}_{\pmb{K},\pmb{L}}\|+\|\pmb{B}\|)\|\pmb{P}_{\pmb{K},\pmb{L}}\|)\cdot \|\Delta_{\pmb{K}}\|.
    \end{align*}
    For $\pmb{L}'$ and $\pmb{L}$, the proof is similar and omitted here.
\end{proof}

\begin{lemma}
    \label{lemma:ef_continuity}
    (Lipschitz continuity of $\pmb{F}_{\pmb{K},\pmb{L}}$) Let $\pmb{K}_0\in\mathcal{K}$ and consider the corresponding sets $\hat{\mathcal{K}}$, $\hat{\mathcal{L}}$ as defined in \eqref{eq:def-hat-K}, \eqref{eq:def-hat-L}. For any $(\pmb{K},\pmb{L})\in \hat{\mathcal{K}}\times\hat{\mathcal{L}}$ and $(\pmb{K}'\in\mathcal{K},\pmb{L}')$ with structures defined in \eqref{def:k_l} that satisfy $\|\Delta_{\pmb{K}}\|\leq D_1$, $\|\Delta_{\pmb{L}}\|\leq D_2$, there exist a positive constants $l_1, l_2$, such that $\|\pmb{F}_{\pmb{K}',\pmb{L}}-\pmb{F}_{\pmb{K},\pmb{L}}\|\leq l_1\|\Delta_{\pmb{K}}\|$, $\|\pmb{F}_{\pmb{K},\pmb{L}'}-\pmb{F}_{\pmb{K},\pmb{L}}\|\leq l_2\|\Delta_{\pmb{L}}\|$.
    where $l_1\coloneqq \sup_{(\pmb{K},\pmb{L})\in\hat{\mathcal{K}}\times\hat{\mathcal{L}}}l_5\|\pmb{B}\|(\|\pmb{A}_{\pmb{K},\pmb{L}}\|+\|\pmb{B}\|)+\|\pmb{R}^u\|+\|\pmb{B}\|^2\|\pmb{P}_{\pmb{K},\pmb{L}}\|$, $l_2\coloneqq \sup_{(\pmb{K},\pmb{L})\in\hat{\mathcal{K}}\times\hat{\mathcal{L}}}l_6\|\pmb{B}\|(\|\pmb{A}_{\pmb{K},\pmb{L}}\|+\|\pmb{D}\|)+\|\pmb{B}\|\|\pmb{P}_{\pmb{K},\pmb{L}}\|\|\pmb{D}\|$.
\end{lemma}

\begin{proof}
The proof is immediate from definition of $\pmb{F}_{\pmb{K},\pmb{L}}$ and Lemma \ref{lemma:p_continuity}.
\end{proof}

\begin{lemma}
\label{lemma:sigma_continuity}
    (Local Lipschitz continuity of $\pmb{\Sigma}_{\pmb{K},\pmb{L}}$) Let $\pmb{K}_0\in\mathcal{K}$ and consider the corresponding sets $\hat{\mathcal{K}}$, $\hat{\mathcal{L}}$ as defined in \eqref{eq:def-hat-K}, \eqref{eq:def-hat-L}. For any $(\pmb{K},\pmb{L})\in \hat{\mathcal{K}}\times\hat{\mathcal{L}}$ and $(\pmb{K}'\in\mathcal{K},\pmb{L}')$ with structures defined in \eqref{def:k_l} that satisfy $\|\Delta_{\pmb{K}}\|\leq D_1$, $\|\Delta_{\pmb{L}}\|\leq D_2$, there exist positive constants $l_7,l_8$ such that $\|\pmb{\Sigma}_{\pmb{K}',\pmb{L}}-\pmb{\Sigma}_{\pmb{K},\pmb{L}}\|\leq l_7\|\Delta_{\pmb{K}}\|$, $\|\pmb{\Sigma}_{\pmb{K},\pmb{L}'}-\pmb{\Sigma}_{\pmb{K},\pmb{L}}\|\leq l_8\|\Delta_{\pmb{L}}\|$ where $l_7\coloneqq \sup_{(\pmb{K},\pmb{L})\in\hat{\mathcal{K}}\times\hat{\mathcal{L}}}2 \Tr(\pmb{\Sigma}_{\pmb{K},\pmb{L}})/\phi\cdot\|\pmb{B}\|(2\|\pmb{\Sigma}_{\pmb{K},\pmb{L}}\|+\|\pmb{B}\|)\|\pmb{\Sigma}_{\pmb{K},\pmb{L}}\|$, $l_8\coloneqq \sup_{(\pmb{K},\pmb{L})\in\hat{\mathcal{K}}\times\hat{\mathcal{L}}}2 \Tr(\pmb{\Sigma}_{\pmb{K},\pmb{L}})/\phi\cdot\|\pmb{D}\|(2\|\pmb{\Sigma}_{\pmb{K},\pmb{L}}\|+\|\pmb{D}\|)\|\pmb{\Sigma}_{\pmb{K},\pmb{L}}\|$.
\end{lemma}
\begin{proof}
    Here we apply Lemma \ref{lemma:unique_sol} and Lemma \ref{lemma:sensitivity_analysis} with $F= \pmb{A}_{\pmb{K},\pmb{L}}^{\top}$, $M= \pmb{\Sigma}_0$, $X= \pmb{\Sigma}_{\pmb{K},\pmb{L}}$, and $\pmb{H}$ is the solution of the Lyapunov equation $I+\pmb{A}_{\pmb{K},\pmb{L}}\pmb{H}\pmb{A}_{\pmb{K},\pmb{L}}^{\top}=\pmb{H}$. Then if let $\|\pmb{H}\|\|\pmb{B}\Delta_{\pmb{K}}\|(2\|\pmb{A}_{\pmb{K},\pmb{L}}\|+2\|\pmb{B}\Delta_{\pmb{K}}\|)<1$, we have
\begin{align*}
    \|\pmb{\Sigma}_{\pmb{K}',\pmb{L}}-\pmb{\Sigma}_{\pmb{K},\pmb{L}}\|&\leq \frac{\|\pmb{H}\|\|\pmb{B}\Delta_{\pmb{K}}\|(2\|\pmb{\Sigma}_{\pmb{K},\pmb{L}}\|+\|\pmb{B}\Delta_{\pmb{K}}\|)\|\pmb{\Sigma}_{\pmb{K},\pmb{L}}\|}{1-\|\pmb{H}\|\|\pmb{B}\Delta_{\pmb{K}}\|(2\|\pmb{\Sigma}_{\pmb{K},\pmb{L}}\|+\|\pmb{B}\Delta_{\pmb{K}}\|)}.
\end{align*}
Hence, we apply Lemma \ref{lemma:bounded_ata} and choose $\pmb{K}'$ such that $\|\pmb{H}\|\|\pmb{B}\Delta_{\pmb{K}}\|(2\|\pmb{\Sigma}_{\pmb{K},\pmb{L}}\|+2\|\pmb{B}\Delta_{\pmb{K}}\|)\leq \|\pmb{H}\|\|\pmb{B}\|\|\Delta_{\pmb{K}}\|(2\|\pmb{\Sigma}_{\pmb{K},\pmb{L}}\|+2\|\pmb{B}\|\|\Delta_{\pmb{K}}\|)<1$, then we have $\|\pmb{\Sigma}_{\pmb{K}',\pmb{L}}-\pmb{\Sigma}_{\pmb{K},\pmb{L}}\|\leq \frac{2 \Tr(\pmb{\Sigma}_{\pmb{K},\pmb{L}})}{\phi}\cdot\|\pmb{B}\|(2\|\pmb{\Sigma}_{\pmb{K},\pmb{L}}\|+\|\pmb{B}\|)\|\pmb{\Sigma}_{\pmb{K},\pmb{L}}\|\|\Delta_{\pmb{K}}\|$. Similarly, when
$
    \|\Delta_{\pmb{L}}\|\leq D_2
$, we have $\|\pmb{\Sigma}_{\pmb{K},\pmb{L}'}-\pmb{\Sigma}_{\pmb{K},\pmb{L}}\|\leq 2 \Tr(\pmb{\Sigma}_{\pmb{K},\pmb{L}})/\phi\cdot\|\pmb{D}\|(2\|\pmb{\Sigma}_{\pmb{K},\pmb{L}}\|+\|\pmb{D}\|)\|\pmb{\Sigma}_{\pmb{K},\pmb{L}}\|\|\Delta_{\pmb{L}}\|$. The proof is concluded by using the definitions of $l_7,l_8$.
\end{proof}

The following lemma is important for controlling the estimation bias caused by using ZO estimation.
\begin{lemma}
\label{lemma:gradient_continuity}
    (Local Lipschitz continuity of $\nabla_{\pmb{K}}\mathcal{G}(\pmb{K},\pmb{L})$) Let $\pmb{K}_0\in\mathcal{K}$ and consider the corresponding sets $\hat{\mathcal{K}},\hat{\mathcal{L}}$ as defined in \eqref{eq:def-hat-K}, \eqref{eq:def-hat-L}. For any $(\pmb{K},\pmb{L})\in \hat{\mathcal{K}}\times\hat{\mathcal{L}}$ and $(\pmb{K}'\in\mathcal{K},\pmb{L})$ with structures defined in \eqref{def:k_l} that satisfy $\|\Delta_{\pmb{K}}\|\leq D_1$,  it holds that $\|\nabla_{\pmb{K}}\mathcal{G}(\pmb{K}',\pmb{L})-\nabla_{\pmb{K}}\mathcal{G}(\pmb{K},\pmb{L})\|\leq (2l_1\|\pmb{\Sigma}_{\pmb{K},\pmb{L}}\|+2l_1 l_7+2l_8\|\pmb{F}_{\pmb{K},\pmb{L}}\|) \|\Delta_{\pmb{K}}\|$ where $l_1$ is defined in Lemma \ref{lemma:ef_continuity} and $l_7,l_8$ are defined in Lemma \ref{lemma:sigma_continuity}.
\end{lemma}

\begin{proof}
    We use the explicit expressions of $\nabla_{\pmb{K}}\mathcal{G}(\pmb{K},\pmb{L})$ and write 
    \begin{align*}
        &\|\nabla_{\pmb{K}}\mathcal{G}(\pmb{K}',\pmb{L})-\nabla_{\pmb{K}}\mathcal{G}(\pmb{K},\pmb{L})\|=\|2\pmb{F}_{\pmb{K}',\pmb{L}}\pmb{\Sigma}_{\pmb{K}',\pmb{L}}-2\pmb{F}_{\pmb{K},\pmb{L}}\pmb{\Sigma}_{\pmb{K},\pmb{L}}\|\\
        &\leq 2\|\pmb{F}_{\pmb{K}',\pmb{L}}-\pmb{F}_{\pmb{K},\pmb{L}}\|\|\pmb{\Sigma}_{\pmb{K}',\pmb{L}}\|+2\|\pmb{F}_{\pmb{K},\pmb{L}}\|\|\pmb{\Sigma}_{\pmb{K}',\pmb{L}}-\pmb{\Sigma}_{\pmb{K},\pmb{L}}\|\\
        &\leq (2l_1\|\pmb{\Sigma}_{\pmb{K}',\pmb{L}}\|+2l_7\|\pmb{F}_{\pmb{K},\pmb{L}}\|)\|\Delta_{\pmb{K}}\|\leq (2l_1\|\pmb{\Sigma}_{\pmb{K},\pmb{L}}\|+2l_1 l_7+2l_8\|\pmb{F}_{\pmb{K},\pmb{L}}\|) \|\Delta_{\pmb{K}}\|,
    \end{align*}
    where in the second inequality we apply Lemma \ref{lemma:ef_continuity}, \ref{lemma:sigma_continuity} and choose $\pmb{K}'$ such that $\Delta_{\pmb{K}}\leq D_1$. In the third inequality above, we utilize the fact that $D_1\leq 1$.
\end{proof}

The following lemma shows that the output $\pmb{L}$ of the inner-loop algorithm that satisfies the accuracy requirement not only implies the boundedness of $\mathcal{G}(\pmb{K},\pmb{L})$ and~$\pmb{L}$ but also the boundedness of $\pmb{P}_{\pmb{K},\pmb{L}}$ and $\pmb{\Sigma}_{\pmb{K},\pmb{L}}$.
\begin{lemma}
\label{lemma:bounded_pkl}(Bounded output of the inner-loop algorithm)
    Let $\pmb{K}_0\in\mathcal{K}$ and consider the corresponding set $\hat{\mathcal{K}}$ as defined in \eqref{eq:def-hat-K}. Let $\pmb{K}\in\hat{\mathcal{K}}$ and $\pmb{L}$ be the output of Algorithm \ref{alg:inner_nested_npg} given $\pmb{K}$ with structures defined in \eqref{def:k_l}, which satisfies $\mathcal{G}(\pmb{K},\pmb{L}(\pmb{K}))-\mathcal{G}(\pmb{K},\pmb{L})\leq \epsilon_1$ with probability at least $1-\delta_1$ for $\delta_1\in(0,1)$. When $\epsilon_1\leq D_3$, we have $\|\pmb{P}_{\pmb{K},\pmb{L}}\| 
        \leq \|\pmb{P}_{\pmb{K},\pmb{L}(\pmb{K})}\|+\epsilon_1/\phi\leq \|\pmb{P}_{\pmb{K},\pmb{L}(\pmb{K})}\|+1/\phi$, $\|\pmb{\Sigma}_{\pmb{K},\pmb{L}}\|\leq \|\pmb{\Sigma}_{\pmb{K},\pmb{L}(\pmb{K})}\|+l_8\sqrt{\lambda_{\min}^{-1}(\pmb{H}_{\pmb{K},\pmb{L}(\pmb{K})})\cdot \epsilon_1}\leq \|\pmb{\Sigma}_{\pmb{K},\pmb{L}(\pmb{K})}\|+l_8$ with probability at least $1-\delta_1$. Here $l_8$ is defined in \ref{lemma:sigma_continuity}.
\end{lemma}

\begin{proof}
Using the triangle inequality, we have 
\begin{eqnarray*}
&\|\pmb{P}_{\pmb{K},\pmb{L}}\| \leq\|\pmb{P}_{\pmb{K},\pmb{L}(\pmb{K})}\| + \|\pmb{P}_{\pmb{K},\pmb{L}(\pmb{K})}-\pmb{P}_{\pmb{K},\pmb{L}}\| \leq  \|\pmb{P}_{\pmb{K},\pmb{L}(\pmb{K})}\| + \Tr(\pmb{P}_{\pmb{K},\pmb{L}(\pmb{K})}-\pmb{P}_{\pmb{K},\pmb{L}}) \\
&\leq \|\pmb{P}_{\pmb{K},\pmb{L}(\pmb{K})}\| + \phi^{-1} \Tr((\pmb{P}_{\pmb{K},\pmb{L}(\pmb{K})}-\pmb{P}_{\pmb{K},\pmb{L}})\pmb{\Sigma}_0)
\end{eqnarray*}
where we use the optimality of $\pmb{P}_{\pmb{K},\pmb{L}(\pmb{K})}$ in the second inequality. For the second result, since the output $\pmb{L}$ of Algorithm \ref{alg:inner_nested_npg} satisfies $\|\pmb{L}(\pmb{K})-\pmb{L}\|_F\leq \sqrt{\lambda_{\min}^{-1}(\pmb{H}_{\pmb{K},\pmb{L}(\pmb{K})})\cdot \epsilon_1}$ with probability at least $1-\delta_1$, we can apply Lemma \ref{lemma:sigma_continuity} and choose $\epsilon_1\leq D_3$ to obtain with probability at least $1-\delta_1$, $\|\pmb{\Sigma}_{\pmb{K},\pmb{L}}-\pmb{\Sigma}_{\pmb{K},\pmb{L}(\pmb{K})}\|\leq l_8\sqrt{\lambda_{\min}^{-1}(\pmb{H}_{\pmb{K},\pmb{L}(\pmb{K})})\cdot \epsilon_1}\,,$
\begin{align*}
\|\pmb{\Sigma}_{\pmb{K},\pmb{L}}\|&\leq \|\pmb{\Sigma}_{\pmb{K},\pmb{L}(\pmb{K})}\|+l_8\sqrt{\lambda_{\min}^{-1}(\pmb{H}_{\pmb{K},\pmb{L}(\pmb{K})})\cdot \epsilon_1}\leq \|\pmb{\Sigma}_{\pmb{K},\pmb{L}(\pmb{K})}\|+l_8.
\end{align*}
\end{proof}

\begin{lemma}
\label{lemma:natural_grad_bound}
    (Bound for Natural Gradients) Let $\pmb{K}_0\in\mathcal{K}$ and consider the corresponding set $\hat{\mathcal{K}}$ as defined in \eqref{eq:def-hat-K}. For $\pmb{K}\in\hat{\mathcal{K}}$ and $\pmb{L}$ be the output of Algorithm \ref{alg:inner_nested_npg} given $\pmb{K}$ with structures defined in \eqref{def:k_l}. For $\delta_1\in(0,1)$ and $\epsilon_1\leq D_3$, there exists a positive constant $B_3$ such that $\|\pmb{F}_{\pmb{K},\pmb{L}}\|\leq B_3$ with probability at least $1-\delta_1$ where $B_3\coloneqq \sup_{\pmb{K}\in\hat{\mathcal{K}}}(\|\pmb{R}^u\|+\|\pmb{B}\|^2(\|\pmb{P}_{\pmb{K},\pmb{L}(\pmb{K})}\|+1/\phi))\|\pmb{K}\|+\|\pmb{B}\|(\|\pmb{P}_{\pmb{K},\pmb{L}(\pmb{K})}\|+1/\phi) \cdot(\|\pmb{A}\|+\|\pmb{D}\|\|\pmb{L}(\pmb{K})\|+\|\pmb{D}\|)$.
\end{lemma}
\begin{proof}
    The proof is immediate from Lemma \ref{lemma:bounded_pkl} and the expression of $\pmb{F}_{\pmb{K},\pmb{L}}$.
\end{proof}

\subsection{Concentration results}
\label{appendix:minibatch_approx}

\begin{lemma}
\label{lemma:bounded_estimated_covariance_mat}
(Bounded estimated covariance matrix) For any sampled trajectory following policies $\pmb{K},\pmb{L}$ defined in \eqref{def:k_l}, we have $\|\widetilde{\pmb{\Sigma}}_{\pmb{K},\pmb{L},\pmb{\xi}}\|\leq \Tr(\pmb{\Sigma}_{\pmb{K},\pmb{L}})\cdot m(N+1)^2\vartheta/\phi$ a.s. for any initial condition $\pmb{\xi}$ that satisfies Assumption \ref{assumption:noises}. Let $\pmb{K}\in\hat{\mathcal{K}}$ and $\pmb{L}$ is the output of Algorithm \ref{alg:inner_nested_npg} given $\pmb{K}$. For $\delta,\delta_1\in(0,1),\epsilon>0$, if we choose $\epsilon_1\leq D_3$ and
 $M_2\geq M_{\pmb{\Sigma}}(\epsilon,\delta)=\mathcal{O}\left(\epsilon^{-2}\cdot\log(\frac{1}{\delta})\right)$ independent trajectories in Algorithm \ref{alg:outer_nested_npg} where 
\begin{align*}
    M_{\pmb{\Sigma}}(\epsilon,\delta)&\coloneqq\sup_{\pmb{K}\in\hat{\mathcal{K}}}\epsilon^{-2}(m^2(N+1)^2\vartheta^2/\phi+\sqrt{m(N+1)})^2(\|\pmb{\Sigma}_{\pmb{K},\pmb{L}(\pmb{K})}\|+l_8)^2\log(\frac{2d_{\pmb{\Sigma}}}{\delta})
\end{align*}
and $l_8$ is defined in Lemma \ref{lemma:sigma_continuity}, we can guarantee $\|\widetilde{\pmb{\Sigma}}_{\pmb{K},\pmb{L}}-\pmb{\Sigma}_{\pmb{K},\pmb{L}}\|\leq \|\widetilde{\pmb{\Sigma}}_{\pmb{K},\pmb{L}}-\pmb{\Sigma}_{\pmb{K},\pmb{L}}\|_F\leq \epsilon$ with probability at least $(1-\delta_1)(1-\delta)$. Setting $\epsilon\leq \phi/2$, we further have $\lambda_{\min}(\widetilde{\pmb{\Sigma}}_{\pmb{K},\pmb{L}})\geq \phi/2 $ and hence $\|\widetilde{\pmb{\Sigma}}_{\pmb{K},\pmb{L}}^{-1}\|\leq \frac{2}{\phi}$.
\end{lemma}

\begin{proof}
For any sampled $\pmb{\xi}$, consider the solution of the Lyapunov equation: $\pmb{\Sigma}_{\pmb{K},\pmb{L},\pmb{\xi}}=\pmb{A}_{\pmb{K},\pmb{L}}\pmb{\Sigma}_{\pmb{K},\pmb{L},\pmb{\xi}}\pmb{A}_{\pmb{K},\pmb{L}}^{\top}+\pmb{\xi}\pmb{\xi}^{\top}$. Using Lemma \ref{lemma:unique_sol} with Assumption \ref{assumption:noises},  we have
\begin{align*}
    \|\pmb{\Sigma}_{\pmb{K},\pmb{L},\pmb{\xi}}\|_F&\leq \|\pmb{\xi}\pmb{\xi}^{\top}\|\cdot \biggl\|\sum_{h=0}^N\pmb{A}_{\pmb{K},\pmb{L}}^h(\pmb{A}_{\pmb{K},\pmb{L}}^{\top})^h\biggr\|_F\leq \Tr(\pmb{\xi}\pmb{\xi}^{\top})\cdot \Tr(\pmb{\Sigma}_{\pmb{K},\pmb{L}})/\phi\\
    &=\Tr(\pmb{\xi}^{\top}\pmb{\xi})\cdot \Tr(\pmb{\Sigma}_{\pmb{K},\pmb{L}})/\phi\leq \Tr(\pmb{\Sigma}_{\pmb{K},\pmb{L}})\cdot m(N+1)\vartheta^2/\phi\quad a.s.,
\end{align*}
where in the second inequality, we apply Lemma \ref{lemma:bounded_ata}.
As an a.s. bounded random variable, we know $\widetilde{\pmb{\Sigma}}_{\pmb{K},\pmb{L}}$ is norm-subGaussian \cite{short_note}. 
Hence w.p. at least $1-\delta$,
\begin{align*}
    \|\widetilde{\pmb{\Sigma}}_{\pmb{K},\pmb{L}}-\pmb{\Sigma}_{\pmb{K},\pmb{L}}\|_F&=\biggl\|\frac{1}{M_2}\sum_{m=0}^{M_2-1}\pmb{\Sigma}_{\pmb{K},\pmb{L},\pmb{\xi}_m}-\pmb{\Sigma}_{\pmb{K},\pmb{L}}\biggr\|_F=\frac{1}{M_2}\biggl\|\sum_{m=0}^{M_2-1}(\pmb{\Sigma}_{\pmb{K},\pmb{L},\pmb{\xi}_m}-\pmb{\Sigma}_{\pmb{K},\pmb{L}})\biggr\|_F \\
    &\leq \frac{\sqrt{M_2}}{M_2}\cdot\left(\Tr(\pmb{\Sigma}_{\pmb{K},\pmb{L}})\cdot \frac{m(N+1)\vartheta^2}{\phi}+\|\pmb{\Sigma}_{\pmb{K},\pmb{L}}\|_F\right)\cdot\sqrt{\log(\frac{2d_{\pmb{\Sigma}}}{\delta})}.
\end{align*}
Since $\pmb{L}$ is the output of Algorithm \ref{alg:inner_nested_npg} given $\pmb{K}$, apply Lemma \ref{lemma:bounded_pkl} and set $\epsilon_1\leq D_3$ to obtain $\|\pmb{L}-\pmb{L}(\pmb{K})\|\leq \|\pmb{L}-\pmb{L}(\pmb{K})\|_F\leq \sqrt{\lambda_{\min}^{-1}(\pmb{H}_{\pmb{K},\pmb{L}(\pmb{K})})\cdot\epsilon_1}\leq D_2$ w.p. at least $1-\delta_1$. We choose
\begin{align*}
    M_2&\geq \sup_{\pmb{K}\in\hat{\mathcal{K}}}\epsilon^{-2}\left(\frac{m^2(N+1)^2\vartheta^2}{\phi}+\sqrt{m(N+1)}\right)^2(\|\pmb{\Sigma}_{\pmb{K},\pmb{L}(\pmb{K})}\|+l_8)^2\log(\frac{2d_{\pmb{\Sigma}}}{\delta}) \\
    &\geq \epsilon^{-2}(\Tr(\pmb{\Sigma}_{\pmb{K},\pmb{L}})\cdot m(N+1)\vartheta^2/\phi+\|\pmb{\Sigma}_{\pmb{K},\pmb{L}}\|_F)^2\log(\frac{2d_{\pmb{\Sigma}}}{\delta})=\mathcal{O}\left(\frac{1}{\epsilon^2}\cdot\log(\frac{2}{\delta})\right),
\end{align*}
where in the first inequality, we apply Lemma \ref{lemma:bounded_pkl} to bound $\|\pmb{\Sigma}_{\pmb{K},\pmb{L}}\|_F$ and $\Tr(M)\leq \sqrt{n}\|M\|_F$ where $M\in\mathbb{R}^{n\times n}$. In conclusion, we have $\|\widetilde{\pmb{\Sigma}}_{\pmb{K},\pmb{L}}-\pmb{\Sigma}_{\pmb{K},\pmb{L}}\|_F\leq \epsilon$ with probability at least $(1-\delta_1)(1-\delta)$. In particular, setting $\epsilon=\phi/2$ yields $\|\widetilde{\pmb{\Sigma}}_{\pmb{K},\pmb{L}}-\pmb{\Sigma}_{\pmb{K},\pmb{L}}\|\leq \|\widetilde{\pmb{\Sigma}}_{\pmb{K},\pmb{L}}-\pmb{\Sigma}_{\pmb{K},\pmb{L}}\|_F\leq \frac{\phi}{2}$. Then we conclude
\begin{align*}
\widetilde{\pmb{\Sigma}}_{\pmb{K},\pmb{L}}&= \pmb{\Sigma}_{\pmb{K},\pmb{L}}-(\widetilde{\pmb{\Sigma}}_{\pmb{K},\pmb{L}}-\pmb{\Sigma}_{\pmb{K},\pmb{L}})\geq \pmb{\Sigma}_{\pmb{K},\pmb{L}}-\|\widetilde{\pmb{\Sigma}}_{\pmb{K},\pmb{L}}-\pmb{\Sigma}_{\pmb{K},\pmb{L}}\|\cdot I
    \geq\frac{\phi}{2}\cdot I.
\end{align*}
\end{proof}

The following lemma describes the relationship between the sample size $M_2$ and the algorithm parameters $r_2,\epsilon_1$, which will be important for determining the total sample complexity.
\begin{lemma}
\label{lemma:bounded_variance}
(Natural gradient estimation variance and sample size) Let $\pmb{K}_0\in\mathcal{K}$ and consider the corresponding set $\hat{\mathcal{K}}$ as defined in \eqref{eq:def-hat-K}. Let~$\pmb{K}\in\hat{\mathcal{K}}$ and $\pmb{L}$ be the output of Algorithm \ref{alg:inner_nested_npg} given $\pmb{K}$. For $\delta\in(0,1)$, $\epsilon>0$, if we choose $r_2\leq D_1$, $\epsilon_1\leq D_3$, 
\begin{align*}
M_2&\geq \max\biggl\{M_{\pmb{\Sigma}}(\phi/2,\delta/2),M_{\pmb{\Sigma}}(\frac{\phi^2\sqrt{2\epsilon}}{4O_1},\delta/2)
    , M_V(\frac{\sqrt{2\epsilon}\phi}{4},\delta/2)\biggr\}=\mathcal{O}\left(\frac{\log(\delta^{-1})}{r_2^2\epsilon}
    \right),
\end{align*}
where $M_V(\epsilon, \delta)\coloneqq \epsilon^{-2}\left(\frac{O_2}{r_2}+O_1\right)^2\cdot \log\left(\frac{2d_{\pmb{K}}}{\delta}\right)=\mathcal{O}\left(\frac{1}{r_2^2\epsilon^2}\cdot\log(\frac{2}{\delta})\right)$,
\begin{align*}
    &O_1\coloneqq \sup_{(\pmb{K},\pmb{L})\in\hat{\mathcal{K}}\times\hat{\mathcal{L}}}(N+1)(d+m)(2l_1\|\pmb{\Sigma}_{\pmb{K},\pmb{L}}\|+2l_1 l_7+2l_7\|\pmb{F}_{\pmb{K},\pmb{L}}\|)+2\|\pmb{F}_{\pmb{K},\pmb{L}}\pmb{\Sigma}_{\pmb{K},\pmb{L}}\|_F,
\end{align*}
$O_2\coloneqq \sup_{\pmb{K}\in\hat{\mathcal{K}}}d_{\pmb{K}}(l_5+\|\pmb{P}_{\pmb{K},\pmb{L}(\pmb{K})}\|+1/\phi)\vartheta^2(N+1).$
Here $M_{\pmb{\Sigma}}(\cdot,\cdot)$ is defined in Lemma \ref{lemma:bounded_estimated_covariance_mat}. Positive constants $l_1,l_5,l_7$ are defined in Lemma \ref{lemma:p_continuity}, \ref{lemma:ef_continuity}, \ref{lemma:sigma_continuity}. Then we have $\|V(\widetilde{\pmb{F}}_{\pmb{K},\pmb{L}})\|\leq \epsilon$ with probability at least $(1-\delta_1)(1-\delta)$. 
\end{lemma}
\begin{proof}
Recall from Lemma~\ref{lemma:ori_descent_inequality} the expression of~$V(\widetilde{\pmb{F}}_{\pmb{K},\pmb{L}})$.  
Then, we use $M^{\top}N+N^{\top}M\preceq M^{\top}M+N^{\top}N$ with  $M\coloneqq \widetilde{\nabla}_{\pmb{K}}\mathcal{G}(\pmb{K},\pmb{L})\widetilde{\pmb{\Sigma}}_{\pmb{K},\pmb{L}}^{-1}-\nabla_{\pmb{K}}\mathcal{G}_{r_2}(\pmb{K},\pmb{L})\widetilde{\pmb{\Sigma}}_{\pmb{K},\pmb{L}}^{-1}$, $N\coloneqq \nabla_{\pmb{K}}\mathcal{G}_{r_2}(\pmb{K},\pmb{L})\widetilde{\pmb{\Sigma}}_{\pmb{K},\pmb{L}}^{-1}-\nabla_{\pmb{K}}\mathcal{G}_{r_2}(\pmb{K},\pmb{L})\pmb{\Sigma}_{\pmb{K},\pmb{L}}^{-1}$ to obtain 
$
    V(\widetilde{\pmb{F}}_{\pmb{K},\pmb{L}})
    \preceq M^{\top}M+ N^{\top}N
$ and 
\begin{align*}
    &\|V(\widetilde{\pmb{F}}_{\pmb{K},\pmb{L}})\|\leq \|M\|^2 +\|N\|^2\\
    &\leq\|\widetilde{\nabla}_{\mathbf{K}}\mathcal{G}(\pmb{K},\pmb{L})-\nabla_{\pmb{K}}\mathcal{G}_{r_2}(\pmb{K},\pmb{L})\|^2\cdot\|
    \widetilde{\pmb{\Sigma}}_{\pmb{K},\pmb{L}}^{-1}\|^2  +\|\nabla_{\pmb{K}}\mathcal{G}_{r_2}(\pmb{K},\pmb{L})\|^2\cdot \|\widetilde{\pmb{\Sigma}}_{\pmb{K},\pmb{L}}^{-1}-\pmb{\Sigma}_{\pmb{K},\pmb{L}}^{-1}\|^2.
\end{align*}
Now we bound the norm of $\pmb{X}\coloneqq\frac{d_{\pmb{K}}}{r_2}\mathcal{G}_{\pmb{\xi}}(\pmb{K}+r_2\pmb{V},\pmb{L})\pmb{V}-\nabla_{\pmb{K}}\mathcal{G}_{r_2}(\pmb{K},\pmb{L})$. Observe first that $\|\pmb{X}\|_F\leq (1)+(2)$ where $(1)=\frac{d_{\pmb{K}}}{r_2}\|\mathcal{G}_{\pmb{\xi}}(\pmb{K}+r_2\pmb{V},\pmb{L})\pmb{V}\|_F$ and $(2)=\|\nabla_{\pmb{K}}\mathcal{G}_{r_2}(\pmb{K},\pmb{L})\|_F.$ Then we use Assumption \ref{assumption:noises} and recall that~$\pmb{V}$ is sampled uniformly from the unit sphere. For term (1), we have
\begin{align*}
    &(1)=\frac{d_{\pmb{K}}}{r_2}\|\mathcal{G}_{\pmb{\xi}}(\pmb{K}+r_2\pmb{V},\pmb{L})\pmb{V}\|_F=\frac{d_{\pmb{K}}}{r_2}|\mathcal{G}_{\pmb{\xi}}(\pmb{K}+r_2\pmb{V},\pmb{L})|\leq \frac{d_{\pmb{K}}}{r_2}\|\pmb{P}_{\pmb{K}+r_2\pmb{V},\pmb{L}}\|\|\pmb{\xi}\pmb{\xi}^{\top}\|\\
    &\leq \frac{d_{\pmb{K}}}{r_2}(\|\pmb{P}_{\pmb{K}+r_2\pmb{V},\pmb{L}}-\pmb{P}_{\pmb{K},\pmb{L}}\|+\|\pmb{P}_{\pmb{K},\pmb{L}}\|)\|\pmb{\xi}\pmb{\xi}^{\top}\|_F\overset{(a)}{\leq} \frac{d_{\pmb{K}}}{r_2}(l_5\cdot r_2+ \|\pmb{P}_{\pmb{K},\pmb{L}}\|)\|\pmb{\xi}\pmb{\xi}^{\top}\|_F\\
    &\leq \frac{d_{\pmb{K}}}{r_2}(l_5\cdot r_2+ \|\pmb{P}_{\pmb{K},\pmb{L}}\|)\vartheta^2(N+1)\overset{(b)}{\leq} \frac{d_{\pmb{K}}}{r_2}(l_5+\|\pmb{P}_{\pmb{K},\pmb{L}(\pmb{K})}\|+1/\phi)\vartheta^2(N+1)
\end{align*}
holds with probability at least $1-\delta_1$, where $(a)$ follows from Lemma \ref{lemma:p_continuity} with $r_2\leq D_1$ and $\epsilon_1\leq D_3$, and $(b)$ is a consequence of Lemma \ref{lemma:bounded_pkl}. For term (2), we have 
\begin{align*}
    (2)
    &\leq \|\nabla_{\pmb{K}}\mathcal{G}_{r_2}(\pmb{K},\pmb{L})-\nabla_{\pmb{K}}\mathcal{G}(\pmb{K},\pmb{L})\|_F+\|\nabla_{\pmb{K}}\mathcal{G}(\pmb{K},\pmb{L})\|_F\\
    &\overset{(c)}{\leq} (N+1)(d+m)(2l_1\|\pmb{\Sigma}_{\pmb{K},\pmb{L}}\|+2l_1\cdot l_7+2l_7\|\pmb{F}_{\pmb{K},\pmb{L}}\|)\cdot r_2+2\|\pmb{F}_{\pmb{K},\pmb{L}}\pmb{\Sigma}_{\pmb{K},\pmb{L}}\|_F\\
    &\leq (N+1)(d+m)(2l_1\|\pmb{\Sigma}_{\pmb{K},\pmb{L}}\|+2l_1\cdot l_7+2l_7\|\pmb{F}_{\pmb{K},\pmb{L}}\|)+2\|\pmb{F}_{\pmb{K},\pmb{L}}\pmb{\Sigma}_{\pmb{K},\pmb{L}}\|_F\leq O_1,
\end{align*}
which holds with probability at least $1-\delta_1$, where $(c)$, stems from using Lemma \ref{lemma:gradient_continuity}. Summing the above inequalities
we obtain 
\begin{align*}
    \|\pmb{X}\|_F&\overset{(d)}{\leq} \sup_{\pmb{K}\in\hat{\mathcal{K}}}\frac{d_{\pmb{K}}}{r_2}(l_5+\|\pmb{P}_{\pmb{K},\pmb{L}(\pmb{K})}\|+1/\phi)\vartheta^2(N+1)+O_1=\frac{O_2}{r_2}+O_1,
\end{align*}
holds with probability at least $1-\delta_1$, where we apply Lemma \ref{lemma:bounded_pkl} in $(d)$. Hence when the output of Algorithm~\ref{alg:inner_nested_npg}, $\pmb{L}$, satisfies the accuracy requirement, term $\pmb{X}$ is bounded, and hence norm-subGaussian w.r.t. random variable $\pmb{V}$. Then we apply Corollary 7 in \cite{short_note} to obtain with probability at least $(1-\delta_1)(1-\delta)$,  
\begin{align*}
    &\|\widetilde{\nabla}_{\pmb{K}}\mathcal{G}(\pmb{K},\pmb{L})-\nabla_{\pmb{K}}\mathcal{G}_{r_2}(\pmb{K},\pmb{L})\|_F=\\
    &\biggl\|\frac{1}{M_2}\sum_{m=0}^{M_2-1}\left(\frac{d_{\pmb{K}}}{r_2}\mathcal{G}(\pmb{K}+r_2\pmb{V}_m,L)\pmb{V}_m-\nabla_{\pmb{K}}\mathcal{G}_{r_2}(\pmb{K},\pmb{L})\right)\biggr\|_F\leq \frac{\sqrt{M_2\log(\frac{2d_{\pmb{K}}}{\delta})} (\frac{O_2}{r_2}+O_1)}{M_2} .
\end{align*}
Hence if we set 
\begin{align}
    \label{eq:mv_sample}
    M_2&\geq \max\left\{M_{\pmb{\Sigma}}(\phi/2,\delta/2),M_V\left(\frac{\sqrt{2\epsilon}\phi}{4},\delta/2\right)\right\} =\mathcal{O}\left(\frac{1}{r_2^2\epsilon}\cdot\log(\frac{1}{\delta})\right),
\end{align}
we have $\|\widetilde{\nabla}_{\pmb{K}}\mathcal{G}(\pmb{K},\pmb{L})-\nabla_{\pmb{K}}\mathcal{G}_{r_2}(\pmb{K},\pmb{L})\|_F^2\cdot\|
    \widetilde{\pmb{\Sigma}}_{\pmb{K},\pmb{L}}^{-1}\|^2\notag\leq \left(\frac{\sqrt{2\epsilon}\phi}{4}\right)^2\cdot \left(\frac{2}{\phi}\right)^2 \leq \frac{\epsilon}{2}$
with probability at least $(1-\delta_1)(1-\delta)$. Moreover, applying Lemma \ref{lemma:bounded_estimated_covariance_mat} with $M_2\geq \max\left\{M_{\pmb{\Sigma}}(\phi/2,\delta/2),M_{\pmb{\Sigma}}(\frac{\phi^2\sqrt{2\epsilon}}{4O_1},\delta/2)\right\}=\mathcal{O}\left(\epsilon^{-1}\cdot\log(\frac{1}{\delta})\right)$, we get 
\begin{align*}
   &\|\nabla_{\pmb{K}}\mathcal{G}_{r_2}(\pmb{K},\pmb{L})\|^2\cdot \|\widetilde{\pmb{\Sigma}}_{\pmb{K},\pmb{L}}^{-1}-\pmb{\Sigma}_{\pmb{K},\pmb{L}}^{-1}\|^2
   \\
   &\leq \|\nabla_{\pmb{K}}\mathcal{G}_{r_2}(\pmb{K},\pmb{L})\|^2 \|\widetilde{\pmb{\Sigma}}_{\pmb{K},\pmb{L}}^{-1}\|^2\|\widetilde{\pmb{\Sigma}}_{\pmb{K},\pmb{L}}-\pmb{\Sigma}_{\pmb{K},\pmb{L}}\|^2\|\pmb{\Sigma}_{\pmb{K},\pmb{L}}^{-1}\|^2\leq \frac{4(O_1)^2\|\widetilde{\pmb{\Sigma}}_{\pmb{K},\pmb{L}}-\pmb{\Sigma}_{\pmb{K},\pmb{L}}\|^2}{\phi^4}\leq \frac{\epsilon}{2}
\end{align*}
with probability at least $(1-\delta_1)(1-\delta/2)$. In conclusion, by sampling 
\begin{align*}
    M_2\geq \max\biggl\{M_{\pmb{\Sigma}}(\frac{\phi}{2}, \frac{\delta}{2}), M_{\pmb{\Sigma}}(\frac{\phi^2\sqrt{2\epsilon}}{4O_1},\frac{\delta}{2}), M_V(\frac{\sqrt{2\epsilon}\phi}{4},\delta/2)\biggr\}=\mathcal{O}\left(\frac{\log(\frac{1}{\delta})}{r_2^2\epsilon}
    \right),
\end{align*}
we have $\|V(\widetilde{\pmb{F}}_{\pmb{K},\pmb{L}})\|_F\leq\epsilon$ with probability at least $(1-\delta_1)(1-\delta)$.
\end{proof}

\begin{lemma}
\label{lemma:bounded_gradient_estimates}
    (Bounded gradient estimates) Let $\pmb{K}_0\in\mathcal{K}$ and consider the corresponding set $\hat{\mathcal{K}}$ as defined in \eqref{eq:def-hat-K}. Let $\pmb{L}$ denote the output of Algorithm \ref{alg:inner_nested_npg} given $\pmb{K}\in\hat{\mathcal{K}}$, choose $r_2\leq D_1$, $\epsilon_1\leq D_3$ and $M_2\geq \max\{M_{\pmb{\Sigma}}(\phi/2,\delta/2),M_V(1,\delta/2)\}=\mathcal{O}\left(\frac{1}{r_2^2}\cdot\log(\frac{1}{\delta})+\log(\frac{1}{\delta})\right)$ where $M_{\pmb{\Sigma}}(\cdot,\cdot)$, $M_V(\cdot,\cdot)$ are defined in Lemma \ref{lemma:bounded_estimated_covariance_mat}, \ref{lemma:bounded_variance} respectively. Then for $\delta\in(0,1)$, we have $\|\widetilde{\pmb{F}}_{\pmb{K},\pmb{L}}\|_F\leq B_4$ with probability at least $(1-\delta_1)(1-\delta)$ where
    \begin{align*}
        B_4\coloneqq \sup_{(\pmb{K},\pmb{L})\in\hat{\mathcal{K}}\times\hat{\mathcal{L}}}\frac{1+ m (N+1)(2l_1\|\pmb{\Sigma}_{\pmb{K},\pmb{L}}\|+2l_1 l_7+2l_7\|\pmb{F}_{\pmb{K},\pmb{L}}\|)  + \|\nabla_{\pmb{K}}\mathcal{G}(\pmb{K},\pmb{L})\|_F}{\phi}.
    \end{align*}
    Here $l_1,l_7$ are defined in Lemma \ref{lemma:ef_continuity}, \ref{lemma:sigma_continuity}.
\end{lemma}

\begin{proof}
    We start by bounding $\|\widetilde{\nabla}_{\pmb{K}}\mathcal{G}(\pmb{K},\pmb{L})\|_F$ with 
    \begin{align*}
        \|\widetilde{\nabla}_{\pmb{K}}\mathcal{G}(\pmb{K},\pmb{L})-\nabla_{\pmb{K}}\mathcal{G}_{r_2}(\pmb{K},\pmb{L})\|_F+\|\nabla_{\pmb{K}}\mathcal{G}_{r_2}(\pmb{K},\pmb{L})-\nabla_{\pmb{K}}\mathcal{G}(\pmb{K},\pmb{L})\|_F+\|\nabla_{\pmb{K}}\mathcal{G}(\pmb{K},\pmb{L})\|_F.
    \end{align*}
    Then we use Lemma \ref{lemma:gradient_continuity} and \eqref{eq:mv_sample} in Lemma~\ref{lemma:bounded_variance}. If we choose $r_2\leq D_1$ and $M_2\geq M_V(1,\frac{\delta}{2})$, then we have with probability at least $(1-\delta_1)(1-\delta/2)$,
    \begin{align*}
        \|\widetilde{\nabla}_{\pmb{K}}\mathcal{G}(\pmb{K},\pmb{L})\|_F
        &\leq 1+ m (N+1)(2l_1\|\pmb{\Sigma}_{\pmb{K},\pmb{L}}\|+2l_1 l_7+2l_7\|\pmb{F}_{\pmb{K},\pmb{L}}\|)  + \|\nabla_{\pmb{K}}\mathcal{G}(\pmb{K},\pmb{L})\|_F.
    \end{align*}
    Applying Lemma \ref{lemma:bounded_estimated_covariance_mat} with $\epsilon_1\leq D_3$ and $M_2\geq \max\bigl\{M_{\pmb{\Sigma}}(\frac{\phi}{2},\frac{\delta}{2}),M_V(1,\frac{\delta}{2})\bigr\}$, we conclude with probability at least $(1-\delta_1)(1-\delta)$ that 
    \begin{align*}
        \|\widetilde{\pmb{F}}_{\pmb{K},\pmb{L}}\|_F&\leq \frac{1+ m (N+1)(2l_1\|\pmb{\Sigma}_{\pmb{K},\pmb{L}}\|+2l_1 l_7+2l_7\|\pmb{F}_{\pmb{K},\pmb{L}}\|)  + \|\nabla_{\pmb{K}}\mathcal{G}(\pmb{K},\pmb{L})\|_F}{\phi}\leq B_4.
    \end{align*}
\end{proof}

\subsection{Useful Technical Lemma}
\label{appendix:technical_lemma}

\paragraph{Basic Results for LQ Problems}
Below we present some well-known results for LQ problems and omit the proofs that can be found in \cite{kaiqing_finite_horizon, Ilyas_lqr}.

\begin{lemma}
\label{lemma:unique_sol}
    The solution to the Lyapunov equation $\pmb{X}=\pmb{A}_{\pmb{K},\pmb{L}}^{\top}\pmb{X}\pmb{A}_{\pmb{K},\pmb{L}}+\pmb{Z}$ (where $\pmb{Z}$ is an arbitrary real matrix with proper dimensions) is unique and has the explicit expression $\pmb{X}=\sum_{i=0}^N(\pmb{A}_{\pmb{K},\pmb{L}}^{\top})^i\pmb{Z}(\pmb{A}_{\pmb{K},\pmb{L}})^i$ for any $\pmb{K},\pmb{L}$ defined in \eqref{def:k_l}.

\end{lemma}

\begin{lemma}\label{lem:two_lyapunov}
    Let $Q_1 \succ Q_2$ and $X_1$, $X_2$ be solutions to the  Lyapunov equations: $X_1 = A^{\top} X_1 A + Q_1 $, $ X_2 = A^{\top} X_2 A + Q_2 $ where $A$ is a stable matrix. Then $X_1 \succ X_2$.
\end{lemma}

\begin{lemma}
\label{lemma:true_matrix_difference}
    (Value matrix difference -\cite[Lemma 5.1]{improved_kaiqing}) For any $\pmb{K},\pmb{K}'\in\mathcal{K}$, $\pmb{L},\pmb{L}'$ defined in~\eqref{def:k_l}, we have the following identity
    \begin{align*}
        \pmb{P}_{\pmb{K}',\pmb{L}'}-\pmb{P}_{\pmb{K},\pmb{L}}&=(\pmb{A}-\pmb{BK}'-\pmb{DL}')^{\top}(\pmb{P}_{\pmb{K}',\pmb{L}'}-\pmb{P}_{\pmb{K},\pmb{L}})(\pmb{A}-\pmb{BK}'-\pmb{DL}')\\
        &\quad +\Delta_{\pmb{K}}^{\top}\pmb{F}_{\pmb{K},\pmb{L}} + \pmb{F}_{\pmb{K},\pmb{L}}^{\top}\Delta_{\pmb{K}} +\Delta_{\pmb{K}}^{\top}(\pmb{R}^u+\pmb{B}^{\top}\pmb{P}_{\pmb{K},\pmb{L}}\pmb{B})\Delta_{\pmb{K}}\\
        &\quad +\Delta_{\pmb{L}}^{\top} \pmb{E}_{\pmb{K},\pmb{L}} + \pmb{E}_{\pmb{K},\pmb{L}}^{\top}\Delta_{\pmb{L}} +\Delta_{\pmb{L}}^{\top}(-\pmb{R}^w+\pmb{D}^{\top} \pmb{P}_{\pmb{K},\pmb{L}}\pmb{D})\Delta_{\pmb{L}}\\
        &\quad +\Delta_{\pmb{L}}^{\top}\pmb{D}^{\top}\pmb{P}_{\pmb{K},\pmb{L}}\pmb{B}\Delta_{\pmb{K}}+\Delta_{\pmb{K}}^{\top}\pmb{B}^{\top}\pmb{P}_{\pmb{K},\pmb{L}}\pmb{D}\Delta_{\pmb{L}}\,.
    \end{align*}
    where~$\Delta_{\pmb{L}} := \pmb{L} - \pmb{L}'$ and~$\Delta_{\pmb{K}} := \pmb{K} - \pmb{K}'$\,. 
\end{lemma}

\begin{proof}
    The proof follows from subtracting the two Lyapunov equations (for the pairs $(\pmb{K}, \pmb{L})$ and $(\pmb{K}',\pmb{L}')$) and performing a few algebraic computations.
\end{proof}

\begin{lemma}
\label{lemma:bounded_ata}
    (Preliminary Lemma for Bounded Perturbation) For any control pair $(\pmb{K},\pmb{L})$ defined in \eqref{def:k_l}, let $\pmb{H},\pmb{H}'$ be the solutions of the following Lyapunov equations\footnote{The solutions of the above Lyapunov equations uniquely exist by applying Lemma~\ref{lemma:unique_sol}. Hence $\pmb{H},\pmb{H}'$ are well-defined.} $I+\pmb{A}_{\pmb{K},\pmb{L}}^{\top}\pmb{H}\pmb{A}_{\pmb{K},\pmb{L}}=\pmb{H}$, $ I+\pmb{A}_{\pmb{K},\pmb{L}}\pmb{H}'\pmb{A}_{\pmb{K},\pmb{L}}^{\top}=\pmb{H}'.$ Then, we have $\|\pmb{H}\|=\|\pmb{H}'\|\leq \Tr(\pmb{\Sigma}_{\pmb{K},\pmb{L}})/\phi$.
\end{lemma}
\begin{proof}
    Consider the dual Lyapunov equation of: $\pmb{\Sigma}_{\pmb{K},\pmb{L}}=\pmb{A}_{\pmb{K},\pmb{L}}\pmb{\Sigma}_{\pmb{K},\pmb{L}}\pmb{A}_{\pmb{K},\pmb{L}}^{\top}+\pmb{\Sigma}_0$ with solution $\pmb{\Sigma}_{\pmb{K},\pmb{L}}$. Applying Lemma \ref{lemma:dual_lyapunov} yields $\Tr(\pmb{\Sigma}_{\pmb{K},\pmb{L}})=\Tr(\pmb{H}\pmb{\Sigma}_0)\geq \lambda_{\min}(\pmb{\Sigma}_0)\cdot \Tr(\pmb{H})\geq \lambda_{\min}(\pmb{\Sigma}_0)\cdot \|\pmb{H}\|_F\geq \lambda_{\min}(\pmb{\Sigma}_0)\cdot \|\pmb{H}\|$. The proof for~$\pmb{H}'$ is similar.
\end{proof}

\section{Proofs of Main Results}
\label{appendix:main_results}
\begin{lemma}
\label{lemma:detailed_descent_inequality}
    (Detailed version of Lemma~\ref{lemma:descent_inequality}) 
Let $\pmb{K}_0\in\mathcal{K}$, $\pmb{K}\in\hat{\mathcal{K}}$
 and let $\pmb{K}'=\pmb{K}-\tau_2\widetilde{\pmb{F}}_{\pmb{K},\pmb{L}}$ where $\pmb{L}$ is the output Algorithm \ref{alg:inner_nested_npg} given $\pmb{K}$. Choose $r_2\leq D_1$, $\epsilon_1\leq D_3$, $\tau_2\leq \min\biggl\{\frac{1}{4G}, \frac{B_{2}}{\sqrt{m(N+1)}B_4},\frac{B_1}{\sqrt{m(N+1)}B_4},1\biggr\}$, $M_2\geq \max\{M_{\pmb{\Sigma}}(\frac{\phi}{2},\frac{\delta}{2}),M_V(1,\frac{\delta}{2})\}$ where $B_1,B_2,B_4$ are defined in Lemma \ref{lemma:slp_continuity}, \ref{lemma:keep_in_k}, \ref{lemma:bounded_gradient_estimates},  and $M_{\pmb{\Sigma}}(\cdot,\cdot)$, $M_V(\cdot,\cdot)$ are defined in Lemma \ref{lemma:bounded_estimated_covariance_mat}, \ref{lemma:bounded_variance}. Then with probability at least $(1-\delta_1)(1-\delta)$, we have $\pmb{K}'\in\mathcal{K}$ and there exist positive constants $c_1,c_2,c_3$ such that $\pmb{P}_{\pmb{K}',\pmb{L}(\pmb{K}')}-\pmb{P}_{\pmb{K},\pmb{L}(\pmb{K})}\preceq \tau_2\cdot (c_1\cdot r_2^2+c_2\cdot \epsilon_1+c_3\cdot \|V(\widetilde{\pmb{F}}_{\pmb{K},\pmb{L}})\|)\cdot I$ where $c_1\coloneqq \sup_{(\pmb{K},\pmb{L})\in\hat{\mathcal{K}}\times\hat{\mathcal{L}}}(4+4G)/\phi^3\cdot (l_1\|\pmb{\Sigma}_{\pmb{K},\pmb{L}}\|+l_1l_7+l_7\|\pmb{F}_{\pmb{K},\pmb{L}}\|)^2 \cdot (\|\pmb{\Sigma}_{\pmb{K},\pmb{L}(\pmb{K})}\|_F+\mathcal{B}_{\pmb{\Sigma}})$, $c_2\coloneqq \sup_{\pmb{K}\in\hat{\mathcal{K}}}(l_2)^2\cdot H(\|\pmb{\Sigma}_{\pmb{K},\pmb{L}(\pmb{K})}\|_F+\mathcal{B}_{\pmb{\Sigma}})/\phi,\, c_3\coloneqq \sup_{\pmb{K}\in\hat{\mathcal{K}}}(2+2G)\cdot(\|\pmb{\Sigma}_{\pmb{K},\pmb{L}(\pmb{K})}\|_F+\mathcal{B}_{\pmb{\Sigma}})/\phi$.
\end{lemma}
\begin{proof}
Using Lemma \ref{lemma:bounded_gradient_estimates} with $M_2\geq \max\{M_{\pmb{\Sigma}}(\frac{\phi}{2},\frac{\delta}{2}),M_V(1,\frac{\delta}{2})\}$, we can ensure $\|\widetilde{\pmb{F}}_{\pmb{K},\pmb{L}}\|\leq B_4$ with probability at least $(1-\delta_1)(1-\delta)$. We can choose $\tau_2\leq \frac{\min\{1, B_{2,\pmb{K}},\mathcal{B}_{1,\pmb{K}}\}}{\sqrt{m(N+1)}B_4}$ to obtain that $\|\Delta_{\pmb{K}}\|_F\leq \min\{1, B_{2},B_{1}\}$. This ensures $\pmb{K}'\in\mathcal{K}$ with probability at least $(1-\delta_1)(1-\delta)$ by applying Lemma \ref{lemma:keep_in_k}. Recall from Lemma \ref{lemma:ori_descent_inequality} that $\pmb{e}_{1,\pmb{K},\pmb{K}'}\succeq 0$, $\pmb{e}_{2,\pmb{K},\pmb{K}'}\succeq 0$, $\pmb{e}_{3,\pmb{K},\pmb{K}'}\succeq 0$. Then we have $\pmb{e}_{1,\pmb{K},\pmb{K}'}\preceq (4\tau_{2}+4\tau_{2}^2\|\pmb{G}_{\pmb{K},\pmb{L}(\pmb{K})}\|)\|\pmb{F}_{\pmb{K},\pmb{L}}^r-\pmb{F}_{\pmb{K},\pmb{L}}\|^2\cdot I$, $\pmb{e}_{2,\pmb{K},\pmb{K}'}\preceq \tau_2\|\pmb{F}_{\pmb{K},\pmb{L}(\pmb{K})}-\pmb{F}_{\pmb{K},\pmb{L}}\|^2 I,\,\pmb{e}_{3,\pmb{K},\pmb{K}'}\preceq(2\tau_2+\tau_2^2\|\pmb{G}_{\pmb{K},\pmb{L}(\pmb{K})}\|)\|V(\widetilde{\pmb{F}}_{\pmb{K},\pmb{L}})\| I$. To bound these errors, we apply Lemma \ref{lemma:ef_continuity}, Lemma \ref{lemma:gradient_continuity} and set $r_2\leq  D_1$, $\epsilon_1\leq D_3$. 
Then we have 
    \begin{align*}
        \pmb{e}_{1,\pmb{K},\pmb{K}'}&\preceq (2\tau_{2}+2\tau_{2}^2\|\pmb{G}_{\pmb{K},\pmb{L}(\pmb{K})}\|)\|\nabla_{\pmb{K}}\mathcal{G}_{r_2}(\pmb{K},\pmb{L})\pmb{\Sigma}_{\pmb{K},\pmb{L}}^{-1}-\nabla_{\pmb{K}}\mathcal{G}(\pmb{K},\pmb{L})\pmb{\Sigma}_{\pmb{K},\pmb{L}}^{-1}\|^2\cdot I \\
        &\preceq (4\tau_{2}+4\tau_{2}^2\|\pmb{G}_{\pmb{K},\pmb{L}(\pmb{K})}\|)/\phi^2 \cdot (l_1\|\pmb{\Sigma}_{\pmb{K},\pmb{L}}\|+l_1\cdot l_7+l_7\|\pmb{F}_{\pmb{K},\pmb{L}}\|)^2r^2\cdot I\,,\\
        \pmb{e}_{2,\pmb{K},\pmb{K}'}&\preceq \tau_2\|\pmb{F}_{\pmb{K},\pmb{L}(\pmb{K})}-2\pmb{F}_{\pmb{K},\pmb{L}}\|^2\cdot I\preceq \tau_2(l_2)^2\cdot H\epsilon_1\cdot I\,.
    \end{align*}
    We obtain with probability at least $(1-\delta_1)(1-\delta)$, \begin{align*}\|\pmb{e}_{\pmb{K},\pmb{K}'}\|&\leq (4\tau_{2}+4\tau_{2}^2\|\pmb{G}_{\pmb{K},\pmb{L}(\pmb{K})}\|)/\phi^2 \cdot (l_1\|\pmb{\Sigma}_{\pmb{K},\pmb{L}}\|+l_1\cdot l_7+l_7\|\pmb{F}_{\pmb{K},\pmb{L}}\|)^2r^2\\
        &\quad +\tau_2(l_2)^2\cdot H\epsilon_1+(2\tau_2+\tau_2^2\|\pmb{G}_{\pmb{K},\pmb{L}(\pmb{K})}\|)\|V(\widetilde{\pmb{F}}_{\pmb{K},\pmb{L}})\|.
    \end{align*}
    Moreover, observe that 
    \begin{align*}
        &\sum_{t=0}^{N}(\pmb{A}_{\pmb{K}',\pmb{L}(\pmb{K}')}^{\top})^t\pmb{e}_{\pmb{K},\pmb{K}'}(\pmb{A}_{\pmb{K}',\pmb{L}(\pmb{K}')})^t
        \overset{(a)}{\preceq}\frac{\|\pmb{e}_{\pmb{K},\pmb{K}'}\|}{\phi}\sum_{t=0}^{N}(\pmb{A}_{\pmb{K}',\pmb{L}(\pmb{K}')}^{\top})^t\pmb{\Sigma}_0(\pmb{A}_{\pmb{K}',\pmb{L}(\pmb{K}')})^t\\
        &\overset{(b)}{\preceq}\frac{\|\pmb{e}_{\pmb{K},\pmb{K}'}\|}{\phi}\|\pmb{\Sigma}_{\pmb{K}',\pmb{L}(\pmb{K}')}\|\cdot I
        \overset{(c)}{\preceq} \|\pmb{e}_{\pmb{K},\pmb{K}'}\|(\|\pmb{\Sigma}_{\pmb{K},\pmb{L}(\pmb{K})}\|_F+B_{\pmb{\Sigma}}\|\Delta_{\pmb{K}}\|_F)/\phi\cdot I
    \end{align*}
    where $(a)$ stems from Lemma \ref{lemma:bounded_ata} and $(b)$ uses $\|AA^{\top}\|=\|A^{\top}A\|$. In $(c)$, we apply Lemma \ref{lemma:slp_continuity}. Hence, with probability at least $(1-\delta_1)(1-\delta)$:
    \begin{align*}
        &\pmb{P}_{\pmb{K}',\pmb{L}(\pmb{K}')}-\pmb{P}_{\pmb{K},\pmb{L}(\pmb{K})}\preceq \sum_{t=0}^{N}(\pmb{A}_{\pmb{K}',\pmb{L}(\pmb{K}')}^{\top})^t\pmb{e}_{\pmb{K},\pmb{K}'}(\pmb{A}_{\pmb{K}',\pmb{L}(\pmb{K}')})^t -\frac{\tau_{2}}{4}\pmb{F}_{\pmb{K},\pmb{L}(\pmb{K})}^{\top}\pmb{F}_{\pmb{K},\pmb{L}(\pmb{K})}\\
        &\preceq \Bigl((4\tau_{2}+4\tau_{2}^2\|\pmb{G}_{\pmb{K},\pmb{L}(\pmb{K})}\|)/\phi^2\cdot (l_1\|\pmb{\Sigma}_{\pmb{K},\pmb{L}}\|+l_1 l_7+l_7\|\pmb{F}_{\pmb{K},\pmb{L}}\|)^2r^2 +\tau_2(l_2)^2 H\epsilon_1\\
        &\quad+\|\pmb{e}_{3,\pmb{K},\pmb{K}'}\|\Bigr) \cdot (\|\pmb{\Sigma}_{\pmb{K},\pmb{L}(\pmb{K})}\|_F+\mathcal{B}_{\pmb{\Sigma}})/\phi\cdot I -\frac{\tau_{2}}{4}\pmb{F}_{\pmb{K},\pmb{L}(\pmb{K})}^{\top}\pmb{F}_{\pmb{K},\pmb{L}(\pmb{K})}\\
        &\preceq \tau_2\cdot (c_1\cdot r_2^2+c_2\cdot \epsilon_1+c_3\cdot \|V(\widetilde{\pmb{F}}_{\pmb{K},\pmb{L}})\|)-\frac{\tau_{2}}{4}\pmb{F}_{\pmb{K},\pmb{L}(\pmb{K})}^{\top}\pmb{F}_{\pmb{K},\pmb{L}(\pmb{K})}.
    \end{align*} 
\end{proof}

\begin{proposition}
\label{proposition:detailed_implicit_regularization}
    (Detailed version of Proposition \ref{proposition:implicit_regularization}) Let Assumption~\ref{assumption:noises} hold.  
Let~$\pmb{K}_0\in\mathcal{K}$ and consider the corresponding $\hat{\mathcal{K}}$ set defined in~\eqref{eq:def-hat-K}. For any~$\delta_1\in(0,1), \epsilon_1>0$ and for any~$\pmb{K}\in\mathcal{K}$, Algorithm~\ref{alg:inner_nested_npg} with single-point estimation outputs~$\pmb{L}$ such that $\mathcal{G}(\pmb{K},\pmb{L}(\pmb{K}))-\mathcal{G}(\pmb{K},\pmb{L})\leq \epsilon_1$ with probability at least $1-\delta_1$ using~$M_1=\widetilde{O}(\epsilon_1^{-2})$ samples. 
Moreover for any~$\delta_2\in(0,1)$ and any integer~$T \geq 1$, if the estimation parameters in Algorithm~\ref{alg:outer_nested_npg} satisfy $\tau_2\leq B_{\tau_2}$, $r_2\leq B_{r_2}$, $\epsilon_1\leq B_{\epsilon_1}$, $\delta_1\leq B_{\delta_1}$, $M_2\geq B_{M_2}=\widetilde{\mathcal{O}}\bigl(T^2\bigr)$ where $B_{\tau_2}\coloneqq \min\left\{\frac{\lambda_{\min}(\pmb{H}_{\pmb{K}_0,\pmb{L}(\pmb{K}_0)})}{6\|\pmb{D}\|^2},\frac{1}{4G}, \frac{B_{2}}{\sqrt{m(N+1)}B_4},\frac{B_1}{\sqrt{m(N+1)}B_4},1\right\}$, $B_{r_2}\coloneqq \min\left\{  D_1,\sqrt{\frac{1}{Tc_1}}\right\}$, $B_{\epsilon_1}\coloneqq \min\left\{D_3, \frac{1}{Tc_2}\right\}$, $B_{\delta_1}\coloneqq \frac{\delta_2}{2T}$, $B_{M_2}\coloneqq \max\Bigl\{M_{\pmb{\Sigma}}\left(\frac{\phi}{2},\frac{\delta_2}{4T}\right)$, $M_{\pmb{\Sigma}}\left(\frac{\phi^2}{4O_1}\cdot\sqrt{\frac{2}{c_3T}},\frac{\delta_2}{4T}\right), M_V\left(\frac{\phi}{4}\cdot\sqrt{\frac{2}{c_3T}},\frac{\delta_2}{4T}\right)\Bigr\}$ where $M_{\pmb{\Sigma}}(\cdot,\cdot),M_{V}(\cdot,\cdot)$ are defined in Lemma \ref{lemma:bounded_estimated_covariance_mat} and \ref{lemma:bounded_variance} respectively, $B_1,B_2,B_4,$ are defined in Lemma \ref{lemma:slp_continuity}, \ref{lemma:keep_in_k}, \ref{lemma:bounded_gradient_estimates}, $c_1,c_2,c_3$ are defined in Lemma \ref{lemma:detailed_descent_inequality}. Then, it holds with probability at least $1-\delta_2$ that $\pmb{K}_t\in\hat{\mathcal{K}}$ for all $t=1,\cdots,T$.
\end{proposition}

\begin{proof}
We firstly consider one step update from $\pmb{K}_0\in\mathcal{K}$ to $\pmb{K}_1$ with $\pmb{K}_1=\pmb{K}_0-\tau_2\widetilde{\pmb{F}}_{\pmb{K}_0,\pmb{L}_0}$ where $\pmb{L}_0$ is the output of Algorithm \ref{alg:inner_nested_npg} given $\pmb{K}_0$.  
\noindent Use Lemma~\ref{lemma:detailed_descent_inequality} with $r_2\leq  D_1$, $\epsilon_1\leq D_3$, $\tau_2\leq \min\Bigl\{\frac{1}{4G}$, $\frac{B_{2}}{\sqrt{m(N+1)}B_4}$, $\frac{B_1}{\sqrt{m(N+1)}B_4},1\Bigr\},$
     $M_2\geq \max\Bigl\{M_{\pmb{\Sigma}}\left(\frac{\phi}{2},\frac{\delta}{2}\right)$, $M_V\left(1,\frac{\delta}{2}\right)\Bigr\}.$ Then with probability at least $(1-\delta_1)(1-\delta)$, we obtain $\pmb{K}'\in\mathcal{K}$ and $\pmb{P}_{\pmb{K}_1,\pmb{L}(\pmb{K}_1)}-\pmb{P}_{\pmb{K}_0,\pmb{L}(\pmb{K}_0)}\preceq \tau_2\cdot (c_1\cdot r_2^2+c_2\cdot \epsilon_1+c_3\cdot \|V(\widetilde{\pmb{F}}_{\pmb{K}_0,\pmb{L}_0})\|)\cdot I.$
To further constrain $\pmb{K}_1$ within the set $\hat{\mathcal{K}}$, we choose small $r_2$, $\tau_2$, and large $M_2$ such that $V(\widetilde{\pmb{F}}_{\pmb{K}_0,\pmb{L}_{\pmb{K}_0}})$ is small: we set $c_1 r_2^2\leq 1$, $c_2 \epsilon_1\leq 1$, $c_3 \|V(\widetilde{\pmb{F}}_{\pmb{K}_0,\pmb{L}_0})\|\leq 1$. Additionally, let
$   \tau_2\leq \frac{\lambda_{\min}(\pmb{H}_{\pmb{K}_0,\pmb{L}(\pmb{K}_0)})}{6\|\pmb{D}\|^2}
$,
then we have $\pmb{P}_{\pmb{K}_1,\pmb{L}(\pmb{K}_1)}\preceq \pmb{P}_{\pmb{K}_0,\pmb{L}(\pmb{K}_0)}+\frac{\lambda_{\min}(\pmb{H}_{\pmb{K}_0,\pmb{L}(\pmb{K}_0)})}{2\|\pmb{D}\|^2}\cdot I.$ Hence $\pmb{K}_1\in\hat{\mathcal{K}}$ is guaranteed. Now we apply this reasoning recursively by choosing $\tau_2\leq B_{\tau_2}$, $r_2\leq B_{r_2}$, $\epsilon_1\leq B_{\epsilon_1}$, $\delta_1\leq B_{\delta_1}$, $M_2\geq B_{M_2}$. 
In particular, we apply Lemma~\ref{lemma:bounded_variance} for the choice of $M_2$ to control $\|V(\widetilde{\pmb{F}}_{\pmb{K},\pmb{L}_0})\|$ and we obtain with probability at least $1-\delta_2$ that $c_1 r_2^2 T\leq 1$, $c_2 \epsilon_1 T\leq 1$, $c_3 \sum_{t=0}^{T-1}\|V(\widetilde{ \pmb{F}}_{\pmb{K}_t,\pmb{L}_t})\|\leq 1$. Summing over $t=0,\cdots,T$,
we obtain with probability at least~$1-\delta_2$ for $t=0,\cdots,T$, we conclude the proof.
\end{proof}

\begin{theorem}
\label{theorem:detailed_last_iterate_stochastic} (Detailed version of Theorem \ref{theorem:last_iterate_stochastic}) Let Assumption \ref{assumption:noises} hold. Let $\pmb{K}_0\in\mathcal{K}$ and consider the corresponding $\hat{\mathcal{K}}$ set defined in \eqref{eq:def-hat-K} and the same inner-loop setting as Proposition~\ref{proposition:implicit_regularization}. 
Moreover for any $\delta_2\in(0,1)$ and any accuracy requirement $\epsilon\geq 0$, let the estimation parameters in Algorithm \ref{alg:outer_nested_npg} satisfy $\tau_2< \min\{\frac{4\mu}{\phi},B_{\tau_2}\}$, $r_2\leq \min\{B_{r_2},\sqrt{\phi\epsilon/(24 c_1\mu\Tr(\pmb{\Sigma}_0))}\}$, $\epsilon_1\leq \min\{B_{\epsilon_1},\phi\epsilon/(24c_2\mu\Tr(\pmb{\Sigma}_0))\}$, $\delta_1\leq B_{\delta_1}$, $M_2\geq\max\Bigl\{B_{M_2}$,$M_{\pmb{\Sigma}}(\frac{\phi}{2},\frac{\delta_2}{4T})$, $M_{\pmb{\Sigma}}\Bigl(\sqrt{\frac{\epsilon\phi^5}{192 \mu c_3\Tr(\pmb{\Sigma}_0)O_1^2}}$,$\frac{\delta_2}{4T}\Bigr)$, $ M_V\Bigl(\sqrt{\frac{\epsilon\phi}{192 \mu c_3\Tr(\pmb{\Sigma}_0)}}$,$\frac{\delta_2}{4T}\Bigr)\Bigr\}$ $=\widetilde{\mathcal{O}}\left(\epsilon^{-2}\right)$, $T\geq \frac{\log (\epsilon/2(\mathcal{G}(\pmb{K}_0,\pmb{L}(\pmb{K}_0))-\mathcal{G}(\pmb{K}^*,\pmb{L}^*)))}{\log(1-\phi\tau_2/4 \mu)} $ where $O_1,M_{\pmb{\Sigma}}(\cdot,\cdot),M_{V}(\cdot,\cdot)$ are defined in Lemma \ref{lemma:bounded_estimated_covariance_mat} and \ref{lemma:bounded_variance}, $B_1,B_2,B_4$ are defined in Lemma \ref{lemma:slp_continuity}, \ref{lemma:keep_in_k}, \ref{lemma:bounded_gradient_estimates}, $c_1,c_2,c_3$ are defined in Lemma \ref{lemma:detailed_descent_inequality}, and $\mu$ is defined in Proposition \ref{proposition:gradient_domination}, $B_{\tau_2},B_{\delta_1}$ are defined in~\ref{proposition:detailed_implicit_regularization}, $\mu$ is defined in Proposition \ref{proposition:gradient_domination}. Then, it holds with probability at least $1-\delta_2$ that $\pmb{K}_t\in\hat{\mathcal{K}}$ for all $t=1,\cdots,T$ and we have $\mathcal{G}(\pmb{K}_T,\pmb{L}(\pmb{K}_T))-\mathcal{G}(\pmb{K}^*,\pmb{L}^*)\leq \epsilon$ with probability at least $1-\delta_2$ using a total sample complexity $\mathcal{O}\left(T(T_{in}M_1+M_2)\right)=\mathcal{O}\left(T\cdot\epsilon^{-2}\right)=\widetilde{\mathcal{O}}\left(\epsilon^{-2}\right)$.
\end{theorem}

\end{document}